\documentclass[a4paper,11pt]{article}
\usepackage{import}
\usepackage{jheppub}
\pdfoutput=1 

\usepackage{jheppub} 

\usepackage[T1]{fontenc} 

\usepackage{amssymb, amsmath, amsthm, amsfonts, mathrsfs}

\usepackage{subfigure}

\title{\boldmath Resurgence in the Bi-Yang-Baxter Model}

\author[a]{Lucas Schepers}
\author[a,b]{and Daniel C. Thompson}
\affiliation[a]{Physics Department, Swansea University, Singleton Campus, SA2 8PP Swansea, United Kingdom}
\affiliation[b]{Theoretische Natuurkunde, Vrije Universiteit Brussel, and the International Solvay Institutes, Pleinlaan 2, B-1050 Brussels, Belgium}

\emailAdd{L.Schepers.988532@swansea.ac.uk}
\emailAdd{D.C.Thompson@swansea.ac.uk}

\abstract{We study the integrable bi-Yang-Baxter deformation of the $SU(2)$ principal chiral model (PCM) and its finite action uniton solutions. Under an adiabatic compactification on an $S^1$, we obtain a quantum mechanics with an elliptic Lam\'e-like potential.   

We perform a perturbative calculation of the ground state energy in this quantum mechanics to large orders obtaining an asymptotic series.  Using the Borel-Pad\'e technique, we determine the expected locations of  branch cuts in the Borel plane of the perturbative series and show that they match the values of the uniton actions.    Therefore, we can match the non-perturbative contributions to the energy with the uniton solutions which fractionate upon adiabatic compactification. 

An off-shoot of the WKB analysis, is to identify the quadratic differential of this deformed PCM with that of an $\mathcal{N}=2$ Seiberg-Witten theory. This can be done either as an $N_f=4$ $SU(2)$ theory  or as an elliptic quiver $SU(2)\times SU(2)$ theory.  The mass parameters of the gauge theory are given by the deformation parameters of the PCM. }

\graphicspath{ {Images/} }

\begin{document} 
\maketitle
\flushbottom

\section{Introduction} \label{sec:intro}

The task of computing exactly the values of observables in an interacting theory is typically, and certainly  in the absence of simplifications afforded by   supersymmetry or integrability, a difficult problem. Perturbation theory may be the only viable recourse to this  and indeed can be capable of making predictions of astonishing accuracy e.g. \cite{Hanneke:2008tm,Aoyama:2017uqe}.    However, a fundamental limitation of such approaches is that the resultant perturbative series will often have a zero radius of convergence. Commonly, we consider some coupling constant $z=g^2$, and perform perturbation theory around $z\approx 0$ for some observable $\mathcal{O}$:
\begin{equation} \label{eq:asex}
    \mathcal{O}(z) = \sum_{n=0}^\infty a_n z^n,
\end{equation}
where $a_n$ will go like $n!A^{-n}$ for a very general class of systems. In QFTs, the origin of this can sometimes be anticipated from the similar factorial growth in Feynman diagrams with the order of perturbation theory\cite{Bender:1971gu,Brezin:1976vw,LeGuillou:1990nq}.   The question then arises what meaning -- if any -- should be ascribed to formal asymptotic perturbative expansions?    

Starting with pioneering work of Bogomolny and Zinn-Justin  \cite{bogomolny1980calculation,ZinnJustin:1980uk}   it has become clear that actually far from being meaningless,  a great deal of information is actually deeply encoded in the asymptotic expansion.  For instance, a growth $a_n = n! (A)^{-n}$ indicates that the theory contains a non-perturbative object (instanton, renormalon, uniton etc) that enters with an action $S_\text{instanton}=A$.   In this scenario, the use of Borel summation to resum the perturbative series will lead to ambiguities (as we shall see later the ray $z\in \mathbb{R}^+$ corresponds to a Stokes line).    Crucially however, this ambiguity can be precisely cancelled by the inclusion of a leading order contribution arising from the non-perturbative saddle.   

Now subleading, in $\frac{1}{n}$, contributions to $a_n$   encode information about fluctuations around this non-perturbative saddle.    With sufficient dedication   one could then establish, from the perturbative saddle alone, that the perturbation series around the non-perturbative saddle will itself typically be asymptotic with a growth indicative of further non-perturbative sectors. Ambiguities in resummation here will be cancelled by a higher non-pertubative sector.   

The cancellations behave in a very specific way. The perturbative sector saddle $[0]$ is void of any instantons, but receives contributions from instanton-anti-instanton events $[\mathcal{I}\bar{\mathcal{I}}]$ and their higher order cousins $[\mathcal{I}^n\bar{\mathcal{I}}^n]$. The single instanton $[\mathcal{I}]$ is to interact with the members of its conjugacy class $\{ [\mathcal{I}^{n+1}\bar{\mathcal{I}}^n] \}$. This information is also often captured using the ``resurgence triangle'' \cite{dunne2012resurgence, dunne2014uniform}. Two instanton configurations that cancel each other's ambiguities are said to be in the same sector and are hence put in the same of column of the resurgence triangle. 

This leads to the idea of resurgence; that deeply encoded in the perturbative expansion lies all the non-perturbative information.   Physical observables appropriately combine contributions from the perturbative sector and relevant non-perturbative sectors into a {\em  trans-series}, introduced by \'Ecalle \cite{ecalle1981iii},  in such a fashion that all ambiguities are  cancelled.  The relative weighting of the contributions to the trans-series can undergo discrete jumps as $z$ is varied - this is similar to the Stokes jump phenomenon.

In the context of quantum mechanics, the interrelation between Borel resummation and the Stokes phenomenon are crucial in the understanding of the WKB approximation. The information captured in the Voros symbols \cite{voros1983return} undergoes jumps encoded by the Delabaere-Dillinger-Pham \cite{DDP93} formula. This information can be visualised by understanding the mutations of Stokes graphs (detailed in section \ref{sec:stokesgraphs})  \cite{delabaere1997exact, delabaere1999resurgent,Bridgeland,iwaki2014exact}.  Algebraically, we can capture this information using a Stokes automorphism. The celebrated work by Gaiotto, Moore and Neitzke \cite{gaiotto2010four, gaiotto2013wall}, connects the very same information to the wall-crossing phenomena in ${\cal N}=2$ four-dimensional gauge theories.  The ideas of resurgence have by now been applied beyond quantum mechanics, to include string theory, gauge theory and matrix models \cite{dunne2012resurgence, dunne2014uniform, dunne2017wkb, basar2015resurgence, aniceto2011resurgence, aniceto2016resurgence, marino2007nonperturbative, marino2008nonperturbative, Dorigoni:2017smz, basar2013resurgence, kontsevich2020analyticity}.
 This sampling of works inevitably does not do justice to the large body of work on this topic and we recommend the reader to consult the  review articles of \cite{dorigoni2014introduction,aniceto2018primer, aniceto2017asymptotics, sauzin2014introduction} both for their pedagogical presentation and wider bibliography.  

Our main focus in this paper is to understand how the ideas of resurgence can be applied in quantum field theories.  To retain a degree of control we choose to work in the setting of 1+1 dimensional field theories,  that happen to be integrable (although in this work integrability will not be employed in a crucial fashion).     The overall aim here is to expose the interrelation between the asymptotic nature of perturbation theory and the non-perturbative sector.   With a direct study of the large order QFT using Feynman diagrams not viable there are two directions one could follow here.  First one could exploit the exact integrability of these models and study the resurgent properties of the TBA system as in \cite{volin2010mass,marino2019renormalons,marino2019resurgence,marino2020renormalons}.  A second approach, first used by \cite{cherman2015decoding} and the one we adopt here, is to consider a reduction of the system to a quantum mechanics where a large order perturbative expansion can be carried out directly. In this approach, adiabaticity, achieved essentially by including a twist in the reduction, is used to argue that the lower dimensional theory still encapsulates the key feature of the higher dimensional one.  Following this approach, it is possible to identify  two-dimensional non-perturbative field configurations (so called unitons rather than instantons in the cases we study) as the origin of the objects that give rise to factorial behaviour in the reduced QM.  This is a crucial first step in establishing the resurgent nature of the QFT.  

In this work we shall specialise to a particular QFT, called the bi-Yang Baxter model.  This theory, introduced by Klim\v{c}ik \cite{klimcik2014integrability}, deforms the principal chiral model (PCM) on a group manifold $G$ with two deformation parameters, denoted by $\eta$ and $
\zeta$, whilst the underlying integrability is preserved.   When $G= SU(2)$, which will be our specific concern here, it was shown in \cite{hoare2014deformations} that the theory is equivalent to one already introduced by Fateev \cite{fateev1996sigma}.  There are a few motivations for studying this particular scenario.  First from a resurgence perspective it offers an access to having multiple parameters that can be dialled to expose interesting features.   Second, we shall see very explicitly that resurgent structure will require consideration of saddle configurations in a complexified field space.   Third, when the two deformation parameters are set equal to each other, $\eta =\zeta = \varkappa$ which we call the critical line, the deformed $SU(2)$ theory is equivalent \cite{hoare2014deformations} to the so-called $\eta$-deformation of $S^3$ viewed as a coset $SO(4)/SO(3)$. This provides an entry point to consider similar deformations of $AdS_5 \times S^5$ \cite{hoare2015towards, delduc2014integrable, hollowood2014integrable} which are of interest since they are thought to encode quantum group deformations in holography.   A resurgence perspective was given in \cite{demulder2016resurgence} for the case with only one parameter, i.e. $\zeta=0$. Here we find whilst some features remain, the inclusion of a further deformation parameter enriches the story quite considerably. 

Let us briefly summarise the findings of our study:
 \begin{itemize}
     \item The bi-Yang Baxter model admits  finite action field configurations that generalise the   uniton configurations  introduced for the PCM by Uhlenbeck \cite{uhlenbeck1989harmonic} and whose role in resurgence was expounded in \cite{cherman2015decoding,demulder2016resurgence}.   In addition there are finite action field configurations that take values in the complexified target space (i.e. consist of complexified field configurations).
     
     \item Upon a certain twisted $S^1$ reduction these configurations are seen, in specific regimes of their moduli space, to break up, or fractionate, into distinct lumps that resemble instanton-anti-instanton pairs or complex instanton configurations.  
     
     \item The twisted spatial reduction of the model results in a quantum mechanics with an elliptic potential 
     \begin{equation}\label{eq:Vintro}
     V(w) = \text{sd}^2(w)(1+(\zeta-\eta)^2\text{sn}^2(w))\,,
     \end{equation}
     in which the modular parameter $m= \frac{4 \eta \zeta}{1+(\eta + \zeta)^2}$.  Taking one of the parameters to zero, the Whittaker-Hill potential studied in \cite{demulder2016resurgence} is recovered. Moreover, along the critical line $\eta=\zeta=\kappa$, the potential reduces to that studied by \cite{basar2013resurgence}. Looking at the co-critical line $\eta=-\zeta$, we recover the potential studied by \cite{basar2015resurgence}. This new system thus interpolates between already known systems.
     
     \item The large order behaviour of the perturbation theory of the ground state energy  gives rise,  using a Borel-Pad\'e transformation, to poles in the Borel plane that are located precisely at the values of the action for the above uniton configurations.    Commensurate to this we find Stokes rays in the $\vartheta= 0,\pi$ directions of the Borel plane, and these are reflected as flip mutations of the corresponding Stokes graph. 
     
     \item The  $\zeta=\eta =\varkappa$ critical line is distinguished by a discontinuous jump in which the Borel pole associated to the one-complex uniton disappears and instead the leading pole in the $\vartheta=\pi$ ray corresponds to a two-complex uniton. At the special point $\varkappa = \frac{1}{2}$, which corresponds to an enhanced $\mathbb{Z}_2$ symmetry,  the real uniton and two-complex uniton have actions of equal modulus indicating a perfect cancellation in which   the perturbative ground state energy becomes a series in $g^4$ rather than $g^2$.  This provides a nice field theory example of resonate behaviour in resurgence\footnote{The potential Equation \eqref{eq:Vintro} at the critical line was indeed used as an proto-typical example to study  such resonances in a 0-dimensional toy example in \cite{aniceto2018primer}. }. 
     
     \item The WKB quadratic differential corresponding to the potential in Equation  \eqref{eq:Vintro} can be equated to the quadratic differential of ${\cal N} =2$ gauge theories in two realisations. First as the elliptic $SU(2)\times SU(2)$ quiver with one of the gauge couplings sent to infinity and with the relative Coloumb branch parameter set to zero.  Second as the $SU(2)$ $N_f =4 $ theory with pairwise equal flavour masses.  In both cases, the masses are described by the quantum-group parameters of the bi-Yang-Baxter model.     
 \end{itemize}
 
The structure of the remainder is as follows: in Section \ref{sec:Model}  we provide a summary of the model under consideration before identifying the uniton configurations in Section \ref{sec:unitons}.  We perform the reduction to quantum mechanics in Section \ref{sec:reduce} and perform a detailed perturbative analysis of this in Section  \ref{sec:WKB}. We end the story by establishing the linkage to the ${\cal N}=2 $ gauge theory in Section \ref{sec:SW}. We close with a discussion of a number of possible future directions.

\section{Defining the Model}
\label{sec:Model}
In this section we shall review some basic properties of the principal chiral model (PCM) and the Yang-Baxter (YB) deformations.

\subsection{Lagrangian }
The action of the undeformed PCM is
\begin{equation}
   S_{PCM} = \frac{1}{2 \pi t}\int d^2\sigma \, \mathscr{L}[g] \,, \quad  \mathscr{L}[g]=\text{Tr}\big(g^{-1}\partial_+g g^{-1} \partial_-g \big)\,.
\end{equation}
Here, $g$ is a map from the world-sheet into a group manifold $G$. The integral is over some world-sheet, which is spanned by lightcone coordinates $\sigma_\pm=\frac{1}{2}(t\pm x)$. We will later transition to a Euclidean signature with holomorphic coordinates $z= \frac{1}{2}(t+ ix)$ and $\bar{z}= \frac{1}{2}(t - ix)$. Derivatives with respect to light cone coordinates are denoted respectively by $\partial_\pm$. Note that $\partial_\pm g$ lives in the tangent space and is a Lie algebra $\mathfrak{g}$ valued form such that  $g^{-1}\partial_\pm g$ is the left-invariant Maurer-Cartan one-form.

We will be considering a system with a bi-Yang-Baxter deformation. To define this theory we introduce the Yang-Baxter operator $R$, which satisfies the modified Yang-Baxter equation
\begin{equation}
    [RA, RB]-R([RA,B] +[A,RB])=[A,B]\,, \quad \forall A,B \in \mathfrak{g}\, . 
\end{equation}
  Its existence implies that we can define a new Lie bracket which satisfies the Jacobi identity and is anti-symmetric (i.e. it defines a homomorphism of Lie algebras)
\begin{equation}
    [A,B]_R:=[RA,B]+[A,RB]\,.
\end{equation}

In this paper, we will specialise to the special case $G=SU(2)$ and we choose a basis $t_i = \frac{1}{\sqrt{2}} \sigma_i$ of the algebra. 
A concrete solution for the Yang-Baxter operator can then be given by
\begin{equation}
    R=\begin{pmatrix}
    0 & -1 & 0 \\
    1 & 0 & 0 \\
    0 & 0 & 0 \end{pmatrix}\,.
\end{equation}
Furthermore, we let $\text{Ad}_g(u)=gug^{-1}$ denote the adjoint operator and we define $R^g = \text{Ad}_{g^{-1}} \circ R \circ \text{Ad}_g$. The action with deformation parameters $\eta$ and $\zeta$, which we sometimes combine into $\chi_\pm= \zeta \pm \eta$,  is given by 
\begin{equation} \label{eq:doubledeflag}
    S_{\zeta, \eta} = \frac{1}{2 \pi t}\int d^2\sigma \, \mathscr{L}[g]\,, \quad \mathscr{L}[g]=\text{Tr}\bigg(g^{-1}\partial_+g\frac{1}{1 -\eta R - \zeta R^g}g^{-1}\partial_-g \bigg)\,.
\end{equation}
 We introduce the notation
\begin{equation}\label{eq:currentnotation}
    J_\pm= \mp(1\pm \eta R \pm \zeta R^g)^{-1} g^{-1}\partial_\pm g \,,
\end{equation}
because the field equations and a the Bianchi identity corresponding to the action \eqref{eq:doubledeflag} can be more easily understood in terms of $J_\pm$.

\subsection{Classical Lax Structure }
Klim\v{c}\'ik \cite{klimcik2014integrability} showed that the following Lax pair with spectral parameter, $\lambda$,
\begin{equation}
    L_\pm(\lambda) = \eta  (R-i) J_\pm +\frac{\left(2 i \eta \pm(1 +\eta ^2-\zeta ^2) \right)}{\lambda \pm 1}J_\pm\,,
\end{equation}
satisfies a zero-curvature condition
\begin{equation}
    \partial_+L_-(\lambda)-\partial_-L_+(\lambda)+[L_-(\lambda),L_+(\lambda)] = 0\,, \qquad \forall \lambda\in\mathbb{C}\,.
\end{equation}
This condition both follows from and implies the equation of motions and the Bianchi identity corresponding to the action \eqref{eq:doubledeflag}. 
 
 \subsection{The Critical Line }
As already noted by Klim\v{c}\'ik \cite{klimcik2014integrability}, the above formulation hides a certain symmetry between $\eta$ and $\zeta$. In particular, when  $\eta=\zeta\equiv \varkappa$, a situation that we shall refer to as the \textit{critical line}, there is a restoration of a $g\rightarrow g^{-1}$ symmetry.  Using the definitions   of $Ad_g$ and $R^g$, it is easy to verify the Lagrangian for the action    \eqref{eq:doubledeflag}, can be written in two equivalent ways.  Either it can be written in terms of left invariant forms, $g^{-1}\partial_\pm g$, as
\begin{equation}
   \mathscr{L}^L_{\zeta,\eta}[g]=\text{Tr}\bigg(g^{-1}\partial_+g\frac{1}{1 -\eta R - \zeta R^g}g^{-1}\partial_-g \bigg)\,, 
\end{equation}
or else in terms of right invariant forms, $\partial_\pm g g^{-1}$, as 

\begin{equation}
   \mathscr{L}^L_{\zeta,\eta}[g]=\mathscr{L}^R_{\zeta,\eta}[g]:=\text{Tr}\bigg(\partial_+gg^{-1}\frac{1}{1 -\eta R^{g^{-1}} - \zeta R}\partial_-g g^{-1} \bigg)\,.
\end{equation}
However, if we perform the transformation $g\rightarrow g ^{-1}$ of the left acting Lagrangian, we see that
\begin{equation}
   \mathscr{L}^L_{\zeta,\eta}[g^{-1}]=\text{Tr}\bigg(\partial_+gg^{-1}\frac{1}{1 -\eta R - \zeta R^{g^{-1}}}\partial_-g g^{-1} \bigg) = \mathscr{L}^R_{\eta,\zeta}[g]= \mathscr{L}^L_{\eta,\zeta}[g]\,.
\end{equation}
Therefore we see that along the critical line $\mathscr{L}^L_{\varkappa,\varkappa}[g^{-1}] = \mathscr{L}^L_{\varkappa,\varkappa}[g]$. This enhanced symmetry has profound effects on the physics, and we shall revisit this scenario many times in the rest of the paper.  We shall see in particular that the perturbative structure changes discontinuously on and off the critical line.   

The critical line also has a second important feature:  the  $SU(2)$ model on the critical line $\eta=\zeta$ is equivalent to the single parameter $\eta$-deformation of the sigma-model on $S^3$ viewed as a coset $SO(4)/SO(3)$ following the construction in   \cite{hoare2015towards, delduc2014integrable}.  This is quite useful since it allows the current study, restricted to the critical line, to have relevance to the behaviour of the deformations of general $\eta$-deformed cosets, and potentially to the full $\eta$-deformation of the $AdS_5\times S^5$ string. 

  The case of $\eta=-\zeta$, which we describe as the co-critical line, will be discussed shortly in the context of the $SU(2)$ model.

\subsection{Classical Symmetries} \label{sec:classicalcurrents}
The undeformed PCM Lagrangian with group $G$ has a global $G_L \times G_R$ symmetry acting as $g\mapsto h_Lgh_R$. The two deformations break this symmetry down to an abelian subgroup.  This is augmented by non-local charges that furnish a Possion-bracket realisation of the quantum group $\mathcal{U}_{q_L}(\mathfrak{g}) \times \mathcal{U}_{q_R}(\mathfrak{g})$.  For the single parameter Yang-Baxter, or $\eta$-deformation, for which only $G_L$ is $q$-deformed and $G_R$ is preserved,  this was demonstrated first in the context of $G= SU(2)$ in  \cite{Kawaguchi:2011pf} and  shown in general  \cite{delduc2013classical}.   The quantum group structure in the case of two-deformation parameters studied here was described in \cite{hoare2015towards}.  Although beyond the current scope, it would be remiss not to mention that Lagrangian descriptions  exist for quantum group deformed symmetries of the full $AdS_5 \times S^5$ superstring  viewed as a $\mathbb{Z}_4$ graded super-coset  \cite{magro2014integrable}.  Here $q$ is real, but somewhat parallel to this have been the construction of $q$ a root-of-unity integrable deformations of the $AdS_5 \times S^5$ superstring  \cite{hollowood2014integrable} which extend the bosonic $\lambda$-deformations introduced in  \cite{Sfetsos:2013wia}.

 Let us study this in detail for $G=SU(2)$ in Minkowskian signature. We will parametrise the group element through Euler angles by
\begin{equation} \label{eq:su2par}
    g= \begin{pmatrix}
    \cos(\theta) e^ {i \phi_1} & i\sin(\theta) e^ {i \phi_2} \\
    i\sin(\theta) e^ {-i \phi_2} & \cos(\theta) e^ {-i \phi_1} \\
    \end{pmatrix} \,,
\end{equation}
where $\theta$, $\phi_1$ and $\phi_2$ are fields taking values in $[0,\pi]$, $[0,\pi]$ and $[0,2\pi]$ respectively. Under the  $U(1)_L \times U(1)_R$ action $\delta g = \epsilon_L  t_3 \cdot g +  \epsilon_R g \cdot t_3 \, ,$ such that $\delta \phi_1+ \delta \phi_2 = \epsilon_L$ and $\delta \phi_1 - \delta \phi_2 = \epsilon_R $. 

The charges are then given by
\begin{equation}
\mathfrak{Q}^3_{L/R} = \int d^\sigma \mathfrak{j}^3_{L/R} \, ,
\end{equation}
with
\begin{equation}
\begin{aligned} \label{eq:minkcurrrents}
   \mathfrak{j}^3_{L} =  \frac{1}{\Delta(\theta)}\left(- \eta\sin(2 \theta)\theta^\prime + \cos(\theta)^2 a_{+}(\theta) \dot\phi_1 +  \sin(\theta)^2 a_{-}(\theta) \dot\phi_2 \right) \,,\\
   \mathfrak{j}^3_{R} =  - \frac{1}{\Delta(\theta)} \left(\zeta \sin(2 \theta)\theta^\prime + \cos(\theta)^2 b_{+}(\theta) \dot\phi_1 +  \sin(\theta)^2 b_{-}(\theta) \dot\phi_2 \right) \,,
\end{aligned}
\end{equation}
the corresponding currents. Here primes and dots denote spatial and temporal derivatives respectively and for convenience we have defined
\begin{equation} \label{eq:abdcoeffs}
\begin{aligned}
    a_\pm(\theta) &=\zeta ^2+\eta  (\zeta \pm \eta ) \cos (2 \theta ) \pm \zeta  \eta 
   +1\,, \\
   b_\pm (\theta)& = \zeta  (\zeta \pm \eta ) \cos (2 \theta )+\zeta  \eta  \pm \eta
   ^2 \pm 1\,, \\
   \Delta(\theta) &= \zeta ^2+\eta ^2+1 +2 \zeta  \eta  \cos (2 \theta )\,.
\end{aligned}
\end{equation}
We will later return to these Noether currents when we perform a twisted reduction of the theory.  

Whilst these $U(1)$ currents define the only {\em local} Noether charges $\mathfrak{Q}^3_{L/R}$,  a crucial property of these models \cite{delduc2013classical} is that they exhibit some non-local conserved charges $\mathfrak{Q}^\pm_{L/R}$ which furnish the algebra under Poisson brackets
\begin{equation}
\begin{aligned}
\{ \mathfrak{Q}_{L/R}^+  , \mathfrak{Q}_{L/R}^-\} = i \frac{q_{L/R}^{\mathfrak{Q}^3} - q_{L/R}^{-\mathfrak{Q}^3} }{q_{L/R} - q_{L/R}^{-1} } \ , \quad \{ \mathfrak{Q}_{L/R}^\pm  , \mathfrak{Q}_{L/R}^3\} = \pm i \mathfrak{Q}_{L/R}^\pm  \, , \\
q_{L } = \exp{ \frac{2 \pi} {\sigma_\zeta}  } \, ,\,  \quad q_{R } = \exp{ \frac{2 \pi} {\sigma_\eta}   } \, , 
\end{aligned}
\end{equation}
where $\sigma_{\eta,\zeta}$ are given by Equation \eqref{eq:renorminv}. In this way the full $G_L \times G_R$ symmetry is recovered, but deformed to have the structure of (a classical version of) a quantum group. 

\subsection{Quantum Integrability}

Although we shall not make direct use of it here, for completeness we briefly recall the quantum S-matrix of the theory.   Based on the above symmetry structure, it is natural to anticipate that the theory has quantum integrability.
Recall that the undeformed PCM on a group has an S-matrix \cite{Wiegmann:1984ec} reflecting the classical $G_L \times G_R$ symmetry with the factorised form 
 \begin{equation}
 \mathbb{S} (\theta) =     S(\theta) \otimes     S(\theta)\, , 
\end{equation} 
in which $\theta$ is the rapidity, and $S(\theta)$ are $G$-invariant S-matrix blocks.   In the deformed theory, the   quantum S-matrix will still take  a factorised form, but with both left and right factors reflecting the  $q$-deformed symmetry.  

For the case of $G=SU(2)$ this was made precise by Fateev \cite{fateev1996sigma} where the S-matrix takes the form 
\begin{equation} \label{eq:smatrix1}
\mathbb{S}_{p_1 , p_2} (\theta) =     S_{\gamma=p_1}^{SG}(\theta) \otimes    S_{\gamma=p_2}^{SG}(\theta) \, .
\end{equation}
Here the building blocks are Sine-Gordon S-matrices \cite{zamolodchikov1977exact}\footnote{The relation to the quantum group structure of these blocks was detailed in \cite{Bernard:1990ys}.}  for which the soliton-soliton scattering phase is given by 
\begin{equation} \label{eq:smatrix2}
   S_{\gamma }^{SG}(\theta)= \exp i \int_{0}^\infty \frac{d \omega}{ \omega}\sin \theta \omega \frac{\sinh( \pi \omega ( \gamma-1) /2) }{\cosh(\pi \omega /2) \sinh (\pi \gamma \omega /2 ) } \, . 
\end{equation}
 A matching of the parameters in \cite{fateev1996sigma}  to those used here is given by\footnote{Here of course the relation between the $p_i$ in the S-matrix and the classical Lagrangian parameters could be renormalised at higher loops.}  $
p_1 = 2\pi  \sigma_\eta \, , \quad p_2 = 2\pi  \sigma_\eta \, .  
$

\subsection{RG Equations}

The sigma-model is renormalisable in the couplings $t,\eta,\zeta$ with RG invariants \cite{sfetsos2015generalised},
\begin{equation} \label{eq:renorminv}
    \sigma_\eta = \frac{1}{t \eta}\,,  \qquad    \sigma_\zeta = \frac{1}{t \zeta}  \,,
\end{equation}
and a non trivial flow\footnote{Here we are presenting the result for $SU(2)$ but the change to $SU(N)$ simply introduces a factor of the the quadratic Casimir on the right hand side of the flow. } (at one-loop) 
\begin{equation} 
    \frac{d}{d\log \mu} t = -\frac{1}{2} t^2 ( 1+ (\eta +\zeta)^2) ( 1+ (\eta -\zeta)^2) \,,
\end{equation}
whose parametric solution  is given by 
\begin{equation}
    \log \mu/\mu_0 = \frac{\sigma_\zeta +\sigma_\eta}{2}  \arctan\left(\frac{\sigma_\eta \sigma_\zeta t}{\sigma_\zeta +\sigma_\eta}   \right)-\frac{\sigma_\zeta -\sigma_\eta}{2}  \arctan\left(\frac{\sigma_\eta \sigma_\zeta t}{\sigma_\zeta -\sigma_\eta}   \right) \, .  
\end{equation}
There is a single real fixed point at the origin $t=\eta =\zeta =0$ but in the complex plane there are   lines of fixed points 
\begin{equation}
    \eta +\zeta = \pm i \, ,\quad  \eta - \zeta = \pm i\, . 
\end{equation}
The critical line is preserved by the RG flow and, analytically continued, intersects these at a special fixed point
\begin{equation}
    \eta= \zeta =  \frac{i}{2}\, . 
\end{equation}

To understand the significance of the RG flows and the imaginary fixed points it is helpful to consider the case of the $SU(2)$ model. The analysis of \cite{hoare2015towards} makes three important observations relevant to us.\footnote{We thank Ben Hoare for communications on these points.}  The bi-Yang-Baxter Lagrangian can be viewed as a non-linear sigma model in a target space equipped with a pure gauge   B-field and metric  given by
\begin{equation}
\begin{split}
      ds^2 = \frac{1}{1+\chi_+^2(1-r^2) + \chi_-^2 r^2}\left[ \frac{dr^2}{1-r^2} + (1-r^2)(1+\chi_+^2 (1-r^2) d\phi_2^2\right. \\ \left.\qquad \qquad   + r^2(1+\chi_-^2 r^2) d\phi_1^2 +2 \chi_-\chi_+ r^2(1-r^2) d\phi_1  d\phi_2 \right]  \, , 
\end{split}
\end{equation}
in which we have used  the Euler angles of Equation \eqref{eq:su2par} and defined $ r = \sin\frac{\theta}{2}$. The first observation is that demanding that the metric be regular and real allows not only $\chi_\pm = \zeta \pm \eta \in \mathbb{R}$ but also pure imaginary\footnote{In general such imaginary parameters would result in an imaginary two-form, but in the SU(2) case this two-form is pure gauge.}  $\chi_\pm = i k_\pm $ with $\mid k_\pm \mid < 1 $.   

Next we can see from the metric that there is, in addition to the $\mathbb{Z}_2$ action $g \rightarrow g^{-1}$ with $\eta \leftrightarrow \zeta$, a second $\mathbb{Z}_2$ invariance 
\begin{equation} \label{eq:PCMflipeta}
\theta \rightarrow \theta + \pi \, , \quad \phi_1 \rightarrow \phi_2 \, , \quad \phi_2 \rightarrow \phi_1 \, , \quad (\zeta, \eta) \rightarrow(\zeta, -\eta) \, .
\end{equation} 
In the case of real parameters,  which we will mostly consider here, this allows us to restrict our attention to    $\eta\in\mathbb{R}^{+}$. Note also that this transformation maps the critical $\eta=\zeta$ line to the co-critical $\eta=-\zeta$ line. 

Finally, and most remarkably, along the imaginary RG fixed points, the target space geometry coincides\footnote{With $\chi_\pm = i k_\pm$ this limit is obtained by setting $k_-=1$ and shifting $\Phi \to \Phi' + k_+ \Psi$, such that $\Psi$ parameterises the free $U(1)$ factor.} with that of an $SU(1,1)/U(1)$ gauged WZW CFT together with a free $U(1)$ boson.  The interpretation of this fixed point is the same on the critical line\footnote{Even if $\eta$ and $\zeta$ are complexified, we will refer to $\eta=\zeta$ as the critical {\em line} in the complex sense, rather than the critical plane.}  (which recall matches the $\eta$-deformation of $S^3$ viewed as a coset) at the point  $  \eta= \zeta =  \frac{i}{2}$.  When considered in the context of the $\eta$-deformation of the $AdS_3 \times S^3$ superstring, the same limit of imaginary deformation parameter is shown to   give rise to the Pohlmeyer reduced theory   \cite{hoare2014deformations}.  
\section{Uniton Solutions} \label{sec:unitons}

We now study non-perturbative field configurations, i.e. exact classical solutions of the Euclidean theory with finite action, analogous to instantons.   At first sight this may seem counter intuitive since there is no obvious topological protection (recall that $\pi_2(G) = 0$) and it is far from obvious that these are good vacua to expand around in a Quantum Field Theory.  However in a seminal early work by Uhlenbeck \cite{uhlenbeck1989harmonic}, classes of such solutions were found and classified for the principal chiral model.  These solutions are known as  {\em unitons} due to the additional constraint $g^2 = - \text{Id}$ and have played a prominent role in recent attempts \cite{cherman2015decoding, demulder2016resurgence} to elucidate the resurgent quantum structure of two-dimensional quantum field theories.  

\subsection{Real Unitons}
 In the Hopf angle parametrisation of the group element \eqref{eq:su2par}, we find a solution to the Euclidean equations of motion given by
\begin{subequations} \label{eq:realuniton}
\begin{align} 
    \label{eq:phiuniton}
    \phi_1=\frac{\pi}{2}, \qquad \phi_2=\pi+ \frac{i}{2} \log \bigg( \frac{f}{\overline{f}} \bigg), \qquad  \theta(f, \overline{f})=\theta(|f|^2), \\
    \label{eq:thetauniton}
    \sin(\theta(|f|^2))^2 = \frac{4 |f|^2}{ (1+|f|^2)^2+(\eta-\zeta)^2(1-|f|^2)^2} =:P(|f|^2) \, , 
\end{align}
\end{subequations}
with  $f(z)$   any holomorhpic function of the Euclidean coordinate $z=x+ti$. Interestingly, the solution can be obtained simply from that of the single deformed case constructed in \cite{demulder2016resurgence} by substituting $\eta^2 \rightarrow (\eta-\zeta)^2$, although this change is not at all apparent from the equations of motion.  The peculiarity of the critical line $\eta =\zeta=\varkappa$ is apparent already at this level; in this situation the uniton solution does not depend on the deformation parameter at all (although the on-shell value of the action will of course depend on $\varkappa$).  

Usually, the topological classification of saddle points in non-linear sigma models with target space $M$ depends on $\pi_2(M)$. However, in the present case we have that $\pi_2 ( SU(2)) = \pi_2 ( S^3) = 0$. From the uniton solutions \eqref{eq:phiuniton}, we see that the uniton is the embedding of a Riemann sphere into a particular $S^2 \subset SU(2)$. The discretisation of the uniton action can be connected to the homotopy group $\pi_1(\mathcal{M})$ of the field space $\mathcal{M}=\{g:S^2 \rightarrow SU(2) \}$. Therefore, the unitons are classified by $\pi_1(\mathcal{M})=\pi_3 ( SU(2) )= \mathbb{Z}$, see also \cite{manton2004topological}. Here this quantisation is reflected in the order of the polynomial $f(z)$. 

Whilst this uniton is not a bona-fide BPS protected solution, for the reasons described above, the solution does satisfy a first order ODE pseudo-BPS condition
\begin{equation} \label{eq:BPS}
    4 x^2(\theta'(x))^2=\sin^2\theta(x)+(\eta-\zeta)^2\sin^4\theta(x), \quad x=|f|^2.
\end{equation}

\begin{figure}[tbp!]
    \centering
        \subfigure[$\eta=0, \zeta=0$]{
        \includegraphics[height=1.2in]{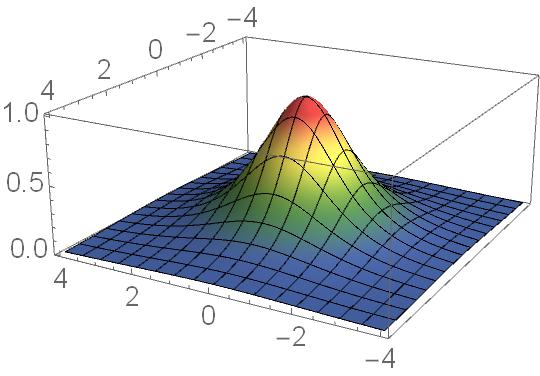}
        }
     \subfigure[$\eta=0.8, \zeta=0$]{
        \includegraphics[height=1.2in]{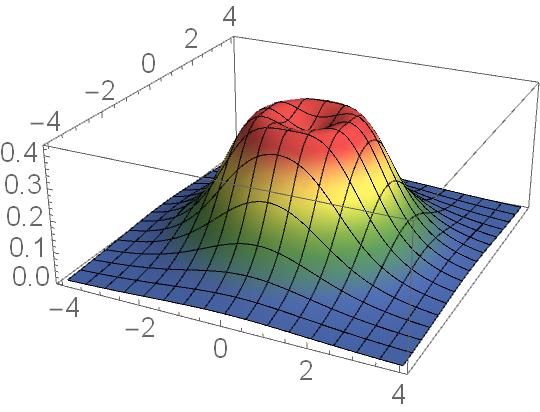}
        }
    \subfigure[$\eta=0.8, \zeta=0.8$]{
        \includegraphics[height=1.2in]{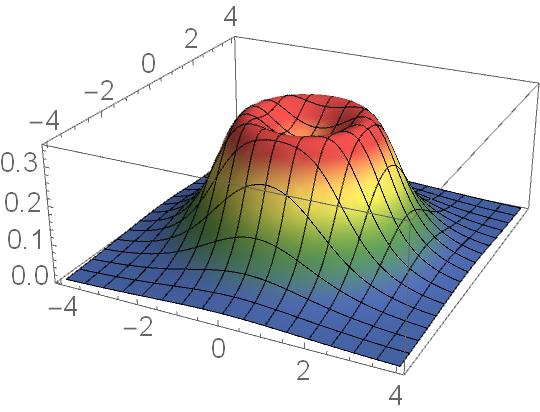}
        }
    \caption{A plot of the Lagrangian density of the $k=1$ real uniton on $\mathbb{R}^2$ illustrating the flattening out as the deformation paramaters are tuned up. The moduli are fixed in these plots such that the uniton is centered at the origin, $\lambda_0=0$, while $\lambda_1=1/2$.}
    \label{fig:realunitonrealpars}
\end{figure}

In Figure \ref{fig:realunitonrealpars}, we illustrate the Lagrangian density of this uniton configuration, for the case that $k$, the degree of $f(z)$, is one.  That is, we take $f(z)=\lambda_0 + \lambda_1z$, where $\lambda_i$ are some moduli that become significant later.  The uniton solution appears as a lump of localised Lagranian density.  The deformation parameters induce some additional structure, qualitatively described by punching a depression and flattening out the lump. 

By substituting this solution into the action \eqref{eq:doubledeflag}, we find
\begin{equation}
    S=\frac{2}{\pi t}\int d^2z \frac{|f(z)|^2|f'(z)|^2 (\theta'(|f(z)|^2))^2}{1 + \eta^2+\zeta^2 +2 \eta\zeta\cos2\theta(|f(z)|^2)} \,.
\end{equation}
To proceed, the integration coordinate is switched from $z$ to $w=f(z)$. The order $k$ of the polynomial $f(z)$ appears as it revolves $k$ times around its integration domain. We can integrate over the argument of $w$, which yields $2\pi$. By changing the  integration variable to $\theta(|w|^2)$, and by making use of  Equation \eqref{eq:BPS} the action evaluates to
\begin{equation}
    S = \frac{2k}{t(1+\chi_+^2)} S_I\,, 
\end{equation}
with\footnote{The notation $S_I$ is perhaps confusing, we have chosen this to be in keeping with other works in the field in which the subscript $I$ is meant to invoke instantons.} 
\begin{equation} \label{eq:realunitonaction}
    S_I=\frac{2}{m} \left( \chi_+ \text{arctan}\chi_+ - \chi_- \text{arctan}\chi_-\right)\,,
\end{equation}
where we recall that $\chi_\pm=\zeta\pm\eta$, and we have defined
\begin{equation} \label{eq:m1}
    m = \frac{4 \eta \zeta}{1+(\eta+\zeta)^2}\,,
\end{equation}
the significance of which will become clear later. Similarly, we have rather artificially extracted a factor of $1+\chi_+^2$ from the action for reasons that will follow later. Observe that $S_I$ is real and positive if $\eta$ and $\zeta$ are real and positive.

Moreover, note that in this formulation, $S_I$ reduces to $1+(\eta+\eta^{-1})\,\text{arctan}(\eta)$ in the single deformation limit $\zeta\rightarrow0$, matching the result of \cite{demulder2016resurgence}.

Another way of of describing the solution is through a projector $\Pi$ obeying $\Pi^2 =\Pi $. We let
\begin{equation}
    g= i ( 2 \Pi - \text{Id} )\,, \quad \implies \quad  g^2= - \text{Id}\,,
\end{equation}
and $\Pi$ given by 
\begin{equation}
\Pi = \frac{ v^\dag \otimes  v}{ v^\dag\cdot v}\,, \quad  
    v = \left( \begin{array}{c}
          1  \\
           \sqrt{\frac{\bar{f}}{f}} \frac{1 + \sqrt{P(|f|^2)}  }{\sqrt{1- P(|f|^2)} } 
    \end{array}\right)\,,
\end{equation}
where $P(|f|^2)$ is as in Equation \eqref{eq:thetauniton}. This approach might be more amenable to higher rank generalisations since it does not require an explicit choice of Hopf coordinates.

\subsection{Complex Unitons}
An important feature of this model is the existence of  a second solution to the equations of motion which lives in the complexified  target space. We shall thus refer to this configuration as a \textit{complex uniton}, and, by contrast, the uniton discussed above shall be referred to as the \textit{real uniton}. For the complex uniton, the configuration of the fields $\phi_i$ shall be the same as for the real uniton given by Equation \eqref{eq:phiuniton}. For $\theta(|f|^2)$, we obtain
\begin{equation}
    \theta(|f|^2)=\frac{\pi }{2}+i \, \text{arctanh} \left(\frac{1}{2} \left(|f|+\frac{1}{|f|}\right) \sqrt{\chi_-^2+1}\right).
\end{equation}
When this is substituted into the action we obtain
\begin{equation}
    S = \frac{2k}{t(1+\chi_+^2)}S_{CI}\,,
\end{equation}
with
\begin{equation} \label{eq:cplxunitonaction}
    S_{CI}=\frac{2}{m}\left(\chi_- \text{arccot}\chi_- - \chi_+ \text{arccot}\chi_+\right) \,,
\end{equation}
$\chi_\pm=\zeta\pm\eta$ and $m$ is as in \eqref{eq:m1}. Interestingly, the action of the real uniton and the complex uniton arise as the integral of the same function. This leads to a surprising connection which is detailed further in Appendix \ref{ap:intaction}. Observe that $S_{CI}$ is real and negative if $\eta$ and $\zeta$ are real and positive.

Readers familiar with the undeformed PCM \cite{cherman2015decoding} might wonder why such complex uniton configurations played no role there.  The answer is simple:  although it is still a solution to the field equations, its action diverges and plays no important role.

\begin{figure}[tbp!]
    \centering
        \subfigure[$\eta=0.2, \zeta=0.1$]{
        \includegraphics[height=1in]{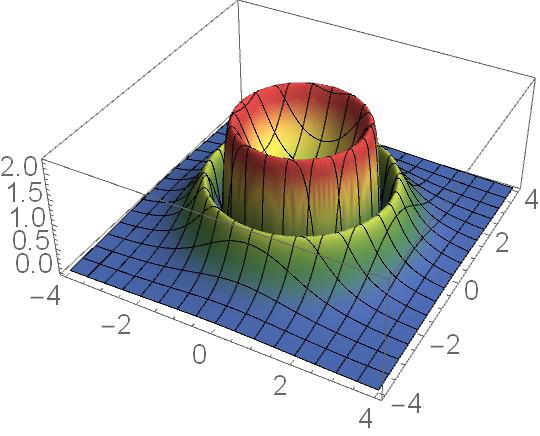}
        }
     \subfigure[$\eta=0.2, \zeta=0.19$]{
        \includegraphics[height=1in]{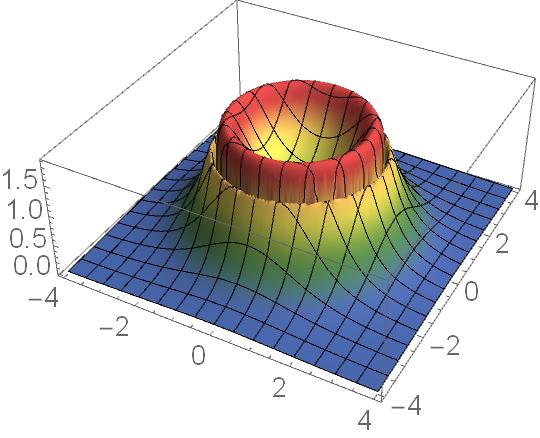}
        }
    \subfigure[$\eta=0.2, \zeta=0.2$]{
        \includegraphics[height=1in]{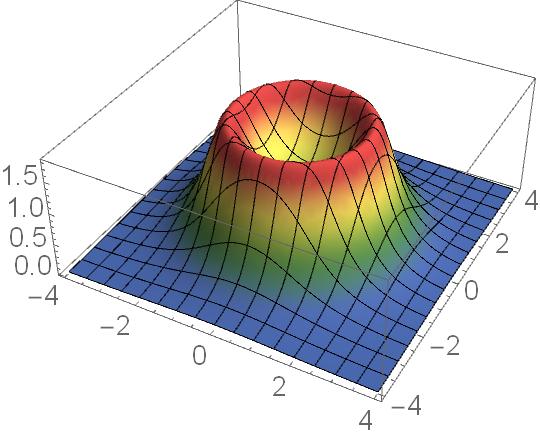}
        }
    \subfigure[$\eta=0.2, \zeta=0.21$]{
        \includegraphics[height=1in]{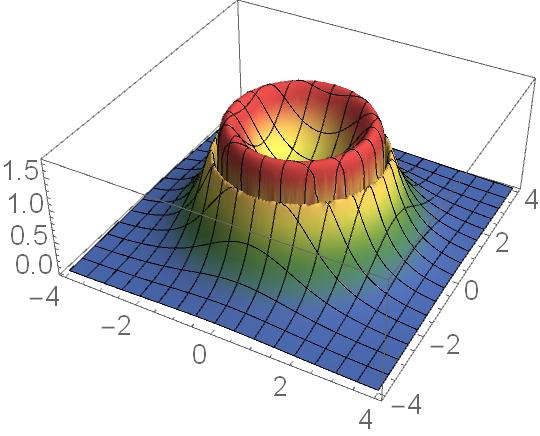}
        }
    \caption{A plot of (the real part of)   the Lagrangian density of the $k=1$ complex uniton on $\mathbb{R}^2$ as the deformation parameters are tuned to cross the critical line. In (a) there is a clear concentric valley structure which is removed precisely at the critical line in (c).   The moduli are fixed in these plots at $\lambda_0=0$,   $\lambda_1=1/2$.}
    \label{fig:criticalcplxuniton}
\end{figure}

In Figure \ref{fig:criticalcplxuniton}, we show the (real part) of the Lagrange density of these complex uniton lumps.  This reveals a peculiar behaviour across the critical line of deformation parameters.  At generic values of deformation parameters, there is a secondary valley in the Lagrangian density.  This structure however disappears discontinuously across the critical line. 

Similarly discontinuous behaviour is visible directly in the value of the complex uniton action eq. \eqref{eq:cplxunitonaction} which exhibits a cusp across the critical line as can be seen from 
\begin{equation}
   \left( \lim_{\eta\rightarrow\zeta^+} -   \lim_{\eta\rightarrow\zeta^-} \right) \partial_\eta S_{CI} = - \frac{4 \pi \zeta^2}{ 1+ 4 \zeta^2}  \,.
\end{equation}
This a strong early hint for a feature that we will later see in detail, namely that the quantum behaviour away from the critical line is rather different from that exactly on the critical line.

\subsection{Uniton Dominance Regimes} 

\begin{figure}[tbp]
        \centering{\includegraphics[height=2.4in]{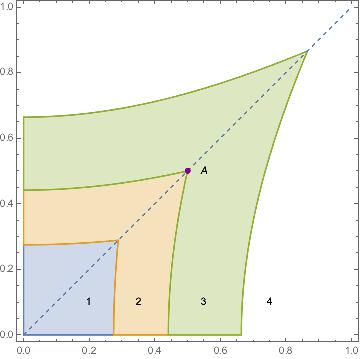}}
         \caption{A plot in the $\eta-\zeta$ plane indicating the hierarchy of the various non-perturbative configurations. In region 1 (blue) $|2S_I|< |S_{CI}|$; in region 2 (yellow) $|S_I|< |S_{CI}|<|2S_I| $; in region 3 (green) $|S_{CI}|<|S_I| <|2S_{CI}| $ and finally in region 4 (white) $|2S_{CI}|<|S_I|$.  The dashed line indicates the critical line $\varkappa:=\eta = \zeta$ and the point $A$ is where $\varkappa=\frac{1}{2}$ and $S_I=-S_{CI}=\pi$ and will be shown to exhibit interesting behaviour. The critical line crosses from region 1 to 2 at $\varkappa=\tfrac{1}{2\sqrt{3}}$, where $2S_I = - S_{CI} = \frac{8\pi}{4\sqrt{3}}$. It crosses from region 3 to 4 at $\varkappa=\sqrt{3}/2$ where $S_I = -2 S_{CI} = \frac{8\pi}{4\sqrt{3}}$.}
    \label{fig:regionplot}
\end{figure}

Whilst discussing the classical aspects of these solutions, let us preempt a little of what is to follow.  We have in the complex and real unitons two types of classical saddles, and one should anticipate that both are important to define the full quantum theory.  However, which (classical) saddle is most important will depend on where we are in (classical) parameter space. Because the configuration with the lowest action yields the biggest contribution in perturbation theory, we divide the parameter space spanned by $\eta$ and $\zeta$ into different regions, based on inequalities among the actions \eqref{eq:realunitonaction} and \eqref{eq:cplxunitonaction}. This is displayed in Figure \ref{fig:regionplot} where one can see that there are demarcations between regimes when the absolute value of the real and complex uniton actions become equal or integer multiples of each other.  One should anticipate that perturbation theory will behave differently in different regimes, and this will indeed be the case as will be seen in Section \ref{sec:WKB}.

\section{Compactification and Fractionation}\label{sec:reduce}

Our primary goal is to expose the quantum resurgent structure of these theories.  We shall do so in a slightly indirect fashion following the arguments proposed in \cite{cherman2015decoding}, to reduce the problem from a full 1+1 dimensional quantum field theory to a tractable quantum mechanics.  This is achieved by performing an adiabatic reduction on a spatial $S^1$ with a twisted boundary condition of the form\footnote{It should be clear from the context if $L$ refers to the compactification radius or if it serves as a label for the left symmetry group, in contrast to $R$ for the right symmetry group.}

\begin{equation}
    g(t, x+L)=e^{iH_L}g(t,x)e^{-iH_R} \, . 
\end{equation}
However, it is more practical, instead, to work with a periodic boundary condition by defining
\begin{equation} \label{eq:gtildedefn}
    \tilde{g}(t,x)=e^{-i H_L x/L} g e^{i H_R x/L} \quad \implies \quad \tilde{g}(t,x+L)=\tilde{g}(t,x).
\end{equation}
Introducing a nonzero $H_L$ and $H_R$ is like turning on an effective background gauge field in the untwisted theory with periodic boundary conditions. This can be subdivided in a contribution from a vectorial twist and an axial twist $H_{V,A}=\frac{1}{L}(H_L \pm H_R)$. 

By an adiabatic compactification, we mean that we are looking for a compactification that has no phase transition as we send the compactification radius $L$ of the $S^1$ from large to small. The contribution of \cite{cherman2015decoding} is the precise analysis of two compactifications: one thermal and one spatial. It is shown that the thermal compactification has a phase transition, whereas the spatial compactification under some additional constraints does not. This is measured by $ \mathcal{F}/N^2$ in the $N\rightarrow\infty$ limit, where $\mathcal{F}$ is the free energy. This quantity has a sharp transition from $\mathcal{O}(1)$ to $0$ for thermal compactification as we go from large $L$ to small $L$, whereas for the spatial compactification it tends to $0$ in the limit for all $L$. Of course here we are at finite $N$ (the target space is $SU(2)$) rendering some of this discussion moot in point but we retain the strategy employed at large $N$ with some post-hoc justification. 

It was shown in \cite{cherman2015decoding} that to achieve adiabatic continuity one must impose two things. Firstly, we need to set $H_A=0$.
Secondly, one must minimise the  contributions of the Wilson line for the background gauge field, $\Omega = \exp(i \oint dx H_V)=\exp(i L H_V)$, to the free energy which occurs when  
\begin{equation}
    \Omega=e^{\frac{\nu i \pi}{N}} \text{diag}\left(1, e^{\frac{2i \pi}{N}}, \ldots, e^{\frac{2i\pi(N-1)}{N}}\right), \qquad \nu=0,1 \, \text{if} \,N=\text{odd}, \text{even}.
\end{equation}
For the $SU(2)$ case this means we require
\begin{equation} \label{eq:minimisepot}
    LH_V=H_L=H_R = \frac{\pi}{2}\begin{pmatrix} 1 & 0 \\ 0 & -1 \end{pmatrix}.
\end{equation}
We will paramterise the effective gauge field as
\begin{equation} \label{eq:vectorialtwist}
    H_L=H_R=\begin{pmatrix} \xi & 0 \\ 0 & -\xi \end{pmatrix},
\end{equation}
so the maximal twist \eqref{eq:minimisepot} is given by $\xi=\pi/2$.

The idea here is that this simplifies the theory considerably, retaining only a small selection of modes from the full theory, but does so in a way that retains the salient perturbative structure.  Whilst this approximation is evidently not complete (for instance the role of the renormalisation in the quantum field theory is somewhat obscured), rather remarkably we will find that we can relate the features of perturbation theory in the resultant quantum mechanics obtained after shrinking the $S^1$ to the non-perturbative saddles found in the full 1+1 dimensional theory. 

To understand the twisted Lagrangian $\mathscr{L}[\tilde{g}]$, we shall consider the currents under both the right and the left acting symmetries $g\rightarrow e^{i \alpha_L \sigma_3} g e^{-i\alpha_R\sigma_3}$ of the untwisted Lagrangian studied in Section \ref{sec:classicalcurrents}. The Minkowkian current are given by Equation \eqref{eq:minkcurrrents}.
In terms of these currents, the twisted Lagrangian obtained by substituting the field \eqref{eq:gtildedefn} with $H_{L/R}$ given by \eqref{eq:vectorialtwist} into  the lagrangian \eqref{eq:doubledeflag} is given by
\begin{equation}
    \mathscr{L}[\tilde{g}] = \mathscr{L}[g] + \frac{\xi}{L}(\mathfrak{j}^3_{L}+\mathfrak{j}^3_{R}) + \frac{8 \xi ^2 }{L^2\Delta(\theta)} \sin ^2(\theta )[ (\zeta-\eta)^2\sin^2(\theta)+1]\,,
\end{equation}
where we recall $\Delta(\theta)=1+\zeta^2+\eta^2+2\zeta\eta\cos(2\theta)$.

We will now perform a Kaluza-Klein reduction and discard all the spatial dependence. Moreover, we will eliminate all total derivatives. In particular, this means the contribution linear in currents $\mathfrak{j}^3_{L/R}$ vanishes. In the resulting Lagrangian, the fields $\phi_i$ become non-dynamic, and thus we can focus on the low energy effective theory by setting all momenta in these directions to zero.  

Following this procedure we thus obtain the reduced Lagrangian 
\begin{equation} \label{eq:lagred}
    \mathscr{L} = \frac{1}{t}\frac{\dot{\theta}^2 - \frac{8\xi^2}{L^2}\sin^2(\theta)[ (\zeta-\eta)^2\sin^2(\theta)+1]])}{\Delta(\theta)} \, . 
\end{equation}
To put the kinetic term into canonical form it is necessary to redefine variables such that the denominator factor $\Delta(\theta)$ can be absorbed.  This is achieved by defining  
\begin{equation}
    \tilde{\theta}=F(\theta, m) \, ,  
\end{equation}
where $F(\phi,m)$ is the elliptic integral of the first kind the modulus $m$ was foreshadowed by Equation \eqref{eq:m1}. Employing Jacobi elliptic functions\footnote{In terms of the indefinite elliptic integral of the first kind $u= \int^\phi_0 d\theta \left(1- m \sin^2 \theta\right)^{-\frac{1}{2}}$, the Jacobi elliptic sine is $\text{sn}(u) = \sin\phi$.  The delta amplitude is $\text{dn}^2(u) = 1 - m\,\text{sn}^2(u)$ and we make use of $\text{sd} (u) = \frac{\text{sn}(u)}{\text{dn}(u)}$.}, the Hamiltonian of the quantum mechanics takes the following form
\begin{equation}
    H = \frac{g^2}{4} p_{\tilde{\theta}}^2 +\frac{1}{g^2} V(\tilde{\theta})\, ,
\end{equation}
with 
\begin{equation} \label{eq:potentialsection4}
    V(\tilde{\theta}) = \frac{4\xi^2}{L^2}\text{sd}^2(\tilde{\theta})(1+\chi_-^2\text{sn}^2(\tilde{\theta})) \, 
\end{equation}
where $g^2= t( 1+\chi_+^2)$.  
Notice in the $\zeta\rightarrow0$ limit, we have that $m\rightarrow 0$ , which implies that $\text{am}(u)\rightarrow u$, so $\text{sn}(u)\rightarrow \sin u$ and $\text{dn}(u)\rightarrow 1$ such that the potential degenerates to a Whitaker--Hill type found for the single deformation in  \cite{demulder2016resurgence}. 

The approach to UV fixed lines, $\eta-\zeta =i$ and $\eta+\zeta = i$ in the complex plane displays further striking behaviour. In elliptic variables these limits correspond to sending $m\rightarrow 1$ and $m \rightarrow \infty$ respectively.  Using the the elliptic variables, when we set $\eta-\zeta=\pm i$, the potential becomes $\text{tanh}^2(\theta)$. Up to a shift, this is a P\"oschl-Teller potential which has an exactly solvable discrete spectrum in terms of Legendre polynomials. The $m\rightarrow\infty$ limit is better understood    without going to elliptic variables, indeed setting $\eta = \zeta = \frac{i}{2}$ we see that $\Delta(\theta) \rightarrow \sin^2\theta$  such that    the  Lagrangian  \eqref{eq:lagred} describes a free particle.  In both cases,  one should not anticipate any asymptotic behaviour to be exhibited.  However, any small deformation away from these  points will   induce a non-trivial potential and a rich resurgent structure will become manifest. This is rather reminiscent of the Cheshire cat resurgence \cite{Kozcaz:2016wvy,Dorigoni:2017smz}, as we obtain a theory that has energy eigenvalues that are not asymptotic in $g^2$, but rather are exact. It would certainly be interesting to understand this directly at the two-dimensional level for which the fixed point is understood as a $SU(1,1)/U(1) + U(1)$ gauged WZW CFT.  

For the remainder of the paper, we shall be studying a quantum mechanical system with potential \eqref{eq:potentialsection4}. Before doing so, let us remark on the fate of the uniton (real and complex) under this twisted reduction.  The first point to remark is that it is straightforward to modify the uniton solutions to accommodate the twisted boundary condition,  this is done by simply by replacing the holomorphic function $f(z)$ entering in the minimal unitons on $\mathbb{R}^2$ with   a twisted version   $f(z)=\lambda_0e^{-\pi z/L}+\lambda_1 e^{\pi z/L}$. 

\begin{figure}[tbp]
    \centering
     \subfigure[$\eta=0.0$]{
        \includegraphics[height=1.in]{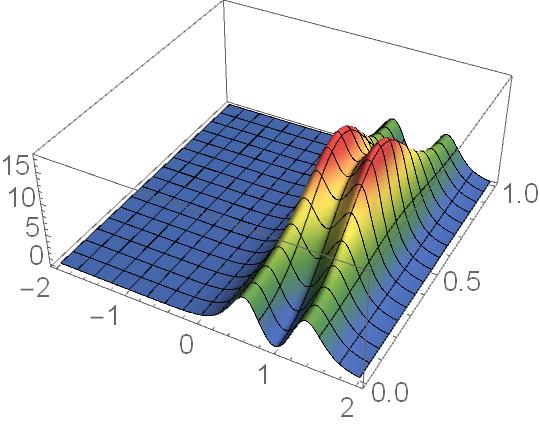}
        }
     \subfigure[$\eta=0.5$]{
        \includegraphics[height=1.in]{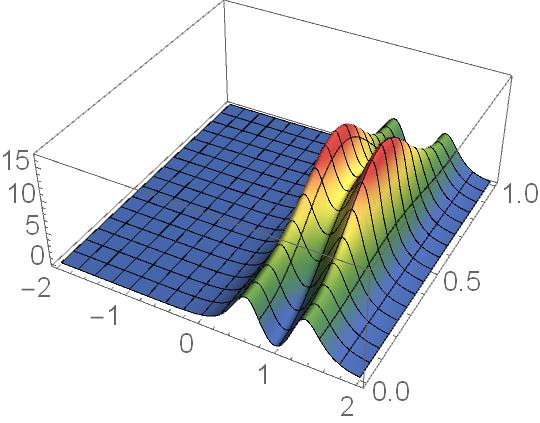}
        }
     \subfigure[$\eta=1.0$]{
        \includegraphics[height=1.in]{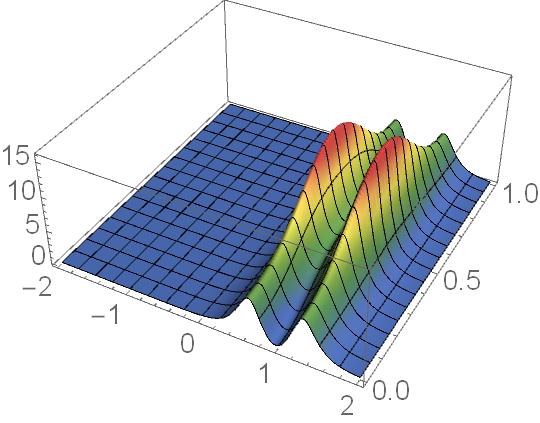}
        }
     \subfigure[$\eta=0.5i$]{
        \includegraphics[height=1.in]{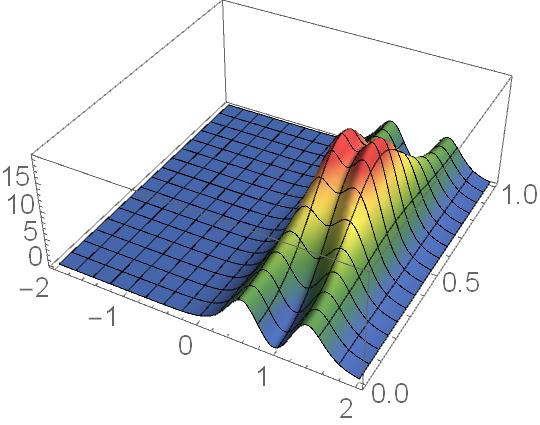}
        }
 
     \subfigure[$\eta=0.0$]{
        \includegraphics[height=1.in]{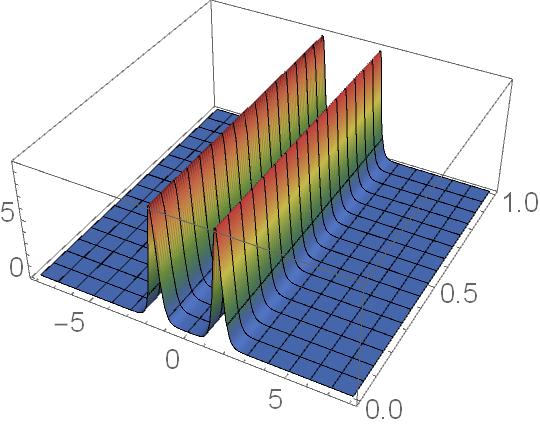}
        }
     \subfigure[$\eta=0.5$]{
        \includegraphics[height=1.in]{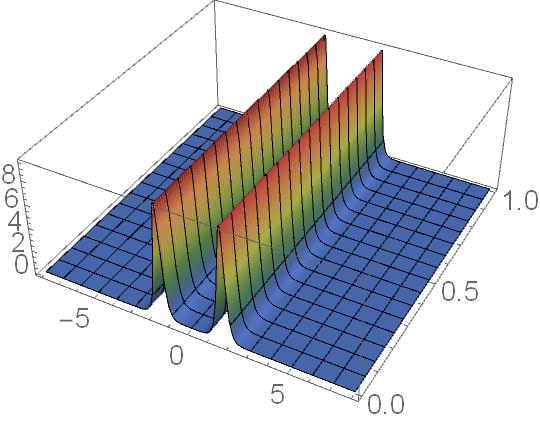}
        }
     \subfigure[$\eta=1.0$]{
        \includegraphics[height=1.in]{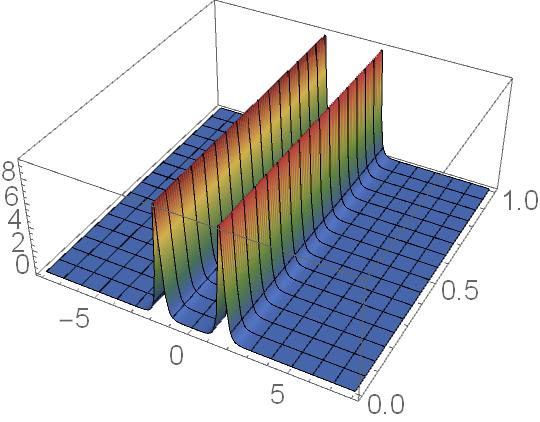}
        }
     \subfigure[$\eta=0.5i$]{
        \includegraphics[height=1.in]{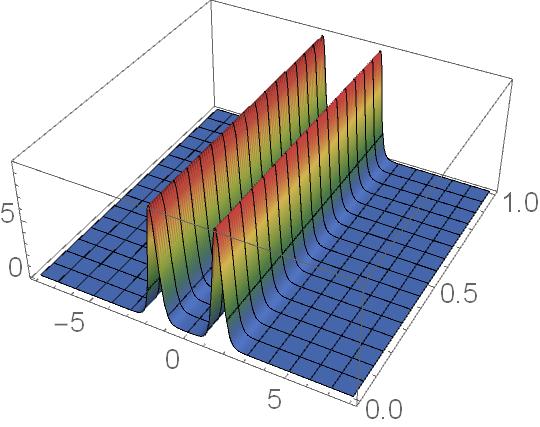}
        }
    \caption{The Lagrangian density of the real uniton on $\mathbb{R}\times S^1$  with twisted periodic boundary conditions. We have set $\zeta=0.5$ everywhere. In the top row, $\lambda_0=e^2$ and $\lambda_1=e^{-4}$ and we cannot see a clear fractionation. In the bottom row we consider $\lambda_0=\lambda_1=e^{-5}$ and there is a clear fractionation.}
    \label{fig:doublerealtwisteduniton}
\end{figure}

\begin{figure}[tbp]
    \centering
     \subfigure[$\eta=0.3$]{
        \includegraphics[height=1.in]{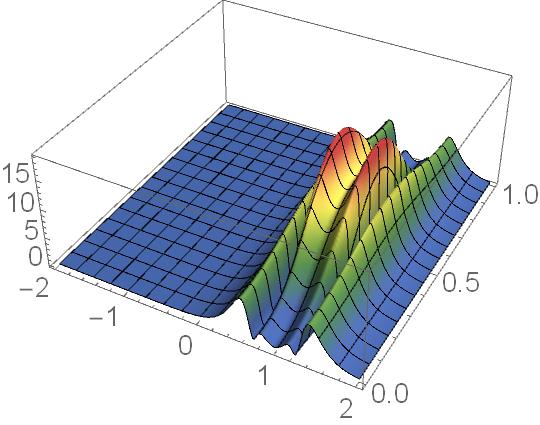}
        }
     \subfigure[$\eta=0.45$]{
        \includegraphics[height=1.in]{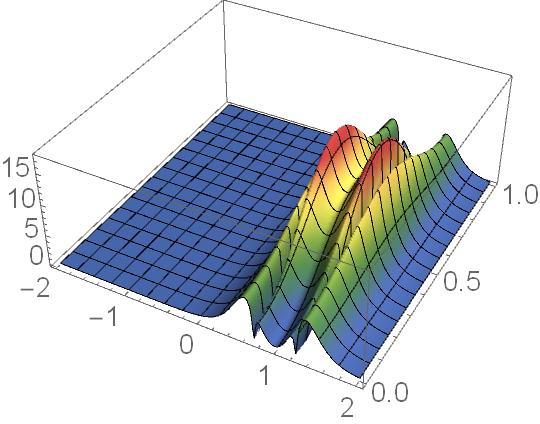}
        }
     \subfigure[$\eta=0.5$]{
        \includegraphics[height=1.in]{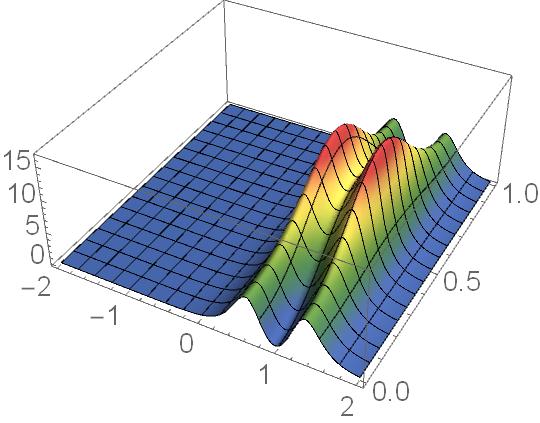}
        }
     \subfigure[$\eta=0.55$]{
        \includegraphics[height=1.in]{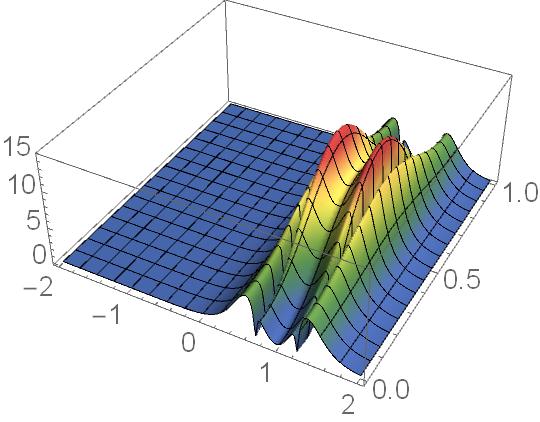}
        }
 
     \subfigure[$\eta=0.3$]{
        \includegraphics[height=1.in]{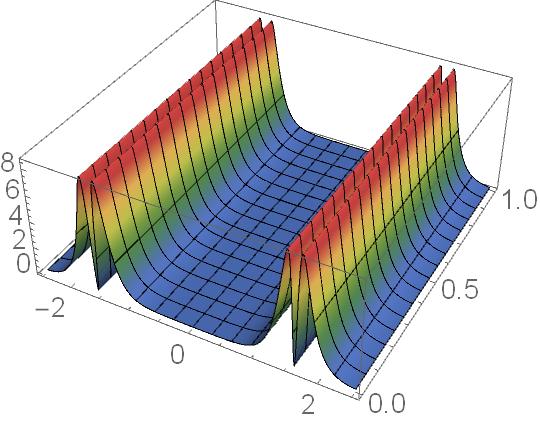}
        }
     \subfigure[$\eta=0.45$]{
        \includegraphics[height=1.in]{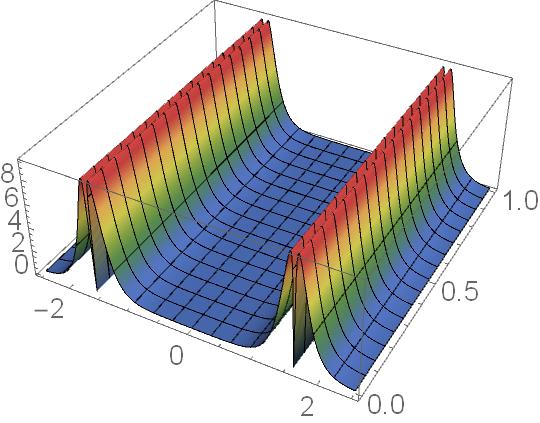}
        }
     \subfigure[$\eta=0.5$]{
        \includegraphics[height=1.in]{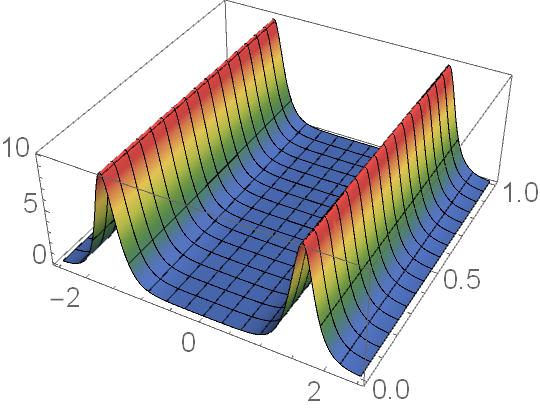}
        }
     \subfigure[$\eta=0.55$]{
        \includegraphics[height=1.in]{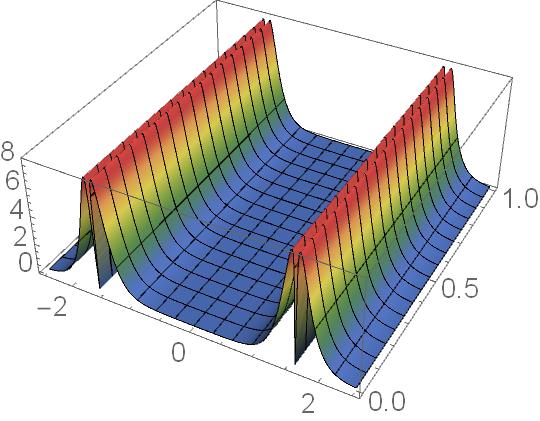}
        }
    \caption{The Lagrangian density of the complex uniton on $\mathbb{R}\times S^1$  with twisted periodic boundary conditions.  Here we have set $\zeta=0.5$ and zoomed in to study the behaviour around the critical line $\eta=\zeta$.   In the top row, we show  $\lambda_0=e^2$ and $\lambda_1=e^{-4}$, which should be contrasted with the bottom row where $\lambda_0=\lambda_1=e^{-5}$ and fractionation is clearly evident.    In both rows we clearly see, in (c) and (g), a sharp change in the profile as the critical line is reached.}
    \label{fig:criticalcomplextwisteduniton}
\end{figure}

Recall that on $\mathbb{R}^2$ the unitons formed localised lumps of Lagrangian density (with some non-trivial profile induced by the deformation parameters) and this is true across the moduli space parameterised by $\{\lambda_0, \lambda_1\}$. In contrast, on the twisted cylinder a different behaviour emerges; there are regions of moduli space for which the real uniton breaks up (or fractionates) into well separated and clearly distinct lumps of Lagrangian density (see Figure \ref{fig:doublerealtwisteduniton}).  In this way we anticipate that a single real uniton makes a contribution to the dimensionally reduced theory much like an instanton anti-instanton pair. The complex uniton exhibits a similar fractionation (see Figure \ref{fig:criticalcomplextwisteduniton}), but in addition we observe a strange phenomenon around the critical line: the additional valley in the uniton density discontinuously vanishes. 

\section{WKB and Resurgence} \label{sec:WKB}

In this section we study a Schr\"odinger Equation
\begin{equation} \label{eq:seqn}
    \left(g^4\frac{\partial^2}{\partial \theta^2 } - V(\theta) + g^2 E \right) \Psi(\theta) = 0 \,,
\end{equation}
with potential (to ease notation we now drop the tilde accent on $\theta$)
\begin{equation} \label{eq:schrodingerpotential}
    V(\theta) =\text{sd}^2(\theta)(1+\chi_-^2\text{sn}^2(\theta))
\end{equation}
and $g^2=t(1+(\zeta+\eta)^2)$. We employ the WKB method to obtain an expansion in $g^2\rightarrow 0$. We make an ansatz
\begin{equation} \label{eq:wkbansatz}
    \Psi(\theta) = \exp \bigg( \frac{i}{g^2} \int_{\theta_0}^\theta d\theta\, S(\theta) \bigg)\,,
\end{equation}
in which, $S(\theta)$ is a function that still depends of $g^2$. This will solve the Schr\"odinger Equation \eqref{eq:seqn} if the function $S(\theta, g^2)$ satisfies the Ricatti Equation
\begin{equation} \label{eq:wkbricatti}
    S^2(\theta)- i g^2 S'(\theta) = p^2(\theta) \,,
\end{equation}
where $p(\theta) = \sqrt{g^2 E - V(\theta)}$ is the classical momentum, as usual. We assume a power series ansatz for $S(\theta)$
\begin{equation}
    S(\theta) = \sum_{n=0} g^{2n} S_{n}(\theta)\,,
\end{equation}
for which there exists a recursive solution widely available in the literature \cite{aniceto2017asymptotics, aniceto2018primer, iwaki2014exact}. At the same time we make a power series ansatz
\begin{equation}
    E = \sum_{n\geq 0} a_n g^{2n}\,.
\end{equation}
Here, $a_n$ of course still depends on the parameters $\eta$ and $\zeta$.

In this section we will compute this perturbative series to a very high order. For explanatory purposes, will mostly restrict our investigation to the behaviour along two trajectories: along the critical line $\varkappa=\eta=\zeta$ and along the line $\zeta=1/5$. We will study how the behaviour transitions as we cross the different regions shown in Figure \ref{fig:regionplot}. Along these trajectories, we compute the Borel-Pad\'e approximant.  We show how its pole structure suggests branch points that precisely match the value of the uniton actions \eqref{eq:realunitonaction} and \eqref{eq:cplxunitonaction}. By looking at the Stokes lines of the quadratic form associated to this potential, we see that these contributions can be associated with saddle trajectories for real values of the coupling.

Next, we use the uniform WKB ansatz \cite{dunne2014uniform} to find an asymptotic form for the perturbative expansion. We show that the perturbative series converges rapidly to its asymptotic form. This asymptotic form, however, depends on which regions of the parameter space we analyse, as different unitons are dominant across the different regions of Figure \ref{fig:regionplot}.

\subsection{Borel Transform}
We use the BenderWu package \cite{sulejmanpasic2018aspects} to compute WKB expansion so that we obtain a perturbative asymptotic expansion of the ground state energy (we will not consider higher level states in this paper). Unfortunately, the script runs too slow for general $\eta$ and $\zeta$ so for most of the asymptomatic analysis to come we will be working with explicit values for the deformation parameters. For specified values of $\eta$ and $\zeta$, we could typically obtain 300 order of perturbation theory in 30 minutes on a desktop computer. The first terms for the deformed model in the expansion come out as
\begin{equation}
\begin{alignedat}{2} \label{eq:EPertubativeExamples}
    E &= 1- \frac{1}{4}g^2 - \frac{1}{16} g^4 - \frac{3}{64} g^6 + \mathcal{O}(g^8), \qquad &\eta=0, \; \zeta=0 \,, \\
    E &= 1- \frac{1}{16}g^2 - \frac{61}{256} g^4  + \frac{777}{4096} g^6 + \mathcal{O}(g^8), \qquad &\eta=\frac{1}{2},\;  \zeta=0 \,, \\
    E &= 1- \frac{69}{1600}g^2 - \frac{360357}{2560000} g^4 + \mathcal{O}(g^6), \qquad &\eta=\frac{1}{2}, \;\zeta=\frac{1}{4} \,, \\
    E &= 1-  \frac{3}{32} g^4 - \frac{39}{2048}g^8 +\mathcal{O}(g^{12}), \qquad  &\eta=\zeta=\frac{1}{2} \,, \\
\end{alignedat}
\end{equation}
The fact that at $\eta=\zeta=1/2$ we obtain a perturbative series in $g^4$ is very specific to this point as is explained further in Figure \ref{fig:asymptotic4}. In essence, it is due to a perfect cancellation of an alternating and a non-alternating series. This can be traced back to the equality $S_I=-S_{CI}=\pi$, see also Figure \ref{fig:regionplot}. 

We compute the Borel transform 
\begin{equation} \label{eq:BorelTransform}
    \hat{E}=\sum_{n\geq 0} \frac{a_n}{n!}\hat{g}^{2n}
\end{equation}
of this series. We would like to understand something about the singularity and branch cut structure of the $\hat{g}^2$-plane, which is also called the Borel plane. We will sometimes use $z=g^2$, while $s=\hat{g}^2$ is the variable in the Borel plane.   The idea, and we will be telegraphic here referring the reader to the excellent reviews e.g. \cite{aniceto2018primer,dorigoni2014introduction},  is that the Borel transform has a finite radius of convergence and the original divergent series can be re-summed by performing a Laplace transformation on  $\hat{E}$.  When the Laplace transformation can be done un-ambiguously this results in a finite re-summed value for the original series.  However, in many interesting cases $\hat{E}(s)$ has poles along the integration path $s\in [0, \infty]$ defining the Laplace transformation.  To give meaning to the integration one can instead deform the integration contour and define the lateral resumation in the direction $\vartheta$ as 
\begin{equation} \label{eq:Borelresum}
{\cal S}_\vartheta E(z)  = \frac{1}{z}\int_0^{e^{i \vartheta}\infty} ds\, e^{-s/z} \hat{E}(s) \,. 
\end{equation} 
A ray, $\vartheta=\vartheta_0$, is said to be a Stokes direction if $\hat{E}(s)$ has singularities along that ray. One can then define two  lateral summations  ${\cal S}_{\vartheta_0+\epsilon} E(z)$ and  ${\cal S}_{\vartheta_0-\epsilon} E(z)$ which have the same perturbative expansion but differ by non-perturbative contributions, a change known as a Stokes jump.  The crucial idea of the resurgence paradigm going back to \cite{bogomolny1980calculation, zinn2004multi1, zinn2004multi2}  is that the inherent ambiguity between these two  perturbative resumations is precisely cancelled by a similarly ambiguous contribution from the fluctuations around an appropriate non-perturbative configurations in the same topological sector. For instance, in quantum mechanic the path integral over the quasi-zero mode separation between an instanton anti-instanton pair has an ambiguous imaginary contribution that cancels that of the ground state energy ambiguity.  The first test of this programme is then that the location of the poles in the Borel plane should be in accordance with the values of the on-shell action for non-perturbative field configurations.    
 
When performing a numerical calculation, the summation defining the Borel transformation has to be cut off at the order to which the perturbative expansion was performed.  Hence $\hat{E}(z)$ becomes a simple polynomial which has no poles. For this reason, we employ the Pad\'e approximant, which is an approximation of the function by the ratio of two polynomials, where the coefficients are determined by demanding that the Taylor series matches the original. By calculating the roots of the denominator of the Pad\'e approximant, we find its poles in the $\hat{g}^2$-plane. These are called the (Borel-)Pad\'e poles. An accumulation of Pad\'e poles suggests a branch point in the Borel plane. These methods are expanded upon further in \cite{aniceto2017asymptotics, aniceto2018primer, costin2019resurgent, costin2020physical}.

\begin{figure}[h!]
    \centering
        \subfigure[$\eta=0$]{
        \includegraphics[height=2.2in]{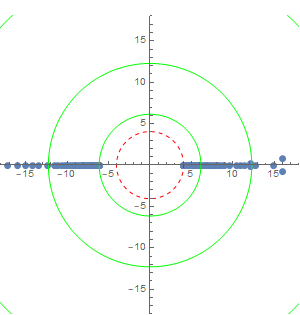}
        }
        \subfigure[$\eta=19/100$]{
        \includegraphics[height=2.2in]{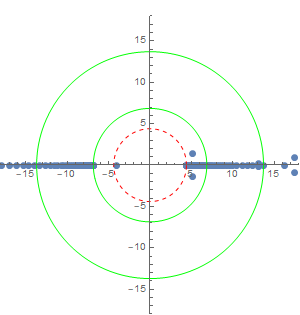}
        } \\
        \subfigure[$\eta=20/100=1/5$]{
        \includegraphics[height=2.2in]{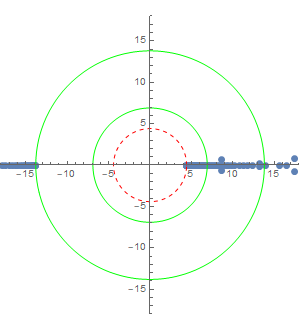}
        }
        \subfigure[$\eta=21/100$]{
        \includegraphics[height=2.2in]{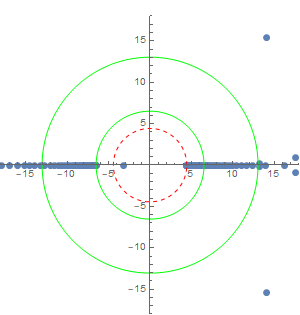}
        }
    \caption{The complex Borel $s$-plane for $\zeta = \frac{1}{5}$ at different values of $\eta$ with blue dots indicating poles of the Borel-Pad\'e approximation obtained from 300 orders of perturbation theory in $g^2$ (hence we computed a total of 150 poles). Accumulations of poles are anticipated to encode branch cuts in the full Borel transform, and isolated poles are expected to be residuals of the numerical approximation. The red dashed circle indicates the magnitude of the the real uniton action located at $|s|  = 2 S_I$. The green dashed circles indicate the magnitude of the complex 1- and 2-uniton actions located at $|s| = |S_{CI}|\, , |2S_{CI}|\,$ respectively. For $\eta$ and $\zeta$ real, the real and complex isntanton action have an complex argument of $0$ and $\pi$ respectively. We see a clear match to the location of expected branch points with these values.  At the critical line $\eta=\zeta$, we observe a curious discontinuous jump; the accumulation of poles at the 1-complex uniton disappears entirely and instead, we get an accumulation point at the complex 2-uniton action $s=2S_{CI}$.}
    \label{fig:borel1}
\end{figure}

\begin{figure}
    \centering
        \subfigure[$\varkappa=0$]{
        \includegraphics[height=1.8in]{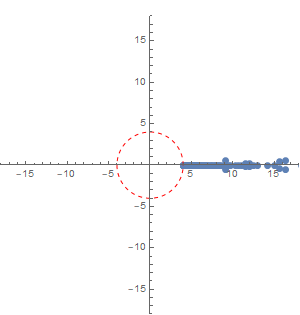}
        }
        \subfigure[$\varkappa=1/5$]{
        \includegraphics[height=1.8in]{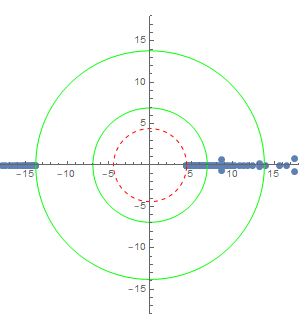}
        } 
        \subfigure[$\varkappa=1/2\sqrt{3}$]{
        \includegraphics[height=1.8in]{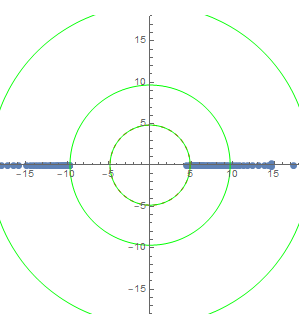}
        } \\
        \subfigure[$\varkappa=2/5$]{
        \includegraphics[height=1.8in]{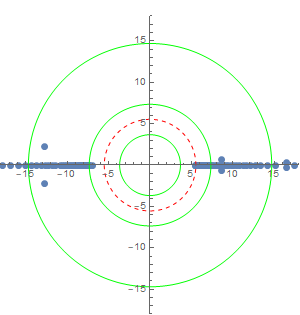}
        } 
        \subfigure[$\varkappa=1/2$]{
        \includegraphics[height=1.8in]{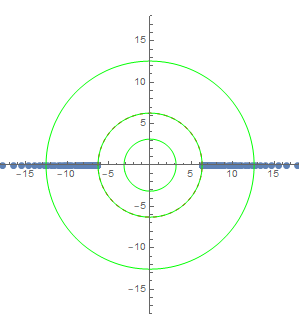}
        }
        \subfigure[$\varkappa=\sqrt{3}/2$]{
        \includegraphics[height=1.8in]{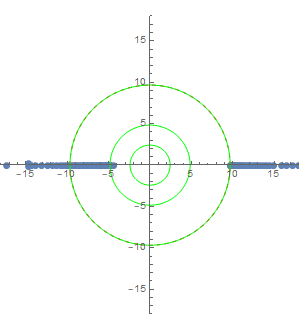}
        }
    \caption{The complex Borel  $s$-plane along the $\zeta = \eta = \varkappa$ critical line as we cross different region of Figure \ref{fig:regionplot}. Colours, key, and numerical approximation as per Figure \ref{fig:borel1}, but we have also plotted the action of the complex 4-uniton $|s|=|4S_{CI}|$ as a green circle.  In the undeformed model $\varkappa=0$ there is not complex uniton \cite{cherman2015decoding} since it has infinite action. When $\varkappa=1/5$, we are in region 1. At $\varkappa=1/2\sqrt{3}$, we have $2S_I=-S_{CI}$ and cross from region 1 to 2. Notice that a dashed red circles coincides with the inner green circle. For $\varkappa=2/5$, we are in region 2. When $\varkappa=1/2$ we cross into region 3 and $S_I=-S_{CI}$. If $\varkappa=\sqrt{3}/2$ we cross from into 4 where $S_I=-2S_{CI}$. Consistent we the results of Figure \ref{fig:borel1}, we note that along the critical line, the branch points along the negative real axis accumulate at $2S_{CI}$, not at $S_{CI}$.}
    \label{fig:borel2}
\end{figure}

Critically, we find that those branch points can be identified precisely with the finite action configurations found previously by the real and complex unitons \eqref{eq:realunitonaction} and \eqref{eq:cplxunitonaction}!  This is illustrated in Figures \ref{fig:borel1} and \ref{fig:borel2} demonstrating the behaviour across the critical line and along it. 
 We are thus able to relate non-perturbative contributions with these instanton configuration.  It is important to emphasise that what we have done is to take a two-dimensional QFT and truncated to a particular quantum mechanics,  but the relevant non-perturbative saddles are coming from finite action solutions in the full two-dimensional theory.  

Beyond the headline matching of poles to non-perturbative saddles lies a more intricate structure.   In Figure \ref{fig:borel1} we show that for generic real values of $\eta$ and $\zeta$, the Borel-Pad\'e approximation suggests the existence of two Stokes rays. 
The first is at $\arg(s) =0 $   for which we see evidence of a branch cut terminating at the value of the real 1-uniton action.  The second is the $\arg(s) =\pi $ ray and with a cut terminating at the complex 1-uniton action.  However, as the parameters are tuned to the critical line $\eta= \zeta$ (see  Figure \ref{fig:borel1} (c)) the location of the cut in the  $\arg(s) =\pi $ direction jumps from the complex 1-uniton to the complex 2-uniton action.    Figure \ref{fig:borel1} confirms that all along the critical $\zeta=\eta = \varkappa $ line that   $\arg(s) =\pi $ branch cut terminates at the complex 2-uniton action.  This implies that for the entire range $0<\varkappa < \frac{1}{2}$ the leading pole (the one nearest to the origin) continues to be that along $\arg(s) =0 $ at the location of the real 1-uniton action. At $\varkappa=\frac{1}{2}$  (see  Figure \ref{fig:borel2} (c)),   the action of the complex 2-uniton coincides with that of the real 1-uniton; this is the non-perturbative feature corresponding to the fact that the   perturbative series in eq. \eqref{eq:EPertubativeExamples} discontinuously jumps to being a series in $g^4$ rather than $g^2$ when $\varkappa=\frac{1}{2}$. 

Having established that it is essential to consider complexified field configurations to understand the Borel pole structure, it is natural to now  analytically continue the deformation parameters $\eta$ and $\zeta$ themselves into the complex plane. 

\begin{figure}
    \centering
        \subfigure{
        \includegraphics[height=3in]{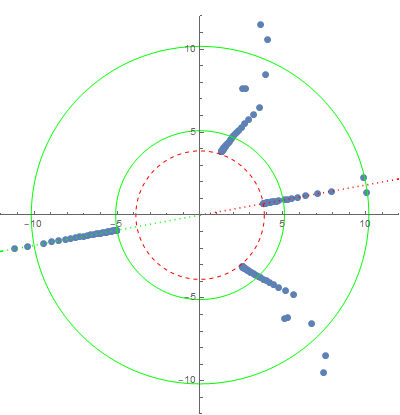}
        }

    \caption{  The complex $s$ Borel plane for $\zeta=1/5$, $\eta=2i/5$.  Colours, key, and numerical approximation as per Figure \ref{fig:borel1} with in addition the argument of the real (complex) uniton indicated by a red (green) dotted ray.  The accumulation points still gravitate towards the uniton actions and are direct with an  argument matching precisely that of the relevant uniton action. In  this particular case, because $\text{Re}(\eta)=\text{Im}(\zeta)=0$, we have that $\chi_+=\overline{\chi_-}$ and therefore the ratio of the actions is real and negative. This explains why the angle between the dotted rays is precisely $\pi$. We were unable to explain the phases of the secondary branch point that have an absolute value equal to that of the real uniton action.}
    \label{fig:genericcomplex}
\end{figure}

\begin{figure}[tbp]
    \centering
        \subfigure[$\theta=0$]{
        \includegraphics[height=2.2in]{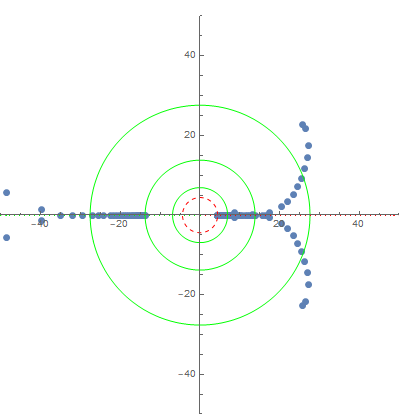}
        }
        \subfigure[$\theta=\pi/3$]{
        \includegraphics[height=2.2in]{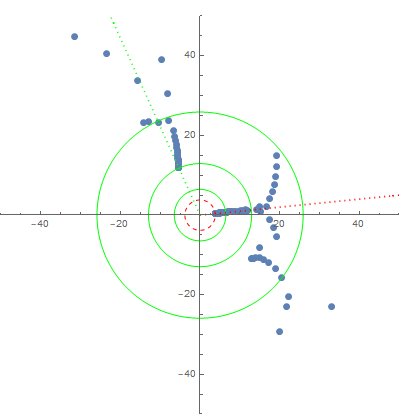}
        } \\
        \subfigure[$\theta=9\pi/20$]{
        \includegraphics[height=2.2in]{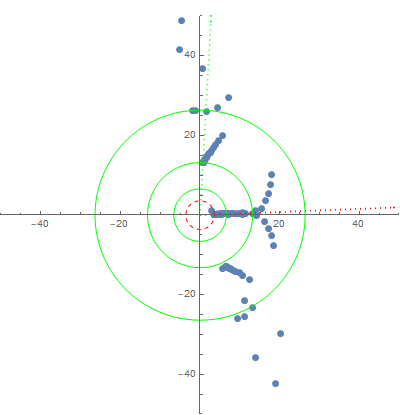}
        }
        \subfigure[$\theta=\pi/2$]{
        \includegraphics[height=2.2in]{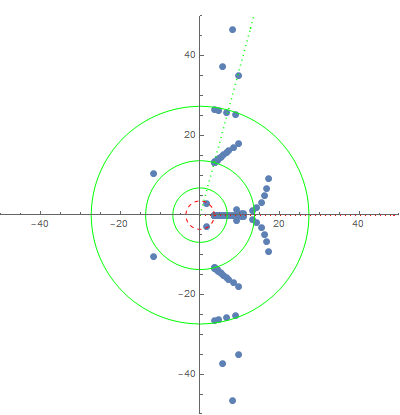}
        }
    \caption{Here, we consider the critical line $\varkappa=\eta=\zeta$ and compute 300 order of perturbation theory. We keep $|\varkappa|=1/5$ fixed, but vary $\theta=\arg(\varkappa)$. We suspect that the tails splitting into 2 ends is due to numerics and could be resolved by going to higher orders. Interestingly, it appears we can see towers of higher order states more easily when $\eta$ and $\zeta$ are analytically continued.}
    \label{fig:critrot}
\end{figure}

Generically, as indicated in Figure \ref{fig:genericcomplex}, the branch cuts continue to match to the values of the uniton actions, and now lie along angles governed by the phase of the uniton action. 
In Figure \ref{fig:critrot} we show what happens as phase of the critical parameter $\varkappa$ is rotated; again we see that the direction of the branch cuts track the phases of the unitons.  These plots also hint, although the numerics are limited, at the existence of a tower of poles located at multiples of the the complex 2-uniton action.  

Finally, we study the potential as it approaches the  point $\eta=\zeta=\frac{i}{2}$ which corresponds to the RG fixed point. Here, $m$ has a pole, so the elliptic potential is not well-defined (but recall that this is a consequence of the Jacobi variables; in the original Euler angle variables this point was simply a free theory). The actions \eqref{eq:realunitonaction} and \eqref{eq:cplxunitonaction} tend to zero\footnote{In general, we have chosen the branch cuts in the Borel plane to run from $2S_I$ to $+\infty$ and from $2S_{CI}$ to $-\infty$; here however a more natural choice would be to take a cut from $2S_I$ to $2S_{CI}$ such that cut is removed entirely as the free theory point is approached.  For this interpretation to make sense it is necessary that the branch points at $2S_{I}$ and $2S_{CI}$ display the same behaviour - which they do (see Equation \eqref{eq:branch} ). }, as do the elliptic periods of the potential. As discussed in the previous section, though a different change of variable this point can be associated to a free theory. 

Firstly, we consider the behaviour as we rotate around $\eta=\zeta=\frac{i}{2}$ on the critical line by looking at
\begin{equation} \label{eq:criticali2}
    \varkappa=\eta=\zeta= \frac{i}{2} + \epsilon e^{i \theta}\,.
\end{equation}
We find that there is an infinite tower of branch points located at
\begin{equation} \label{eq:2scibranchpoles}
    2 S_{CI} + 2 n (S_{I} - S_{CI} ), \qquad n\in\mathbb{Z}\,.
\end{equation}
In particular, for $n=1$ and $n=0$ there are branch poles at the real and complex uniton actions respectively. This is consistent with the previous analyses.

In addition we consider the behaviour as we rotate around $\eta=\zeta=i/2$ slightly off the critical line, that is, let
\begin{equation} \label{eq:offcriticali2}
    \eta= \frac{i}{2}, \quad \zeta = \frac{i}{2} + \epsilon e^{i \theta}\,.
\end{equation}
In this case we find a tower of branch points located at
\begin{equation} \label{eq:1scibranchpoles}
    S_{CI} + 2 n (S_{I} - S_{CI} ), \qquad n\in\mathbb{Z}\,.
\end{equation}
This in particular reproduces the the branch point at $S_{CI}$ for $n=0$, which is consistent with off-critical line behaviour. There are also hints off branch point of the tower given by Equation \eqref{eq:2scibranchpoles}, but the numerics are not as clean.

\begin{figure}[tbp]
    \centering
        \subfigure[$\theta=0$]{
        \includegraphics[height=2.2in]{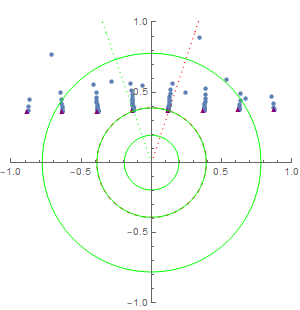}
        }
        \subfigure[$\theta=\pi/4$]{
        \includegraphics[height=2.2in]{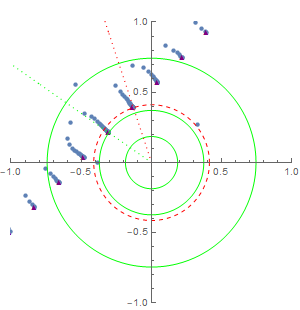}
        } \\
        \subfigure[$\theta=2\pi/4$]{
        \includegraphics[height=2.2in]{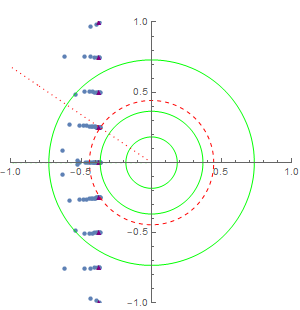}
        }
        \subfigure[$\theta=3\pi/4$]{
        \includegraphics[height=2.2in]{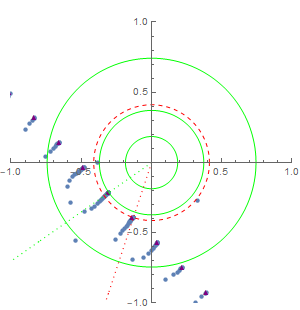}
        }
    \caption{Here, we look at the behaviour around the special point $\varkappa=\frac{i}{2}$, paramatrised by Equation \eqref{eq:criticali2} with $\epsilon=0.01$. We observe that the branch poles, indicated by purple triangles, are given precisely by Equations \eqref{eq:2scibranchpoles}. Note also that we have zoomed relative to other Borel plots shown since  both the real and the complex uniton action tend to $0$ as $\varkappa\rightarrow\frac{i}{2}$.}
    \label{fig:criticali2}
\end{figure}

\begin{figure}[tbp]
    \centering
        \subfigure[$\theta=0$]{
        \includegraphics[height=2.2in]{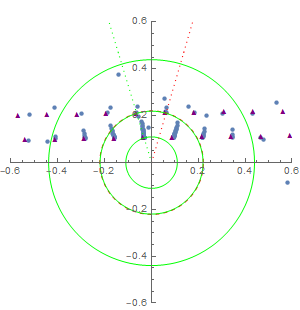}
        }
        \subfigure[$\theta=\pi/4$]{
        \includegraphics[height=2.2in]{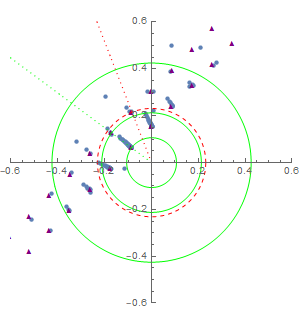}
        } \\
        \subfigure[$\theta=2\pi/4$]{
        \includegraphics[height=2.2in]{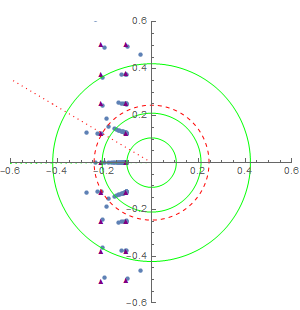}
        }
        \subfigure[$\theta=3\pi/4$]{
        \includegraphics[height=2.2in]{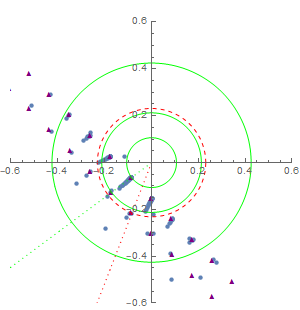}
        }
    \caption{Here, we look at the behaviour around the special point $\eta=\zeta=\frac{i}{2}$, paramatrised by Equation \eqref{eq:offcriticali2} with $\epsilon=0.01$. We find a very clear set of inner branch point given by Equations \eqref{eq:1scibranchpoles}. In addition, there are traces of the outer tower given by Equation \eqref{eq:2scibranchpoles}.}
    \label{fig:offcriticali2}
\end{figure}

The relevant Borel plots are shown in Figures \ref{fig:criticali2} and \ref{fig:offcriticali2}. We emphasise that perturbations of the form $\epsilon e^{i\theta}$ are not relevant for generic values of $\eta$ and $\zeta$. Only at $\varkappa=i/2$ do these have a substantial effect on the Borel poles.

\subsection{Uniform WKB}
We will also consider the problem through the lens of uniform WKB. The construction by Dunnel and \"Unsal \cite{dunne2014uniform} will be followed closely. We make an ansatz for the Schr\"odinger equation \eqref{eq:seqn}
\begin{equation}
    \Psi(\theta) = \frac{D_\nu(\frac{1}{g}u(\theta))}{\sqrt{u'(\theta)}} \,,
\end{equation}
where $D_\nu(\theta)$ is the parabolic cylinder function which satisfies the Schr\"odinger equation of the harmonic oscillator with energy $B:=\nu+1/2$. Contrary to ordinary analysis, $\nu$ is not an integer. However, in the $g^2\rightarrow 0$ limit, it is exponentially close to an integer. The difference with the energy level $N$ is denoted by $\delta\nu = \nu - N$. The energy eigenvalue in uniform WKB will be denoted by $\mathcal{E}$.  $u(\theta)$ and $\mathcal{E}$ are again expanded as a power series in $g^2$:
\begin{equation} \label{eq:uniformexpansionceoffs}
    u(\theta) = \sum_{n=0}g^{2n} u_n(\theta), \qquad \mathcal{E}(B) = \sum_{n=0} g^{2n} \mathcal{E}_n (B) \,.
\end{equation}
They will now satisfy a slightly modified Ricatti Equation (Equation (18) of \cite{dunne2014uniform}) which can be solved perturbatively.  Integration constants are determined by demanding that $u(\theta)$ is regular around $\theta=0$. $\mathcal{E}_n(B)$ is a polynomial of order $n$ in $B$ of definite parity: $\mathcal{E}_n(B) = (-1)^{n+1}\mathcal{E}_n(-B)$. Of course, in our problem, it also depends on $\eta$ and $\zeta$.

For $u_0(\theta)$ we find
\begin{equation}
\begin{aligned}
   (u_0(\theta))^2 &=  4 \int_0^\theta d\theta \,\sqrt{V(\theta)} \\
   &=\frac{4}{m}\Bigg(\chi_+  \arctan(\chi_+ ) - \chi_+ \arctan \left( \frac{\chi_+  \text{cn}(\theta)}{\sqrt{\chi_-^2 \text{sn}(\theta)^2+1}} \right)+ \\
   &i \chi_-\left(\log (1+i \chi_-)-\log \left(\sqrt{\chi_-^2
   \text{sn}(\theta)^2+1}+i \chi_-\text{cn}(\theta)\right)\right)\Bigg)\,,
 \end{aligned}
\end{equation}
where $\chi_\pm=\zeta\pm \eta$. For $n>0$, we use a power series ansatz of $u_n(\theta)$ in $\theta$ which results in the following coefficients for the expansion of the energy at level $B$
\begin{equation}
\begin{aligned}
    \mathcal{E}_0 &= 2B \,, \\
    \mathcal{E}_1 &=  \frac{\left(4 B^2+1\right) ( -1 + \chi_-^2+ \chi_+^2 + 3\chi_-^2\chi_+^2)}{8 ( 1+ \chi_+^2)} \,, \\
    \mathcal{E}_2 &= \frac{-1}{8} B^3 \left(17 \chi_-^4+16m  \chi_-^2+2 \chi_-^2+1\right)- \frac{B}{32} \left(8m(1-m+7\chi_-^2)+67 \chi_-^4+22 \chi_-^2+3\right)\,,
\end{aligned}
\end{equation}
where we recall $m$ is given by Equation \eqref{eq:m1}. We also found $\mathcal{E}_3$, but the expression is too long to be displayed usefully. As a consistency check we note coefficients match up perfectly with \cite{demulder2016resurgence} upon setting $\zeta=0$. 

\subsection{Asymptotic Analysis} \label{sec:asymp}
We now have the ingredients to investigate the asymptotic behaviour of the perturbative series for the ground state energy. Let us first split the behaviour into three contributions
\begin{equation} \label{eq:collectasympexpansion}
    E_n \sim E_n^{S_I} + E_n^{S_{CI}} + E_n^{2S_{CI}} + \ldots \,,
\end{equation}
where $E_n^{k S_{(C)I}}$ is a contribution due to the (complex) k-uniton. For the real uniton, this contribution will look like $E_n^{kS} \propto (2 k S) ^{-n} \Gamma(n+a)$.

It is possible to use the uniform WKB ansatz to determine the precise asymptotic form for $E_n^{S_I}$. The procedure is detailed in \cite{dunne2014uniform} but we shall give a brief overview here. The first step is to impose a global boundary condition based on the periodicity of the potential 
\begin{equation}
    \Psi(\theta + L) = e^{i\alpha}\Psi(\theta),
\end{equation}
where $L$ is the periodicity and $\alpha\in[0,\pi]$ is the Bloch angle. In addition we demand a Bloch condition that relates the values of the wave function at some midpoint of the potential $\theta_\text{midpoint}$. In the potential \eqref{eq:schrodingerpotential}, this would be the half period $\theta_\text{midpoint}=\mathbb{K}(m)$. We shall therefore need to compute $u(\theta_\text{midpoint})$. By using the periodicities of the Jacobi elliptic functions we find\footnote{Note that because the Jacobi functions appear squared in the potential, we need not worry about the fact that Jacobi functions are strictly speaking anti-periodic across the interval $2K(m)$.}
\begin{equation} \label{eq:u0midpoint}
    u_0(\theta_\text{midpoint}) = \sqrt{2 S_I} \,,
\end{equation}
and
\begin{equation} \label{eq:u1midpoint}
    u_1(\theta_\text{midpoint}) = \frac{\log [S_I(1+\chi_-^2)/4]}{\sqrt{2 S_I}} \,,
\end{equation}
where $S_I$ is given by \eqref{eq:realunitonaction}. Expanding the boundary condition in terms of $\nu=N+\delta\nu + (\delta\nu)^2+\ldots$ allows us to determine $\delta\nu$ in terms of $g^2$. This can be used to compute the $N^\text{th}$ energy level
\begin{equation}
    E^\text{perturbative}_N (g^2) = \mathcal{E}(N,g^2) + \delta\nu \left[\frac{\partial \mathcal{E}(\nu, g^2)}{\partial \nu}\right]_{\nu=N} + \mathcal{O}((\delta\nu)^2)\,.
\end{equation}

The first ambiguity of $\mathcal{E}(N=0, g^2)$, located in the instanton-anti-instanton sector, is the imaginary part of $\delta\nu \left[\frac{\partial \mathcal{E}(\nu, g^2)}{\partial \nu}\right]_{\nu=N}$. By considering dispersion relations\footnote{$\mathcal{C}$ denotes a counter-clockwise closed contour around $g^2=0$. The first equality is simply a restatement of \eqref{eq:uniformexpansionceoffs} using Cauchy's theorem. Next we deform the contour up and down the positive real axis and around infinity to obtain the second equality.}
\begin{equation}
\begin{aligned}
    \mathcal{E}_k(N=0) &= \oint_\mathcal{C} \frac{\mathcal{E}(N=0, g^2)}{(g^2)^{k+1}} d(g^2) \\
    &= \frac{1}{i \pi} \int_0^{+\infty} \frac{\text{Disc}_0 \mathcal{E}(N=0, g^2)}{(g^2)^{k+1}} d(g^2)\,
\end{aligned}
\end{equation}
for the coefficients \eqref{eq:uniformexpansionceoffs}, we can determine an asymptotic form \cite{dunne2014uniform}. We calculate Stokes discontinuities more carefully in Section \ref{sec:stokesdiscontinuities}.

The resulting asymptotic expansion from the uniform WKB method are as follows. In the regime where $| S_{I}| < |S_{CI}|$, the perturbative energy coefficients are dominated by the following behaviour
\begin{equation} \label{eq:realasymptotic}
    E_n^{S_I} \approx A(\eta, \zeta)\left(\frac{1}{2 S_I}\right)^{n+1}\Gamma(n+1) \left( 1 + a^1_{I}(\eta, \zeta)\frac{2 S_I}{n} + \mathcal{O}\left(\frac{1}{n^2}\right)\right)\,,
\end{equation}
where
\begin{equation} \label{eq:realstokes1}
    A(\eta,\zeta) = - \frac{1}{\pi}\frac{16}{1+ \chi_-^2}\,.
\end{equation}
Because Equation \eqref{eq:u1midpoint} is an $\eta\rightarrow\eta-\zeta$ substitution compared to the single deformation case, the same follows for Equation \eqref{eq:realstokes1}.  Working in higher order in the wave function allows a determination of the sub-leading contributions.  E.g.  $a_I^1(\eta,\zeta)$, which is a correction due to an instanton-anti-instanton $[I\overline{I}]$ event, is determined from $u_2(\theta_\text{midpoint})$ which did not however prove easy to analytically evaluate. 

Furthermore, from our numerical analysis, we predict that the 1-complex uniton and the 2-complex uniton behave as
\begin{equation} \label{eq:complexasymptotic}
\begin{aligned}
      E_n^{S_{CI}} &\approx B(\eta, \zeta) \left(\frac{1}{ S_{CI}}\right)^{n+1/2}\Gamma(n+1/2) ( 1 + a^1_{CI}(\eta, \zeta) \frac{2 S_{CI}}{n} +  \mathcal{O}(n^{-2}) ) \,, \\
      E_n^{2S_{CI}} &\approx - A(\eta, \zeta)\left(\frac{1}{2 S_{CI}}\right)^{n+1}\Gamma(n+1)( 1 + a^1_{2CI} (\eta, \zeta) \frac{4 S_{CI}}{n} +  \mathcal{O}(n^{-2}) ) \,,
\end{aligned}
\end{equation}
where
\begin{equation}
    B(\eta, \zeta) = -\frac{\sqrt{A( \eta, \zeta)}}{\pi} = \frac{ - 4i }{\sqrt{\pi^3(1+ \chi_-^2)}}\,.
\end{equation}
We emphasise that these predictions for the asymptotic behaviour are not derivable from any conventional uniform WKB, but are  based on  empirical evidence.  

\begin{figure}[tbp]
    \centering
        \subfigure{
        \includegraphics[height=1.6in]{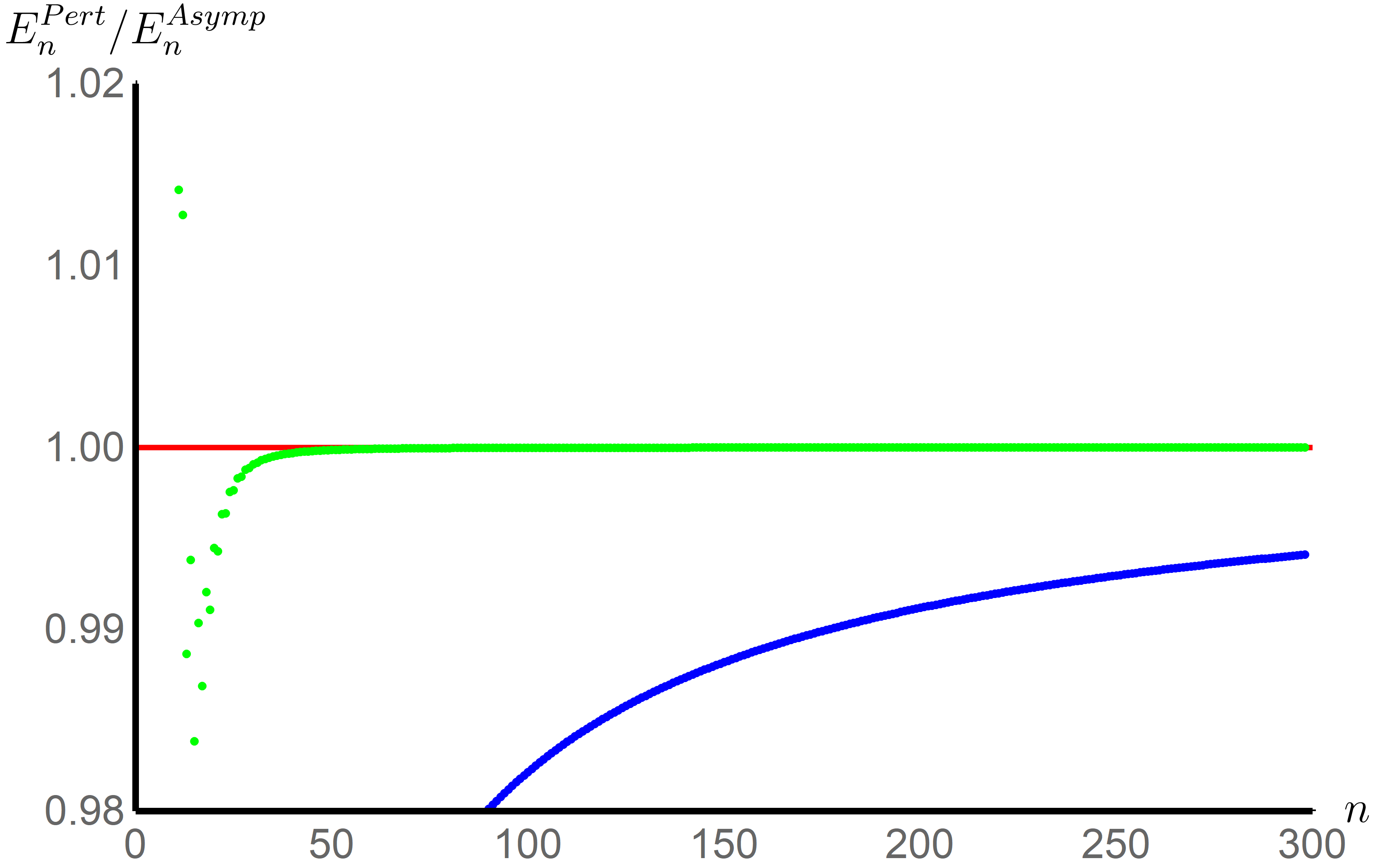}
        }
        \subfigure{
        \includegraphics[height=1.6in]{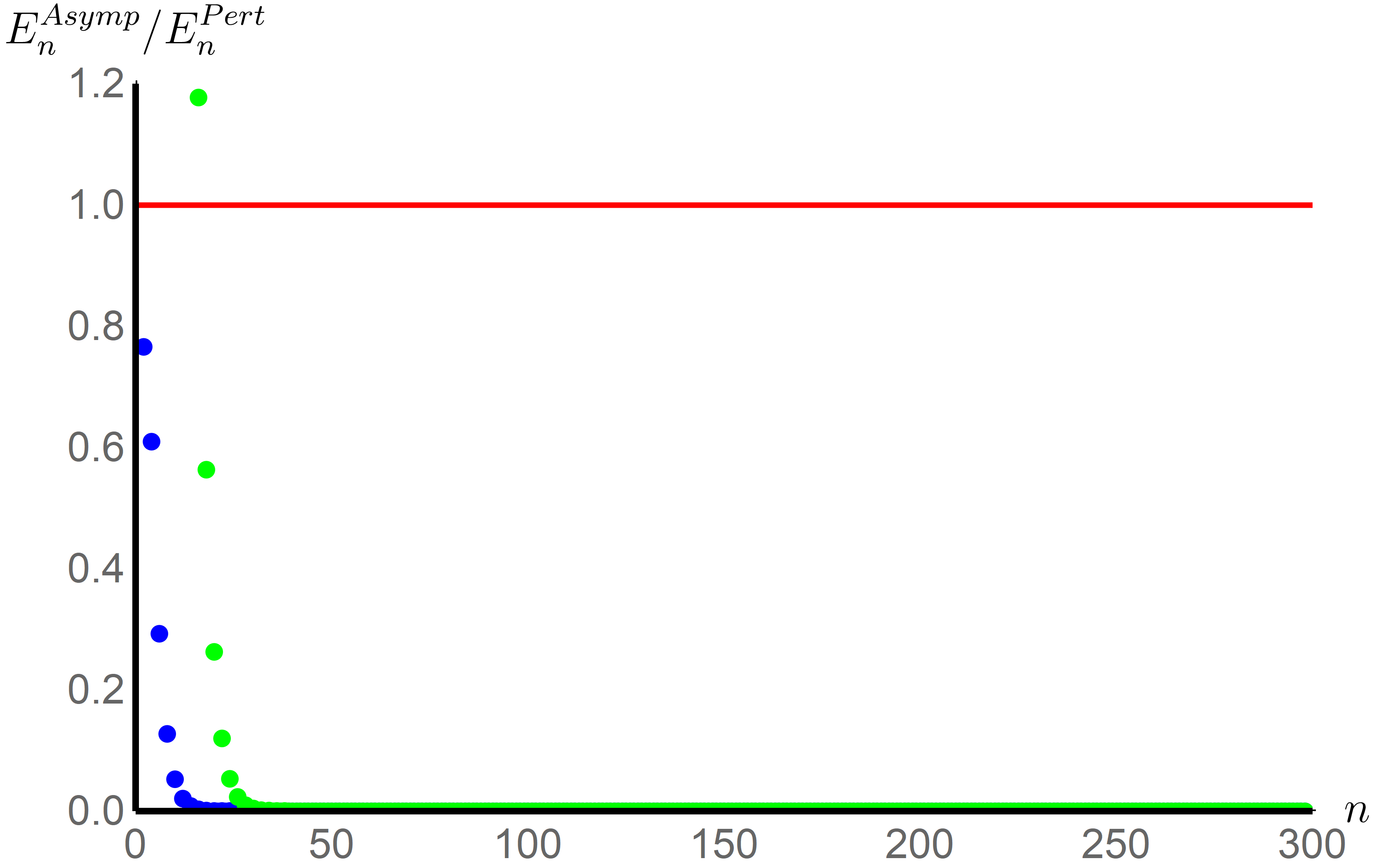}
        }  
    \caption{Here we study the convergence of the perturbative coefficients to the  asymptotic prediction \eqref{eq:realasymptotic}. Their ratio is given by the blue dots. To accelerate the convergence we employ the second Richardson Transformation, here given in green. In both plots we follow the trajectory where $\zeta=1/5$. In the left plot $\eta=19/100$, we obtain virtually the same results for $\eta=1/5$. Here, we are in the first region of Figure \ref{fig:regionplot} where $| 2S_{I}| < |S_{CI}|$. Therefore, the real uniton is dominant, both on and off the critical line. In the right plot we show $\eta=2/5$, which is in region 2. Using the same asymptotic expansion, we see that the approximation fails, because the real uniton is dominant anymore.}
    \label{fig:asymptotic1}
\end{figure}

\begin{figure}[tb]
    \centering
        \subfigure{
        \includegraphics[height=1.6in]{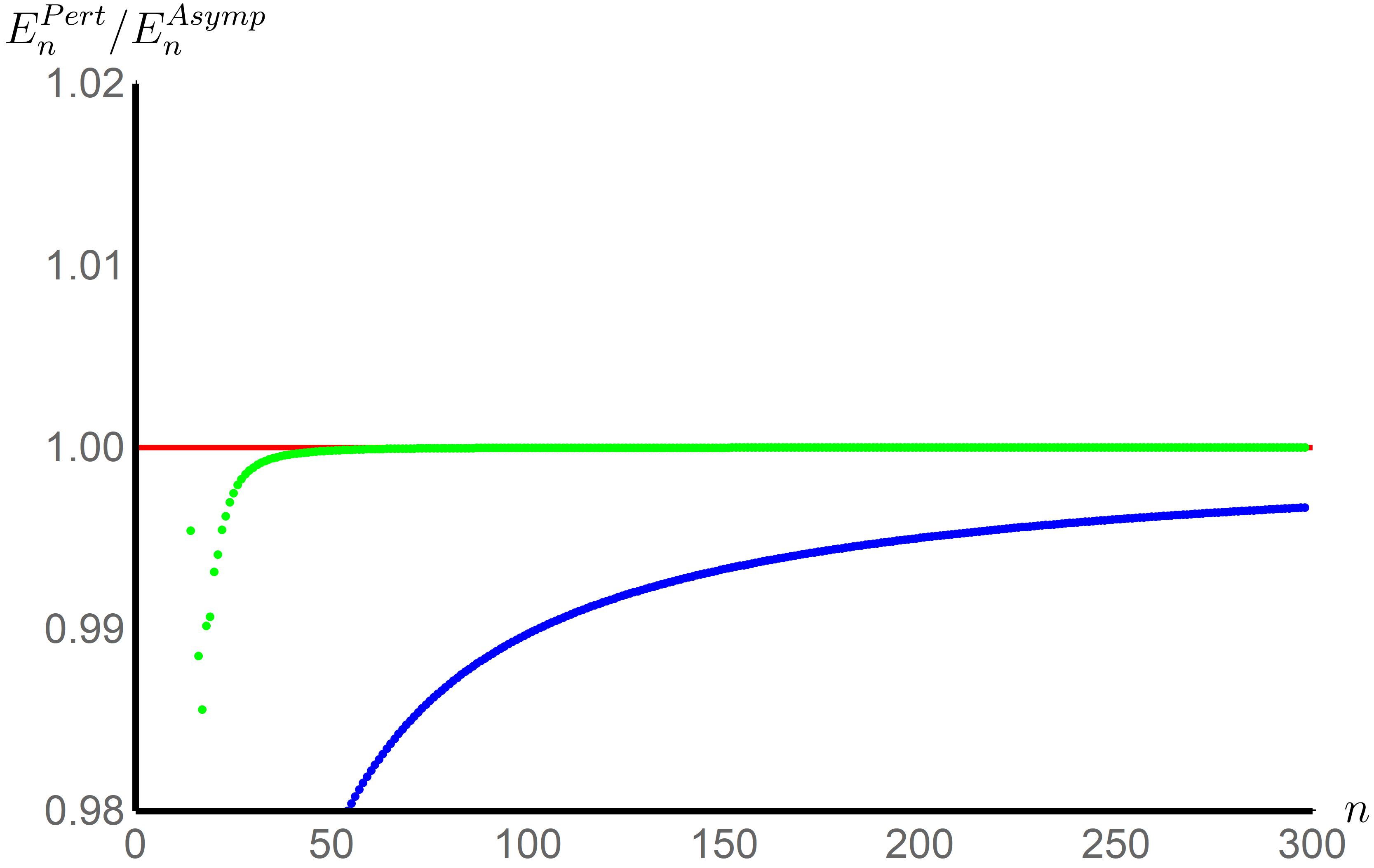}
        }
        \subfigure{
        \includegraphics[height=1.6in]{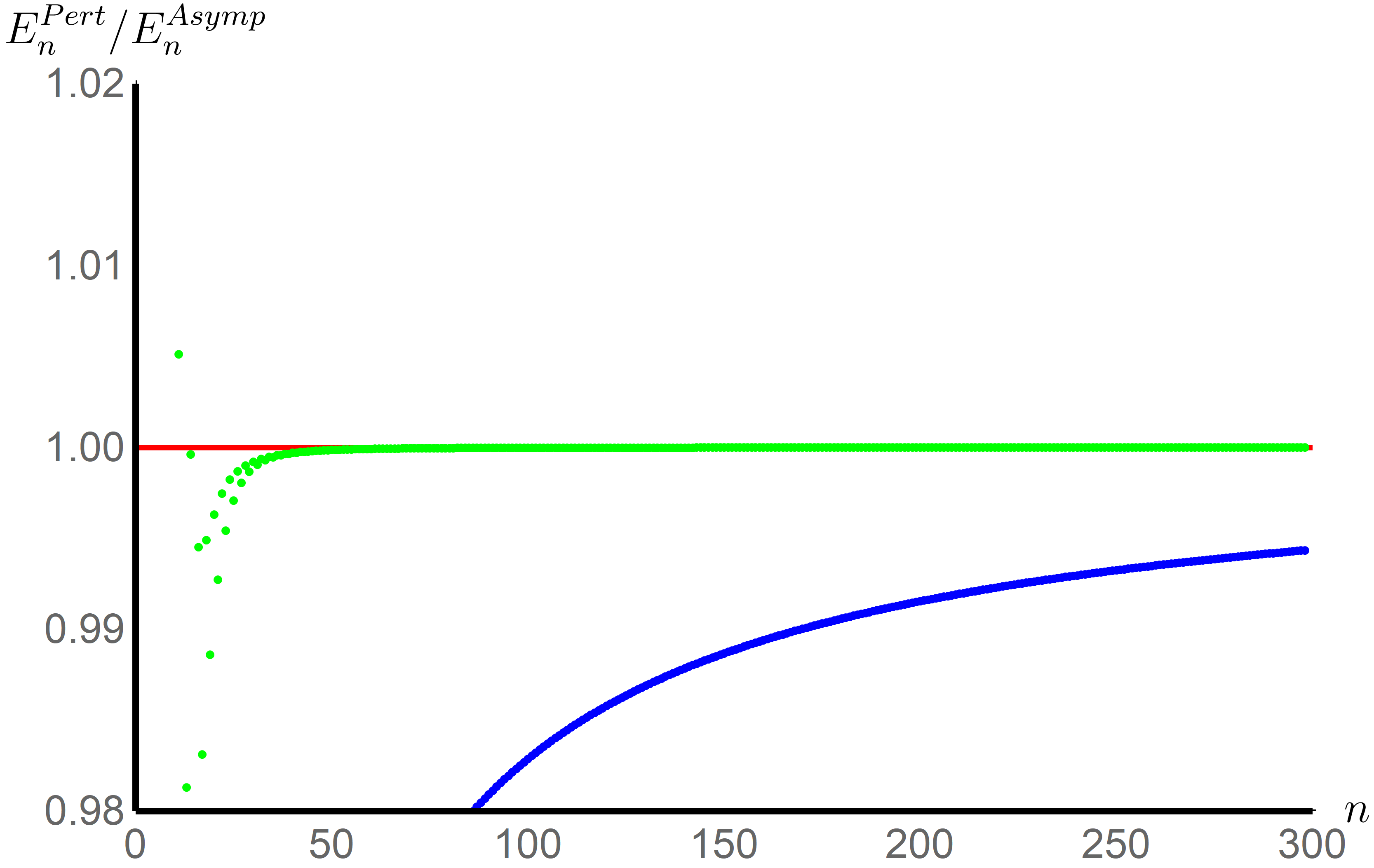}
        } 
    \caption{Colours are as in Figure \ref{fig:asymptotic1}.  We follow the critical line $\varkappa=\zeta=\eta$. In the first plot $\varkappa=1/2\sqrt{3}$, which is on the border of regions 1 and 2 of Figure \ref{fig:regionplot} where $S_I=-S_{CI}=8\pi/3\sqrt{3}$. In the second plot $\varkappa=2/5$, which is firmly in region 2. In both cases $|S_I|<|2S_{CI}|$. Because along the critical line there is no complex 1-uniton contribution, the real uniton is dominant.}
    \label{fig:asymptotic2}
\end{figure}

\begin{figure}[tb]
    \centering
        \subfigure{
        \includegraphics[height=1.6in]{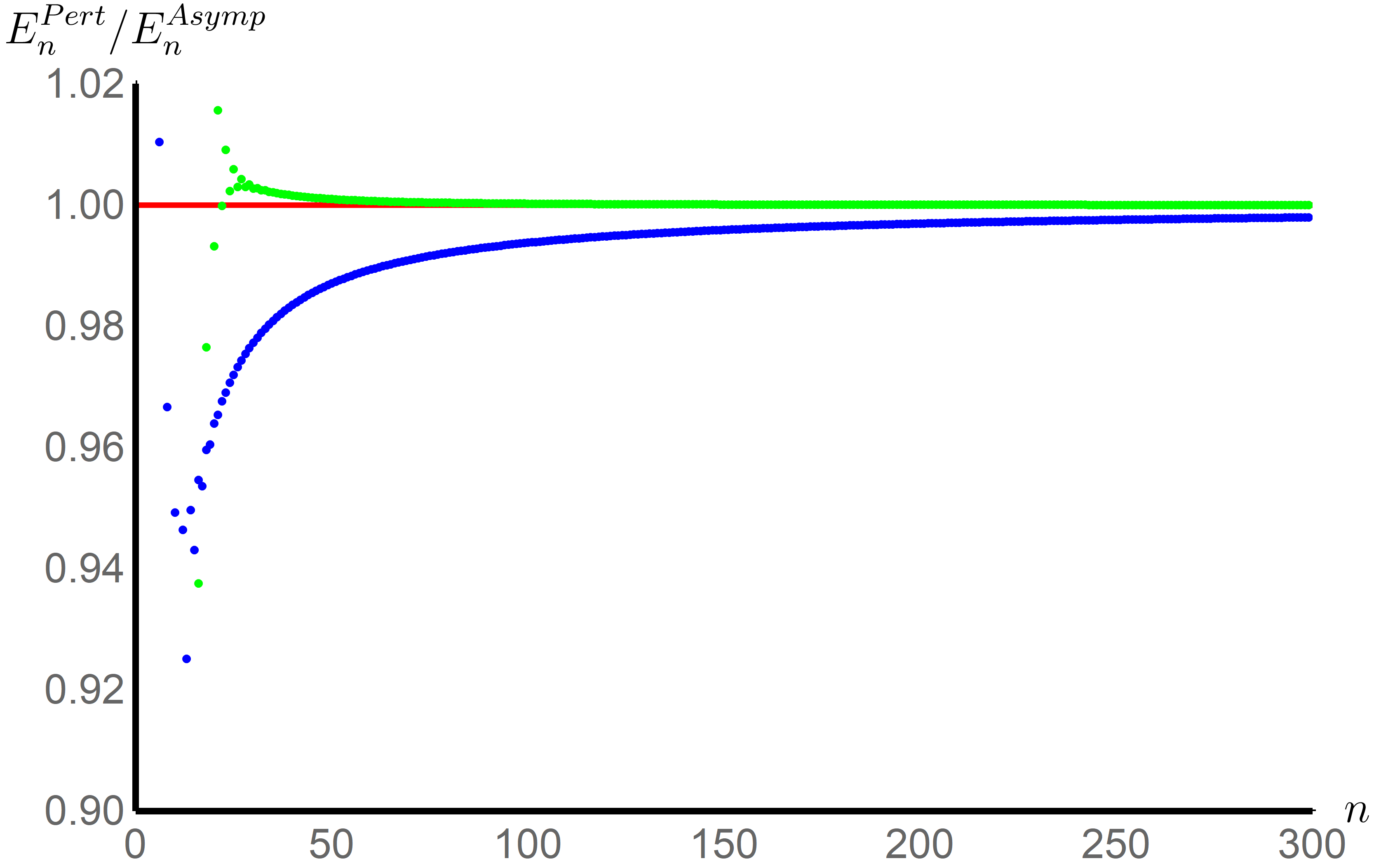}
        }
        \subfigure{
        \includegraphics[height=1.6in]{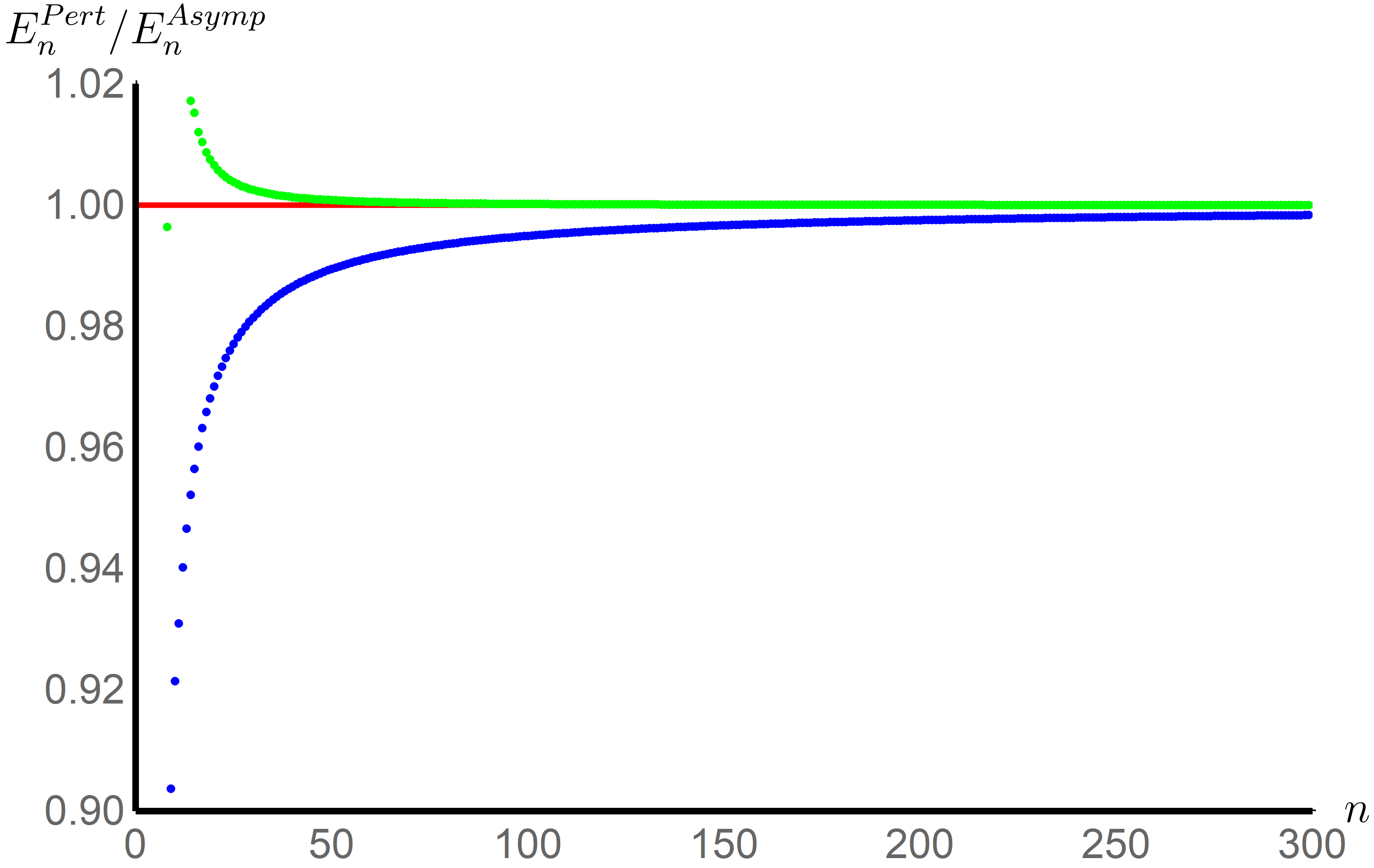}
        } 
    \caption{Colours are as in Figure \ref{fig:asymptotic1}.  In both plots we follow the trajectory where $\zeta=1/5$. In the first plot $\eta=2/5$, in the second plot $\eta=1/2$. We are thus in the second and third region of Figure \ref{fig:regionplot}. Because $| 2S_{I}| > |S_{CI}|$, the complex uniton is dominant.}
    \label{fig:asymptotic3}
\end{figure}

\begin{figure}[tb]
    \centering
        \subfigure{
        \includegraphics[height=1.6in]{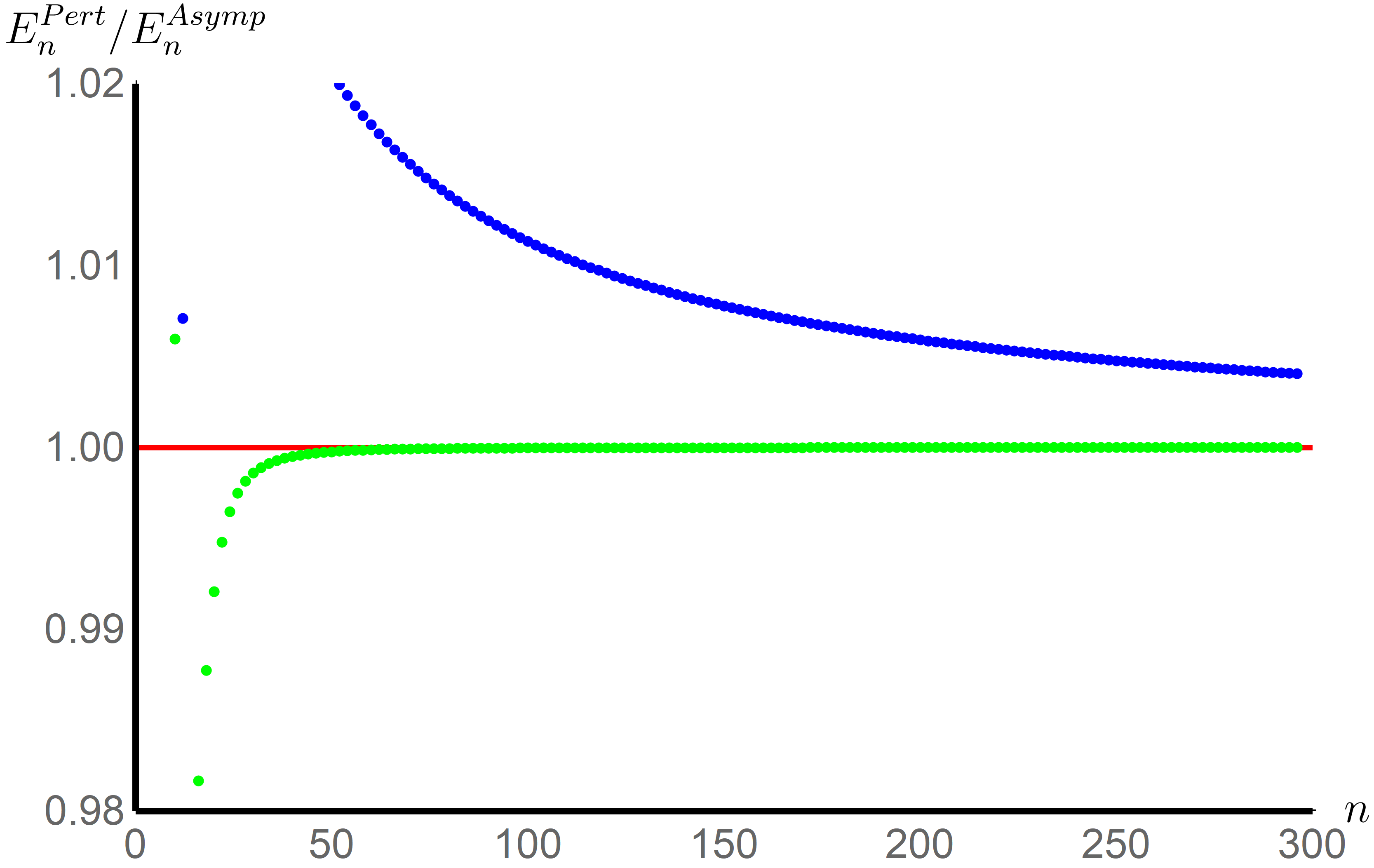}
        }
        \subfigure{
        \includegraphics[height=1.6in]{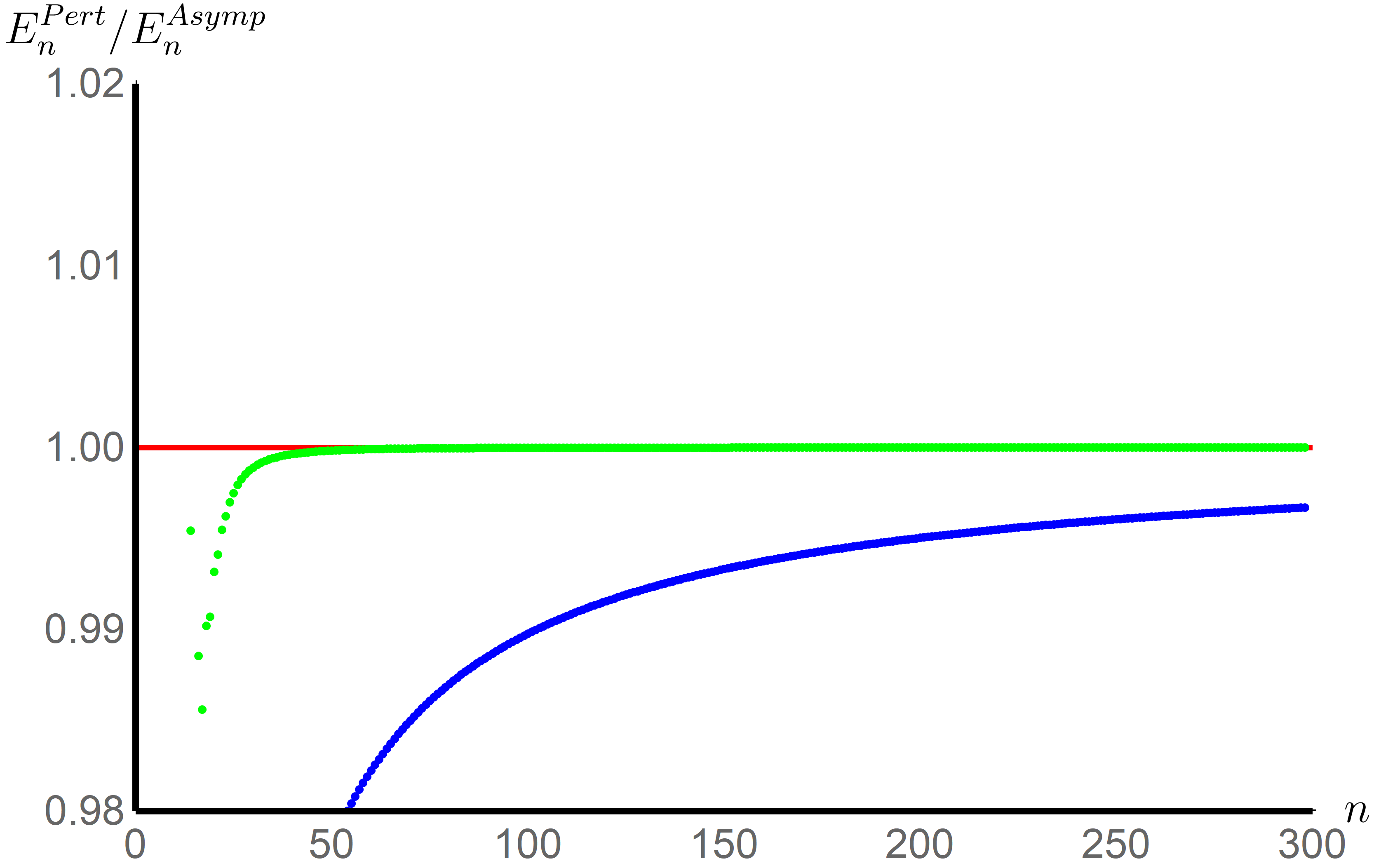}
        } 
    \caption{Colours are as in Figure \ref{fig:asymptotic1}. Here, we study the behaviour along the critical line $\varkappa=\eta=\zeta$. In the first plot, $\varkappa=1/2$, the second plot $\varkappa=2/\sqrt{3}$. We know that in regions 3 and 4 of Figure \ref{fig:regionplot} along the critical line the complex 2-uniton is dominant. This is verified by the second figure. However, $\varkappa=1/2$ is a very special point indeed as it acquires equal contributions from the complex 2-uniton and the real uniton. Because $S_I=-S_{CI}=\pi$, the only difference is that these contributions are non-alternating and alternating respectively. These precisely cancel out, leading to a series in $g^4$, as already foreshadowed in Equation \eqref{eq:EPertubativeExamples}.}
    \label{fig:asymptotic4}
\end{figure}

In Figures \ref{fig:asymptotic1}, \ref{fig:asymptotic2}, \ref{fig:asymptotic3} and \ref{fig:asymptotic4} we compare the asymptotic expression $E_n^{S_I}$ from Equation \eqref{eq:realasymptotic} with the actual values $E_n^\text{pert}$ obtained from the perturbative calculation with the BenderWu package. We plot the ratio and study its convergence to 1. Doing so in Figure \ref{fig:asymptotic1}, we numerically verify Equation \eqref{eq:realasymptotic}.   The convergence of the raw data (shown in blue in Figure \ref{fig:asymptotic1}) is somewhat slow - a situation that could be improved by determining $a_I^1(\eta, \zeta)$.

However, convergence can be improved spectacularly by  using a Richardson transform (see e.g. \cite{aniceto2018primer, aniceto2017asymptotics}). Indeed, with just the second Richardson transform (shown in green in Figure \ref{fig:asymptotic1})   we see convergence between the 300\textsuperscript{th} order perturbative data and asymptotic predictions with a typical accuracy of between $4\cdot 10^{-7}$ and $9\cdot 10^{-7}$. This is an impressive agreement approaching the theoretical uncertainty resulting from using the second Richardson transformation (results should be accurate to $O(1/n^3)$, hence for $n=300$ this is  $1/300^3 \approx 4\cdot 10^{-8}$ ).  Further theoretical uncertainty arises from the undetermined sub-leading terms in the asymptotic prediction. For the single deformed potential in  \cite{demulder2016resurgence} we have $a_I^1(\eta, \zeta=0) = \frac{1}{24}\left(-23 +77 \eta^2 + \frac{8}{1+\eta^2}\right)$.  Under the assumption that $a_I^1(\eta, \zeta)$ is of the same order as $a_I^1(\eta, 0)$, we can estimate the magnitude of this uncertainty, which also matches well with the measured accuracy\footnote{To give an impression of the magnitude of this discrepancy, $a_I^1(0, 0) = -15$, $a_I^1(1/5, 0) \approx -12.2$, $a_I^1(1/2, 0) = 2.65$ and $a_I^1(1, 0) = 58$. }.

As an additional remark, in Figure \ref{fig:borel1} we saw that the single complex instanton contribution disappears at the critical line $\eta=\zeta$. We suspect that a consequence of this is that the 1-uniton behaviour of Equation \eqref{eq:realasymptotic} remains dominant until $|2 S_{I}| < |2 S_{CI}|$ if $\eta=\zeta$. Therefore, the real uniton is dominant not only in region 1 of Figure \ref{fig:regionplot}, but also in region 2 along the critical line. This is corroborated by the numerical analysis displayed in Figure \ref{fig:asymptotic2}. 

 In \cite{basar2013resurgence}, the potential along the critical line is studied. It is observed that the potential respect a symmetry that sends $m\rightarrow m'$, $g^2\rightarrow-g^2$ and $\theta\rightarrow i \theta$. This $\mathbb{Z}_2$ duality interchanges the real and the complex instanton solutions and therefore also interchanges their actions. It follows that $m=\frac{1}{2}$ is the fixed point of the duality, which can be traced back to $\varkappa=\frac{1}{2}$. We can also reformulate the $m\rightarrow m'$ transformation in terms of $\varkappa$ by sending $\varkappa\rightarrow\frac{1}{4\varkappa}$. Note that the asymptotic expansion of the energy \eqref{eq:collectasympexpansion}, \eqref{eq:realasymptotic}, \eqref{eq:complexasymptotic}, respects this symmetry only if we ignore the $E_n^{S_{CI}}$ contribution, which is precisely what happens on the critical line. Moreover, at the fixed point $m=\frac{1}{2}$, or $\varkappa=\frac{1}{2}$, we have that $E_n^{S_{I}}$ and  $E_n^{S_{2CI}}$ contribute equally. 

The computations that support the predictions given by Equation \eqref{eq:complexasymptotic} are exhibited in Figures \ref{fig:asymptotic3} and \ref{fig:asymptotic4}. Here, we investigate the regimes in which the 1- and 2-complex unitons are dominant. This corresponds to regions 3 and 4 and region 2 off the critical line of Figure \ref{fig:regionplot}.

At the boundary between region 1 and 2 in Figure \ref{fig:regionplot}, we would expect from the asymptotic expansions \eqref{eq:realasymptotic} and \eqref{eq:complexasymptotic} that the real 1-uniton and the complex 1-uniton interact approximately at the same order. For example, the point $\zeta=0$, $\eta_c=0.274$, considered in \cite{demulder2016resurgence}, belongs to this family. However, because the asymptotic expansions do not precisely match, there is not a perfect cancellation of alternating and non-alternating terms like there is at $\varkappa=\eta=\zeta=1/2$. The perturbative series  along this border is thus in $g^2$ and not in $g^4$.

Combining all the information in the analyses of Equations \eqref{eq:realasymptotic} and \eqref{eq:complexasymptotic} and Figures \ref{fig:asymptotic1}, \ref{fig:asymptotic2}, \ref{fig:asymptotic3} and \ref{fig:asymptotic4}, we thus arrive at the following picture:  across   the $\zeta=1/5$ trajectory, varying $\eta$, we find that the real uniton is dominant in region 1 of \ref{fig:regionplot}, while the complex 1-uniton is dominant in regions 2, 3, and 4. Along the critical line, there is no 1-complex uniton, thus the real uniton is dominant in regions 1 and 2, while the complex 2-uniton is dominant in regions 3 and 4. 

Lastly, let us compare the perturbative calculation with the asymptotic expansion \eqref{eq:realasymptotic} to say something about $a^1_{I}(\eta, \zeta)$. Equating the predicted asymptotic to the perturbative expansion and rearranging implies that  
\begin{equation}
    \frac{(2 S_I)^{n+1}}{\Gamma(n+1) A(\eta, \zeta)} E^\text{pert}_n -1 \approx a^1_{I}(\eta, \zeta) \frac{2 S_{I}}{n}\,.
\end{equation}
By performing a Richardson transformation on the left hand side we can make predictions about $a^1_{I}(\eta, \zeta)$ in the regime where the real uniton dominates. The same can be done for $a^1_{CI}(\eta,\zeta)$. Example results are given in Tables \ref{table:reala1} and \ref{table:clpxa1}. In addition, we can predict  $a^1_{2CI}$ along the critical line for $\varkappa>1/2$. For example, we expect $a^1_{2CI}=-0.0581325$ for $\varkappa=\sqrt{3}/2$.   Whilst the $a^1_{I}(\eta, \zeta)$ can in principle be determined from uniform WKB, there is not yet a systematic understanding of how to determine the $a^1_{CI}$ and $a^1_{2CI}$ . 

\begin{table}[tb]
\begin{center}
\begin{tabular}{| l|l| } 
\hline
 $\eta$ & $a_{I}^1(1/5, \eta)$ \\ 
 \hline
 0 & -0.509487 \\ 
 1/100 & -0.497592  \\ 
 1/20 & -0.444087 \\
 1/5 & -0.157644 \\
 \hline
\end{tabular}
\caption{Numerical predictions for $a_{I}^1(1/5, \eta)$ for selected values of $\eta$ and $\zeta$. We used the 10th Richardson transform and 300 perturbative coefficients. The $\eta=0$ result agrees with the exact result from \cite{demulder2016resurgence}.}
\label{table:reala1}
\end{center}
\end{table}

\begin{table}[tb]
\begin{center}
\begin{tabular}{ |l|l| } 
\hline
 $\eta$ & $a_{CI}^1(0.4, \eta)$ \\ 
 \hline
 0.2 & 0.204395 \\ 
 0.38 & 7.20539  \\ 
 0.39 & 14.9317 \\
 0.395 & 34.06471 \\
 0.4 & 431.158 \\
 0.41 & 15.3672 \\ 
 \hline
\end{tabular}
\caption{Numerical predictions for $a_{CI}^1(0.4, \eta)$ for selected values of $\eta$ and $\zeta$. We used the 10th Richardson transform and 150 perturbative coefficients. Notice the sudden jump at the Critical point $\eta=\zeta$, because the 1-uniton approximation brakes down at this point. Had we used the $E_{I}$ approximation, we would have obtained $a_{I}^1(0.4, 0.4)=54.9459$. This might suggest the coefficients $a_{I}^1$ and $a_{CI}^1$ have a simple pole at $\eta=\zeta$. However, it should be noted the numerics are quite unstable around the critical point as the asymptotic series approximates the perturbative series much slower.}
\label{table:clpxa1}
\end{center}
\end{table}

\subsection{Stokes Discontinuities} \label{sec:stokesdiscontinuities}
In this section we will make a schematic attempt to show the significance of our results and how this might be implemented to expose the resurgent structure of the system. We make a simplification to further explain the significance of the coefficients $A$ and $B$ in the asymptotic forms in Equations \eqref{eq:realasymptotic} and \eqref{eq:complexasymptotic}.  Let us consider new asymptotic expansions in $z=g^2$ whose coefficients $ E_n^{S_{I}}$, $ E_n^{S_{CI}}$ and $ E_n^{S_{2CI}}$ are, for all $n$ and not just large enough $n$, given by the leading behaviour of Equations \eqref{eq:realasymptotic} and \eqref{eq:complexasymptotic} (the sub-leading behaviour will be discussed later):
\begin{equation}
\begin{aligned}
    \tilde{E}_I(z) =\sum_{n=0}^\infty E_n^{S_I} z^n \, ,  \quad 
    \tilde{E}_{CI}(z) =\sum_{n=0}^\infty E_n^{S_{CI}}z^n \, , \quad 
    \tilde{E}_{2CI}(z)= \sum_{n=0}^\infty E_n^{S_{2CI}}z^n\, . 
\end{aligned}
\end{equation} 
Their Borel transforms, using   Equation \eqref{eq:BorelTransform} with   $s=\hat{g}^2$, are given by
\begin{equation}\label{eq:branch}
\begin{aligned}
    \hat{E}_I(s) =\frac{A(\eta, \zeta)}{2 S_I - s} \, , \quad 
    \hat{E}_{2CI}(s)&=\frac{-A(\eta, \zeta)}{2 S_{CI} - s}  \, , \quad 
    \hat{E}_{CI}(s)&=\frac{B(\eta, \zeta)\sqrt{\pi}}{\sqrt{ S_{CI} - s}} \,  .
\end{aligned}
\end{equation}
We remind the reader that $S_{CI}$ is a negative real number if $\eta$ and $\zeta$ are real  whereas $S_I$ will be positive real, thus explaining the locations of the Borel poles in our preceding Borel analysis. 

Recalling the re-summation in a direction $\vartheta$  of a  series $\tilde{\psi}(z)$ is given by
\begin{equation}\label{eq:directionalsum}
    \mathcal{S}_\vartheta \tilde{\psi}(z) = \frac{1}{z} \int_0^{e^{i \vartheta} \infty} ds \,e^{-s/z} \hat{\psi} (s)\, ,
\end{equation}
we can also see that the Borel resummation of $\tilde{E}_I$ is singular only along the positive real axis (i.e. there is a Stokes ray along $\vartheta=0$), whilst the Borel resummations of $\tilde{E}_{CI}$ and $\tilde{E}_{2CI}$ are singular only along the negative real axis (i.e. a Stokes ray along $\vartheta=\pi$).  Resummations along these rays are inherently ambiguous. To study these ambiguities we adopt lateral Borel resummations $\mathcal{S}_{\vartheta^\pm} \tilde{\psi}(z) = \mathcal{S}_{\vartheta \pm \epsilon} \tilde{\psi}(z)$. We thus compute that non-perturbative ambiguity due to the 1-uniton is
\begin{equation} \label{eq:leadingambiguity}
    (\mathcal{S}_{0^+}- \mathcal{S}_{0^-})\tilde{E}_{I}(z)= - \frac{2 \pi i}{z} \text{Res}_{s=2S_I}\left[ e^{-s/z} \frac{A(\eta, \zeta)}{2 S_I - s}\right] = \frac{2 \pi i}{z} A(\eta, \zeta)  e^{- 2S_I/z}.
\end{equation}
The sign after the first equality is due to the clockwise integration contour. Similarly
\begin{equation}
    (\mathcal{S}_{\pi^+}- \mathcal{S}_{\pi^-})\tilde{E}_{2CI}(z) = - \frac{2 \pi i A(\eta, \zeta)}{z}  e^{-2 S_{CI}/z}.
\end{equation}
To resum $\hat{E}_{CI}(z)$, we choose the branch cut to go from $z=S_{CI}$ to negative infinity. (Hence the branch cut of the square root function lies along the positive real axis). The integral from $0$ to $S_{CI}$ does not contribute. For the remaining bit, we switch to an integration variable $x=S_{CI}-s$, and solve the integral. Performing the outlined procedure then gives
\begin{equation}
      (\mathcal{S}_{\pi^+}- \mathcal{S}_{\pi^-})\tilde{E}_{CI}(z) = \frac{1}{z} \int_\gamma ds\, e^{-s/z} \frac{B(\eta, \zeta) \sqrt{\pi}}{\sqrt{S_{CI}-s}} \\
      =  \frac{2  B(\eta, \zeta)\sqrt{\pi}}{\sqrt{z}}  e^{- S_{CI}/z} 
\end{equation}

The reason we are interested in computing quantities such as  $(\mathcal{S}_{\vartheta^+}- \mathcal{S}_{\vartheta^-})\tilde{E}(z)$ is that this might shed light on the nature of the Stokes automorphism $\mathfrak{S}_\vartheta$ which is defined by 
\begin{equation} \label{eq:stokesauto}
     \mathcal{S}_{\vartheta^+}- \mathcal{S}_{\vartheta^-} = - \mathcal{S}_{\vartheta^-} \circ \text{Disc}_{\vartheta} = \mathcal{S}_{\vartheta^-} \circ (\mathfrak{S}_\vartheta - \text{Id}).
\end{equation}
The Stokes automorphism describes the analytic structure of the ambiguities as a Stokes ray is crossed \cite{aniceto2017asymptotics, dunne2017wkb}. 

For the undeformed model \cite{cherman2015decoding}, it was conjectured that the Stokes automorphism of the perturbative sector is due to a contribution $\mathcal{E}_{[I \overline{I}]}(z)$ of the intantin-anti-instanton sector. This means there would be some expansion around a secondary saddle point that impacts the perturbative series $E_{[0]}(z)$ of the perturbative sector $[0]$ which was calculated above. This intricate interplay of sectors from different saddle point is part of the rich study of resurgence as it is the starting point of establishing large-order relations.

On the field theory side, different contributions are ascribed to the fractons which constitute the unitons. Although typically these contributions are combined in sectors classified by $\pi_2$, we re-emphasise that for the $SU(2)$ PCM this group is trivial. Instead we classify the sectors through $\pi_3$. It is expected within the resurgence paradigm \cite{cherman2015decoding, dunne2017wkb, aniceto2017asymptotics, aniceto2018primer, zinn2004multi1, zinn2004multi2, bogomolny1980calculation} that ambiguities should cancel within each sector. That means that the fracton-anti-fracton event should carry an ambiguity that matches the ambiguity obtained by resumming the perturbative sector given by Equation \eqref{eq:leadingambiguity}. 

The contributions due to discontinuities along individual (branch) singularities $w$ are often described in terms of Alien derivatives $\Delta_w$ defined by
\begin{equation}
    \mathfrak{S}_\vartheta = \exp \left(\sum_{\omega\in\text{sing}_\vartheta} e^{-w/z} \Delta_w \right),
\end{equation}
where $\text{sing}_\vartheta$ is the set of singular (branch) points in the direction $\vartheta$. Typically it is of the form $\text{sing}_\vartheta = \{ n A, | n\in \mathbb{Z}_{n\geq 1} \}$, and $A=2S_I$ might be some action. The Alien derivatives hence generate the Stokes automorphism  (for a modern review see \cite{dorigoni2014introduction}). The alien derivative is then expected \cite{cherman2015decoding} to look like
\begin{equation}
    \Delta_{2S_I}E_{[0]}(z) = s_1 E_{[I \overline{I}]}(z),
\end{equation}
where $s_1$ is the Stokes constants which might be related to $A(\eta,\zeta)$.

\subsection{Stokes Graphs} \label{sec:stokesgraphs} 
 
Stokes graphs provide a graphical method to understand the Borel resumability and jumping phenomena  associated to the WKB solutions of a Schr\"odinger equation as encoded by the DDP forumla \cite{DDP93} for the behaviour of Voros symbols \cite{voros1983return} across Stokes rays.  As parameters in the Schrodinger potential are varied, the Stokes graph can undergo topology changes, or mutations, which have a rich mathematical structure  \cite{Bridgeland,iwaki2014exact} and are captured by the Stokes automorphism \eqref{eq:stokesauto} described above.  From a physics perspective,  the seminal work \cite{Gaiotto:2009ma} showed that the mutations of Stokes graphs are intimately related to BPS spectrum of ${\cal N}=2$ four-dimensional gauge theory, where the Stokes automorphism describes wall-crossing phenomena. 

Let us review some terminology required to explain what is meant by Stokes graphs.  We consider a  Schr\"odinger equation defined over a Riemann surface $\Sigma$ with local coordinate $w$, 
\begin{equation}
  \left( \frac{d}{dw^2}  - \frac{1}{g^4}Q(w,g^2)\right) \Psi(w) = 0 \, , 
\end{equation}
where $g^2$ is a small parameter in which we construct formal perturbative expansions. In a general theory $Q(w,g^2)$ itself can be expanded in $g$, though we are interested here in the case where $Q(w,g^2) \equiv Q_0(w)$ is given by the classical momentum $p(w) = \sqrt{  E -V(w)}$.   Under coordinate transformations $w \rightarrow \tilde{w}(w)$, $Q_0$ transforms holomophically with weight 2 and thus defines a meromorphic quadratic differential 
\begin{equation}\label{eq:quaddiffsch}
    \phi_{Sch} = p(w)^2 dw \otimes dw \, .  
\end{equation}
Trajectories of $\phi_{Sch}$ are defined as curves $\gamma$ of constant phase   in the sense that if $\partial_t$ is tangent to $\gamma$ then $\lambda\cdot \partial_t = e^{i \vartheta}$ where $\phi_{Sch}= \lambda \otimes \lambda$.  Equivalently they can be defined by 
\begin{equation}
    \text{Im} \left[ \int^w dw \,p(w) \right] = \text{constant}\, , 
\end{equation}
and these provide a foliation of $\Sigma$.  Generically these trajectories will start and end at poles of $p(w)$, but a special role is played by   {\em Stokes trajectories} satisfying
\begin{equation} \label{eq:stokeslines}
    \text{Im} \left[ \int^w dw\, p(w) \right] = 0\, , 
\end{equation} 
which have at least one end point at a zero of $p(w)$, which is also called a turning point.  A Stokes trajectory is a {\em saddle} if both end points are located at zeros. It is {\em regular} if these zeros are different and it is {\em degenerate} if it is a loop.  Given $ \phi_{Sch}(w)$, we define the associated  {\em Stokes graph}, $G[ \phi_{Sch}]$, as a graph with vertices comprised of zeros and poles of $ \phi_{Sch}$ and edges comprised of Stokes trajectories. 
 
It is useful to consider the effect on the Stokes graph of  rotating $g^2$ into the complex plane. An equivalent way to see this is to define the Stokes graph in a direction $\vartheta$, $G_\vartheta[\phi_{Sch}] = G[e^{2i \vartheta}\phi_{Sch}]$ whose edges  satisfy
\begin{equation}
    \text{Im}\left[ e^{i\vartheta}  \int_a^w  dw \, p(w) \right] = 0  \, ,
\end{equation}
where $a$ is a zero of $p(w)$.  The crucial linkage   is that, if   $G_\vartheta$ has  no saddles,   then the formal WKB solutions to the Schrodinger system are Borel summable in the direction $\vartheta$ in the sense of Equation \eqref{eq:directionalsum} (this is explained for general surfaces $\Sigma$ in \cite{iwaki2014exact} reporting on a result attributed to Koike and Sch\"afke \cite{Koike}). Along Stokes rays, however, a saddle will emerge. As $\vartheta$ is varied across the ray, the topology of $G_\vartheta$ will undergo a transition (known as a flip for a regular saddle or a pop for a degenerate saddle).

Let us sketch the schematic structure of the Stokes graphs applied to the case at hand for which we have 
\begin{equation}
    p(w)^2  =  E- \text{sd}^2(w)(1+ \chi_-^2\text{sn}^2(w)) \, . 
\end{equation}
 Because $p(w)$ is an elliptic function with periodic identification $w\sim w + 2 \mathbb{K}(m) \sim w + 2 i \mathbb{K}(m') $, it will suffice to study it in its fundamental domain. For $\eta \neq \zeta$ there are two distinct poles located at $w =i \mathbb{K}(m')  $ and $ w=   \mathbb{K}(m) + i \mathbb{K}(m')  $. For $E\neq 0$ and $\eta \neq \zeta$ there are generically four zero's which are given by solutions of 
\begin{equation}
    r^4 (\zeta - \eta) + r^2 (1+m E) - E = 0 \, , \quad  r=\text{sn}(w \mid m) \, . 
\end{equation}
 In the range\footnote{Here we view $E$ as a parameter that can be continuously varied, and we find taking a small positive $E$ helps in regulating the diagrams.} $0<E< E_c=  1+ (\eta + \zeta)^2$, two of these zeros are located along the $\text{Im}(w)=0$ axis symmetrically distribute about the half period $w= \mathbb{K}(m) $, with the two remaining zeros in the $\text{Re}(w)=0$ axis symmetrically distributed about $w = i\mathbb{K}(m') $. When $E= 1+ (\eta + \zeta)^2$, the two reals zeros coalesce at $w=\mathbb{K}(m)  $  and if the energy increases still further this single zero proceeds to acquire an imaginary part and approach the pole at   $\mathbb{K}(m) + i \mathbb{K}(m')$  
 
\begin{figure}[tb]
    \centering
     \subfigure[$\vartheta=0$]{
        \includegraphics[height=1.8in]{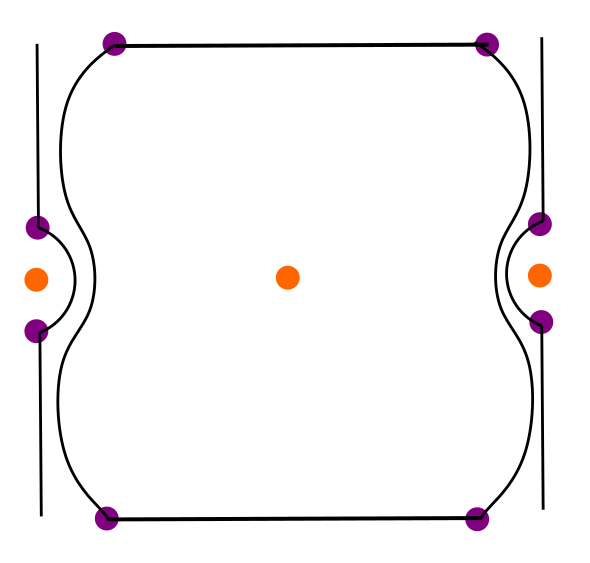}
        } 
        \subfigure[$0 < \vartheta < \pi$]{
        \includegraphics[height=1.8in]{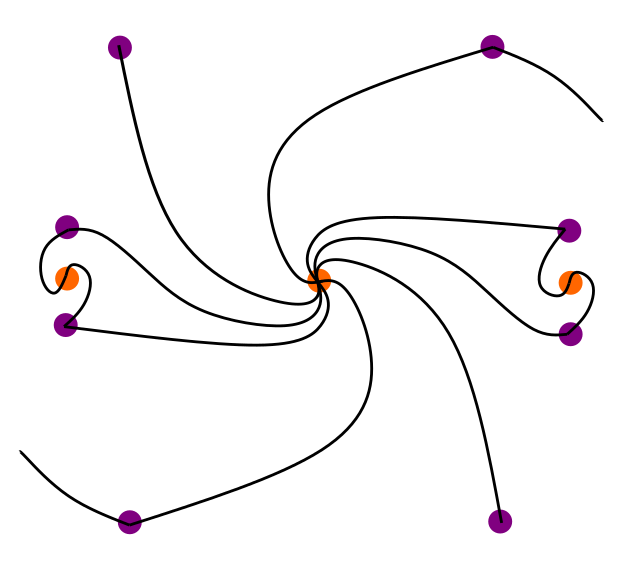}
        } \\
        \subfigure[$\vartheta=\pi$]{
        \includegraphics[height=1.8in]{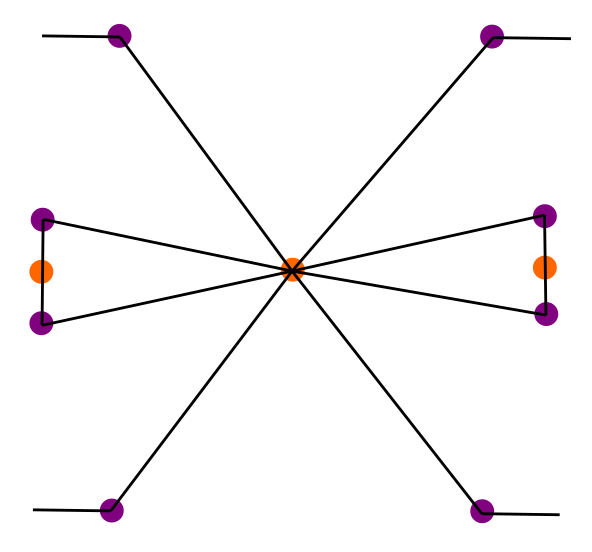}
        }
        \subfigure[$\pi<\vartheta<0$]{
        \includegraphics[height=1.8in]{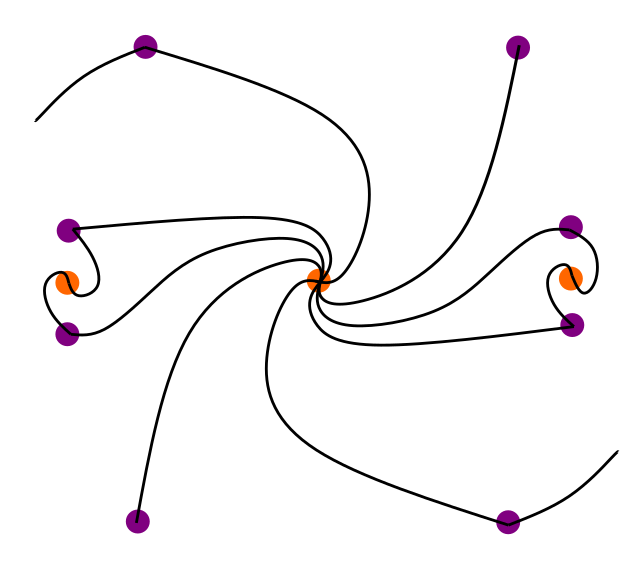}
        }
    \caption{Sketches of the directional Stokes graphs for generic values $\eta \neq \zeta$ with $0<E<E_{c}$. Poles and are shown in orange and zeros in purple.    We have shown one fundamental domain per Figure, but note that the trajectories can of course cross into neighbouring domains. In particular, in (a) and (c), horizontal and vertical trajectories form saddles with the images of zero in the next domain.}
    \label{fig:stokesphenomena}
\end{figure}

\begin{figure}[tb]
    \centering
        \subfigure{
        \includegraphics[height=2.5in]{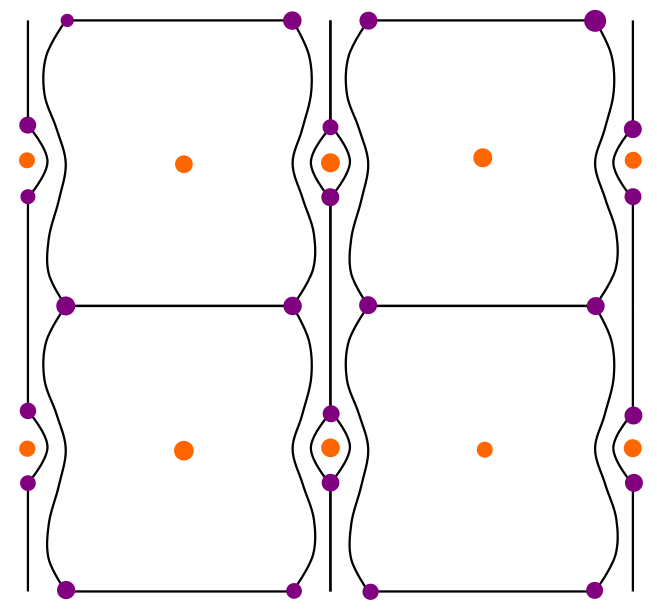}
        }
        \subfigure{
        \includegraphics[height=2.5in]{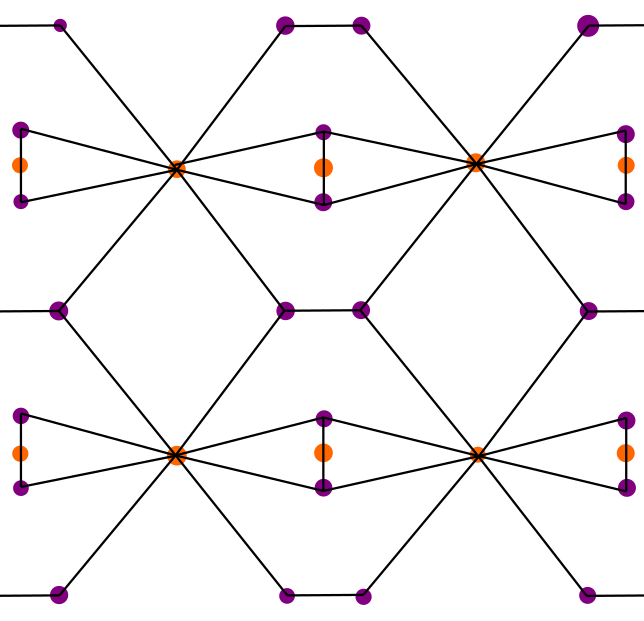}
        }
    \caption{The lattice formed in 4 fundamental domains  by saddles in the Stokes graph with  $\eta \neq \zeta$ with $0<E<E_{c}$  for  $ \vartheta=0$ (left) and $\vartheta= \pi$ (right). }
    \label{fig:phi0stokesline}
\end{figure}

 Looking at $E<E_c$  we sketch the directional Stokes graphs in Figure \ref{fig:stokesphenomena} and  \ref{fig:phi0stokesline}. In complete agreement with the discussion of the   Borel pole structure,  we see two directions $\vartheta = 0 ,\pi$ for which the graphs contain saddles and over which the graphs undergo flip transitions.

\begin{figure}[tb]
   \centering
      \subfigure[$\vartheta=0 $]{
        \includegraphics[height=1.8in]{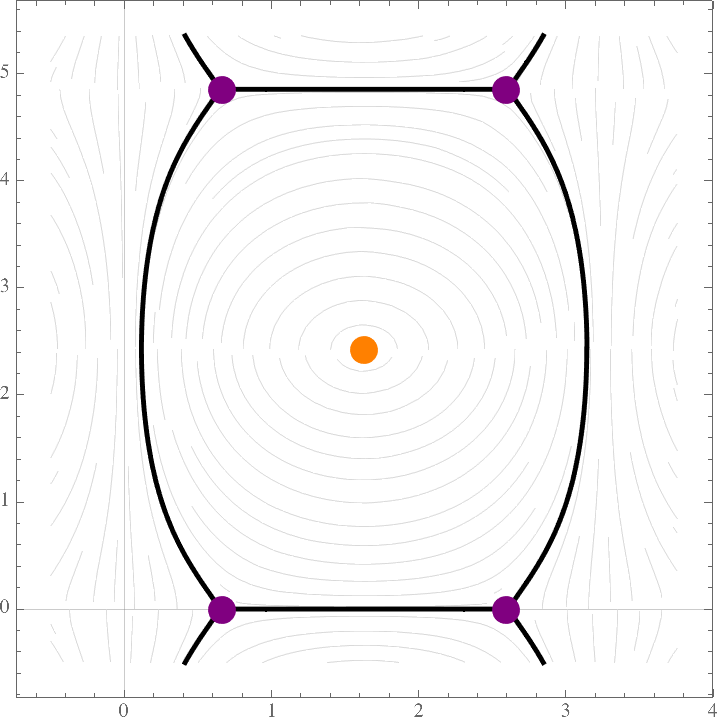}
        }
      \subfigure[$\vartheta=\frac{\pi}{2}$]{
      \includegraphics[height=1.8in]{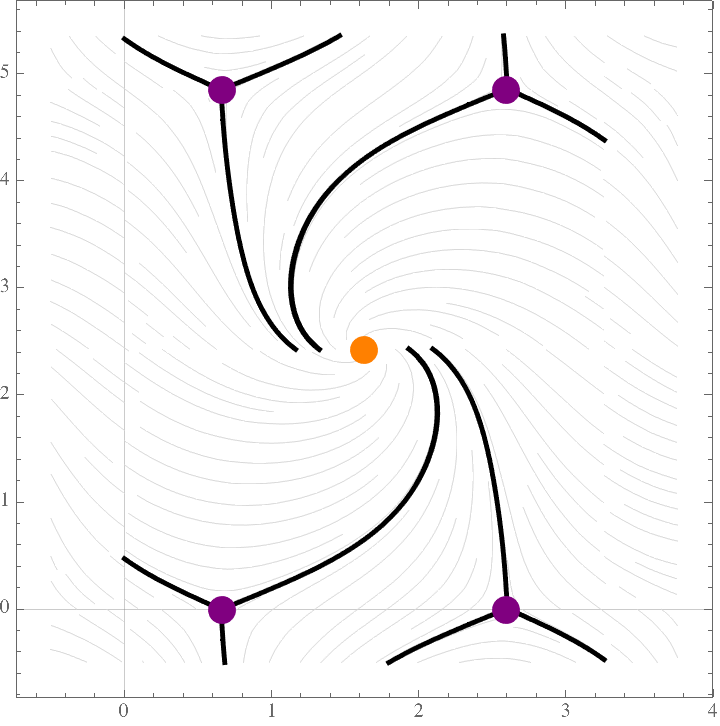}
      } \\
      \subfigure[$\vartheta=\pi$]{
     \includegraphics[height=1.8in]{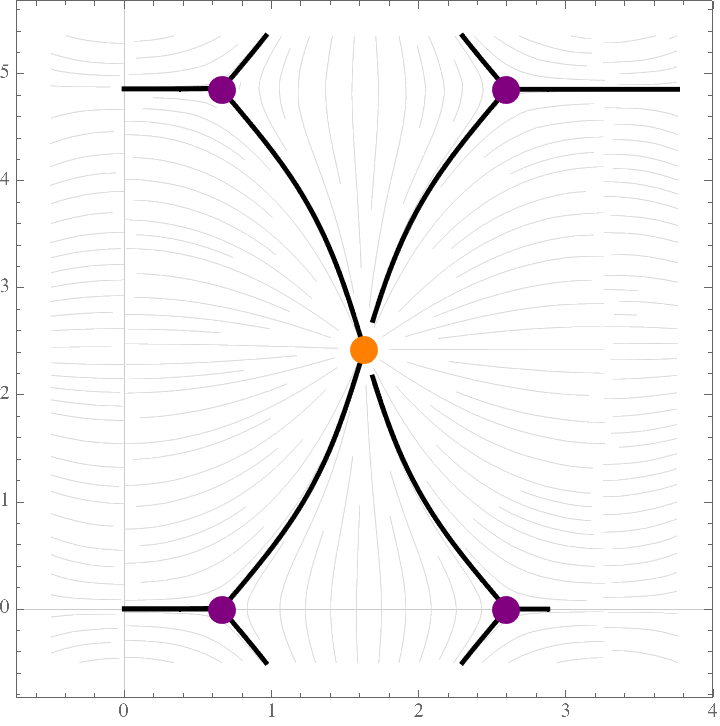}
      } 
     \subfigure[$\vartheta=\frac{3\pi}{2}$]{
    \includegraphics[height=1.8in]{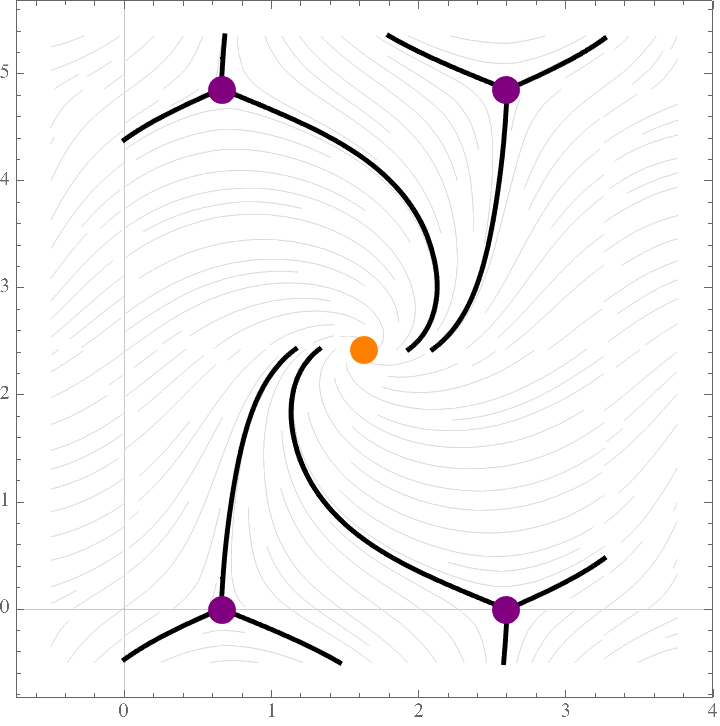}}
 \caption{Here we plot the Stokes graphs in the directions $\vartheta = 0, \frac{\pi}{2}, \pi, \frac{3\pi}{2}$.  Here we display the critical line $\varkappa=0.2$ and we set $E=0.4$.  Poles are shown in orange and zeros in purple.   As the direction crosses $\vartheta = 0,\pi$ saddles manifest themselves and a flip mutation is seen.  }
   \label{fig:stokesphenomena1}
\end{figure}

 In the critical case of $\eta =\zeta$ an important  modifications occurs.  The two zeros on the imaginary axis coincide at, and annihilate against,  the pole at $w = i\mathbb{K}(m') $   leaving just two remaining zeros   situated on the real axis (for $E<E_c$) and the double pole at the centre of the fundamental domain.  This topology change is the graphical reason behind the jump in critical line behaviour such that the complex 1-uniton makes no contribution.  In this case however still saddles persist in  the two directions $\vartheta = 0 ,\pi$ as shown in Figure  \ref{fig:stokesphenomena1}.
 
\section{Connection to $\mathcal{N}=2$ Seiberg-Witten Theory} \label{sec:SW}

From the WKB treatment above we saw that Stokes graphs are a elegant way of visualising the structure of the Borel plane.   In a seminal work, Gaiotto, Moore and Neitzke \cite{gaiotto2010four, gaiotto2013wall} explained how the same structure plays a crucial role in the spectrum of BPS states of $d=4$, $\mathcal{N}=2$ gauge theories.   The essential idea (going back to the construction of Klemm et al \cite{Klemm1996bj} for $SU(2)$ theories that will be relevant  here) is that BPS states on the Coulomb branch associated with M2 branes stretched on a curve $\gamma$ between sheets of the M5 brane carry charge $Z =\frac{1}{\pi} \int_\gamma \lambda_{SW} $ but have mass given given by $M =\frac{1}{\pi} \int_\gamma |\lambda_{SW}| $.  The BPS bounds is saturated providing that $\lambda_{SW}$ has constant phase along the curve, i.e. $\lambda_{SW}\cdot \partial_t = e^{i \vartheta}$.  For certain values of $\vartheta$  these Stokes curves become finite and start and end at the zero's of $\lambda_{SW}$  and the BPS state, in this case a hypermultiplet,  has finite mass. 

It is natural to wonder if the integrable theories we consider here have an analogue description in gauge theory. Stated more precisely, we are led to ask if there is a gauge theory for which the  quadratic differential obtained as the square of the Seiberg Witten differential,  $\phi_{SW} = \lambda_{SW} \otimes \lambda_{SW} $,  matches that defined by the quantum mechanics arising from the reduction of the two-dimensional non-linear sigma model we consider.   

This has been shown to be the case first for the undeformed PCM on $S^3$. The corresponding quantum mechanics had a trigonometric Mathieu potential \cite{cherman2015decoding}  and for which the corresponding gauge theory is $SU(2)$, $N_f =0$. The resurgent structure of the Schr\"odinger equation corresponding to this quadratic differential was studied in \cite{dunne2017wkb}. An interesting connection with the TBA equations of the corresponding integrable SG field theory was made by \cite{grassi2019non}.

For   the single parameter $\eta$-deformed theory it was shown in \cite{demulder2016resurgence} that the quantum mechanics has a Whittaker-Hill (or double sine-Gordon)  potential and the corresponding gauge theory is  $SU(2)$, $N_f =2$ (in the first realisation of \cite{Gaiotto:2009ma}) with equal masses for the flavours.  In this scenario an interesting connection is made between the masses of the flavours and the RG invariant combination of tension and deformation parameter  parameter, namely that $M=m_1=m_2  \propto \sigma_{\eta} = \frac{1}{ t \eta} $. 

Here we shall provide a similar correspondence for the potential with two deformation parameters. We shall do so in two related ways, first linking to an $SU(2)\times SU(2)$ quiver theory and secondly linking to $SU(2)$ $N_f = 4$ theory.  To begin it is convenient to understand the form of the potential of the quantum mechanics considered above as a generalised Lam\'e potential. 

\subsection{The Generalised  Lam\'e Potential }

First, we will rewrite the potential, $V(w)$,  in terms of Weierstrass functions $\wp(z)$. We shall denote the periods of the Weierstrass function as $2\omega_1$ and $2\omega_2$. The elliptic invariants are given by $g_2$, $g_3$ and the constants $e_i$ denote the roots of corresponding cubic. The modular parameter of the torus $\tau=\omega_2/\omega_1$ is given by
\begin{equation} \label{eq:tau}
    \tau=\frac{i\mathbb{K}(m')}{\mathbb{K}(m)}\,,
\end{equation}
where $m = \frac{e_2 - e_3}{e_1 - e_3} $  is the Jacobi elliptic parameter and $m' = 1- m $. They are related to the invariant cross-ratio as 
 \begin{equation} \label{eq:crossratio}
    \omega= \frac{(e_3- e_1)^2 - 9 e_2^2}{(e_3 -e_1)^2} = 4 m m' \, . 
\end{equation}
In terms of\footnote{At this point we are using $z$ as a coordinate on the torus which we trust will not be confused with the earlier usage as $g^2$.}  $w=z\sqrt{e_1-e_3}$, the potential can then be rewritten as
\begin{equation} \label{eq:weierstrassform}
    V(z) = (e_1 - e_3) \frac{\frac{1}{3} ((e_1 - e_2)(e_1 - e_3) - 3 e_1) + \wp(z)}{(e_2 - \wp(z))(e_3 - \wp(z))} \, ,
\end{equation}
when the elliptic moduli are fixed by

\begin{equation} \label{eq:p-invariants}
\begin{aligned}
  e_1  = \frac{2-m}{m'} (1+ \chi_-^2 ) \, , \quad 
    e_2  = \frac{2m-1}{m'}  (1+ \chi_-^2 ) \, , \quad 
    e_3   = -\frac{1+m}{m'} (1+ \chi_-^2 ) \, . 
\end{aligned}
\end{equation}
In particular, this means a relation between the $z$ and $w$ coordinate 
\begin{equation}\label{eq:ztow}
 w= z \sqrt{  (1+\chi_+^2)}\,  .  
\end{equation} 
We can further rewrite Equation \eqref{eq:weierstrassform} as a generalised Lam\'e potential 
\begin{equation} \label{eq:generalisedlame}
    V(z) = h + \sum _{i=0}^3 c_i \wp(z+\omega_i)\, ,
\end{equation}
where
\begin{equation}
\begin{aligned}
    h &= \frac{\left(e_3-e_1\right) \left(e_1^2-3 e_1+2 e_2 e_3\right)}{3 \left(e_2-e_3\right){}^2} \\ 
    c_0 &= c_1  = 0 \\
    c_2 &= \frac{\left(e_1-e_3\right) \left(-e_1+e_3+3\right)}{3 \left(e_2-e_3\right){}^2} \\
    c_3 &= \frac{\left(e_1-e_2-3\right) \left(e_3-e_1\right)}{3 \left(e_2-e_3\right){}^2}\, ,
\end{aligned}
\end{equation}
and $\omega_3 = \omega_1+\omega_2$ and $\omega_0=0$. 

The Weierstrass form of the potential is also revealing about the nature of the $(\eta, \zeta)\rightarrow (- \eta, \zeta)$ transformation, which was given an interpretation in the PCM context in Equation \eqref{eq:PCMflipeta}. From Equation \eqref{eq:p-invariants} it is clear that in the Weierstrass description this corresponds to interchanging $e_2 \leftrightarrow e_3$. This has the effect of rescaling the coordinate by $z\rightarrow \sqrt{\frac{e_1-e_3}{e_1-e_2}} z$ and an overall scaling of the potential $V\rightarrow \frac{e_1-e_2}{e_1-e_3} V$. Such transformations are easily absorbed in a rescaling of the coupling constant and do not alter the physics. 

Let us study two special cases in this formulation. First, the critical line $\eta=\zeta\equiv \varkappa$ corresponds to $  e_1 - e_2 - 3 = 0$, which implies that $c_3=0$. In this situation, equation \eqref{eq:weierstrassform} simplifies to
\begin{equation}
    V(z)=\frac{e_1-e_3}{\wp(z)-e_2}.
\end{equation}
However, it is perhaps more natural to think about a co-critical point where $\eta=-\zeta$.  In this situation the potential reduces to
\begin{equation}
    V(z) = \frac{1}{3} - \frac{1}{12\eta^2}(1-\wp(z)),
\end{equation}
which is similar to the Lam\'e potential studied in \cite{basar2015resurgence}, identifying $-4\eta^2 = k^2$. We know that this potential governs the WKB curve of the vacuum structure of $SU(2)$ $\mathcal{N}=2^\ast$ Seiberg-Witten theory, which is a mass deformation of an $\mathcal{N}=4$ theory \cite{dorey2002exact}.   

The $\zeta\rightarrow 0$ limit is quite delicate in this description as can immediately be seen from the fact that the Jacobi elliptic parameter $m \rightarrow 0 $ and correspondingly the modular parameter diverges as $\tau \rightarrow i \infty$.  In particular, in this regime not all $e_i$ are distinct which is forbidden in the generic Weierstrass setting, because the determinant
\begin{equation}
    \Delta = g_2^3 - 27 g_3^2 = 11664(1 + \chi_+^2)^2(1 + \chi_-^2)^2(\chi_-^2 - \chi_+^2)^2
\end{equation}
of the polynomial $4t^3 - g_2 t - g_3$ vanishes. However, if we consider the case in which we blow up one of the periods $\omega_2\rightarrow \infty$, we see that
\begin{equation}
\begin{aligned}
    g_2 &= 2 \times 60 \sum_{n=1}^\infty \frac{1}{(n \omega_1)^4} = \frac{4 \pi^4}{3 \omega_1^4}, \\
    g_3 &= 2 \times 140 \sum_{n=1}^\infty \frac{1}{(n \omega_1)^6} = \frac{8 \pi^6}{27 \omega_1^6},
\end{aligned}
\end{equation}
where we have used that the Riemann $\zeta$-functions takes the following values: $\zeta(4)=\frac{\pi^4}{90}$ and $\zeta(6)=\frac{\pi^6}{945}$. This leads to $\Delta=0$, thus we may identify the two limits. To conclude, in this regime, $\omega_2\rightarrow \infty$ and we break the finite double periodicity, i.e. the length of one side of the torus has a pole. Moreover, $e_2$ and $e_3$ are not distinct anymore. In particular this leads to a pole in $c$, $c_2$ and $c_3$.

\subsection{$N_f=2$ Elliptic $SU(2)\times SU(2)$ Quiver Theory}
We now consider the $SU(2)\times SU(2)$ quiver gauge theory with two flavours. To extract the relevant differential, we employ Witten's string theory construction of the Seiberg-Witten theories \cite{seiberg1994electric, witten1997solutions}. On the M-theory side, we compactify along the $x^6$ and the $x^{10}$ direction which creates a base torus $E$ with modular parameter $\tau$ which is the base Riemann surface. Let $z$ be a coordinate on the torus. The Seiberg-Witten curve takes the form \cite{witten1997solutions}
\begin{equation} \label{eq:quiverF}
    F(v, z) = (v-v_1(z))(v-v_2(z)) \equiv 0 \,,
\end{equation}
in which roots in $v$ are the locations of the D4-branes. Let the locations $z_1$ and $z_2$ of the 2 NS5-branes be marked points on the base torus. We require that $v_i(z)$ has a pole at $z_i$ with residue $m_i$ parametrising the masses of the hypermultiplets. In addition, we allow a fibration of the $v$-space over the base torus $E$ around $z=0$
\begin{equation}
\begin{aligned}
    z  \rightarrow z  + 2\pi R, \qquad      v\rightarrow v +m_1+m_2 \, . 
\end{aligned}
\end{equation}
Using the double periodicity and the singularity structure of $v_i(z)$, we can completely fix the form of the coefficients:
\begin{equation}
    \begin{aligned}
    v_1(z)+v_2(z) &= m_1\zeta(z-z_1) + m_2\zeta(z-z_2) - (m_1 + m_2) \zeta(z) + c_0 \\
    v_1(z)v_2(z) &= \frac{1}{4} (m_1+m_2)^2 \wp(z ) + B(\zeta(z-z_1) - \zeta(z-z_2)) + C,
    \end{aligned}
\end{equation}
where $B$, $C$ and $c_0$ are some moduli and $\zeta(z)$ is the quasiperiodic Weierstrass function defined by $\zeta'(z) = -\wp(z)$ such that the combination $\sum_i a_i \zeta(z'-z'_i) = 0$ is doubly periodic if $\sum_i a_i = 0$ and has a simple pole around $z=0$ with a residue of $1$.  

The Seiberg-Witten differential is given by
\begin{equation}
    \lambda_{SW} = \hat{v} dz \, , 
\end{equation}
in which 
\begin{equation}
    \hat{v} = v- \frac{1}{2}(v_1(z )+ v_2(z )) \, .
\end{equation}
We can now use the definition of the curve equation \eqref{eq:quiverF} to determine that  
\begin{equation}
    \hat{v}^2 = \frac{m_1^2}{4} \wp (z - z_1) + \frac{m_2^2}{4} \wp (z- z_2) + u_-(\zeta(z-z_1) - \zeta(z-z_2)) + u_+, 
\end{equation}
where
\begin{equation}
    u_\pm = \langle \text{Tr} \Phi_1^2 \pm \text{Tr} \Phi_2^2 \rangle.
\end{equation}
are Coulomb branch moduli.

We would like to match the  quadratic differential  
\begin{equation}
\phi_{SW} = \lambda_{SW}  \otimes \lambda_{SW} 
\end{equation}
  to that of Schrodinger system given in Eq. \eqref{eq:quaddiffsch}.
 
By inspection this identification is achieved when the coordinates $z$ and $w$ are related exactly as described in Equation \eqref{eq:ztow}
  
and when the Coulomb branch parameter $u_-=0$ with the locations of the five branes are fixed to the half-periods $z_1 = \omega_2$ and $z_2 = \omega_3$.  To complete the identification we must match the hypermultiplet masses to the parameters of the Schrodinger the system and the result is quite striking; we find that they are directly given by the parameters that control the underlying quantum group symmetry of the YB deformed PCM
\begin{equation}
\sigma_\eta = \frac{m_1 + m_2}{2 \nu } \ , \quad \sigma_\zeta = \frac{m_1 - m_2}{2 \nu } \,  ,
\end{equation} 
in which we have reinstated   chemical potential and compactification radius in the combination $\nu^2 = \frac{4\xi^2}{L^2}$. 
The final Coulomb branch parameter, $u_-$, is related linearly to the energy of the Schrodinger system (the exact coefficients do not appear very insightful at this stage).

The two gauge couplings of the quiver are given in terms of the torus modular parameter by \cite{dorey2002exact}
\begin{equation}
\begin{aligned}
z_1 - z_2 = \tau_1 = \frac{4\pi i }{g_1^2} + \frac{\theta_1}{2\pi } \, , \\
\quad \tau- (z_1 - z_2) = \tau_2 = \frac{4\pi i }{g_2^2} + \frac{\theta_2}{2\pi } \, .
\end{aligned}
\end{equation}
 Now since the roots of the elliptic curve are all real, and the five branes are located at the half periods we concluded that $z_1- z_2= \omega_1\in \mathbb{R}$ and that $\tau$ is pure imaginary.  As a result we see that the coupling $g_1 \rightarrow \infty$ with $\frac{4\pi i}{g^2_2} = \tau$  finite whist the theta angles obey $\theta_1= -\theta_2= 2 \pi \omega_1$.   In this language the critical line is approached in the limit that the mass  $m_2 \rightarrow 0$. 
 
\subsection{$N_f=4$ $SU(2)$ Theory }
In \cite{tai2010triality, tai2010uniformization}, Ta-Sheng Tai recovers curves with the form \eqref{eq:generalisedlame} in some SW curve setting via a duality to the Heun equation. We will now show how one obtains the appropriate SW curve.

Let us now connect the Schrodinger system obtained above to the quadratic differential for the Seiberg-Witten curve of 4d $\mathcal{N}=2$ supersymmetric Yang-Mills with  $SU(2)$ gauge group and $N_f = 4$ flavours.   The theory is specified by a Coulomb branch parameter $u$, the four flavour masses $m_i$ and the marginal coupling $\tau_{YM} = \frac{4 \pi i}{g_{YM}^2} + \frac{\theta}{2\pi}$. The UV curve of the theory is given by 
\begin{equation}
F(t, v) =  t^2 ( v- m_1) (v-m_2) + b (v^2 -u)t + c( v-m_3) (v-m_4) = 0  \,,
\end{equation}
in which the parameters $b$ and $c$ are related to the elliptic invariants $g_2 = \frac{1}{12}\left( b^2 -3 c \right)$ and $g_3 = \frac{1}{432} \left( 9 b c - 2 b^3 \right)$ and moreover the roots of the polynomial $(c + b t + t^2 )  = (t-t_+)(t -t_-)$ can be understood as relating to the M-theory lifting of the NS5 branes in the IIA picture.  The SW differential is obtained as 
\begin{equation}
\lambda_{SW} = \hat{v} \frac{dt}{t} \,, \quad \hat{v} = v - \frac{c(m_3 + m_4) + (m_1 + m_2) t^2 }{2 (c + b t + t^2 ) }\,,
\end{equation}
in which the shifted quantity $\tilde{v}$ is used to factor out the overall U(1) degree of freedom.   We can define a change of coordinates
\begin{equation}
    t = 4 \wp(z) - \frac{b}{3} \, , 
\end{equation}
by which we can bring the quadratic differential to the form
\begin{equation}
\phi_{SW} = \lambda_{SW}  \otimes \lambda_{SW}  =  \left[ h  + \sum_{i=0}^3 c_i  \wp(z - w_i) \right] dz \otimes dz \,,
\end{equation}
where $w_i$ with $i=1\dots3$ are the half-periods and $w_0 = 0$.  The coefficients $c_i$ in this expression are slightly unedifying expressions depending to $t_\pm$, the mass parameters and for $h$ also on $u$,  but in particular  $c_0 = (m_1 - m_2)^2$ and $c_1 = (m_3 -m_4)^2$.   

We would like to match this to Schrodinger system of eq. \eqref{eq:quaddiffsch}.
 
Using the relation between coordinates $z$ and $w$ given by eq.  \eqref{eq:ztow}, we find that the matching is achieved with setting the flavor masses pairwise equal
\begin{equation}
       m_1 =m_2= M  \, , \quad     m_3 =m_4= \tilde{M} \, ,
\end{equation}
and relating them to the bi-Yang-Baxter parameters according to 
\begin{equation}
    \sigma_\eta = \frac{2}{\nu}\left(\tilde{M}- M\right) \, , \quad  \sigma_\zeta =  \frac{2}{\nu}\left(\tilde{M}+ M\right) \frac{1+m'}{m}\, . 
\end{equation}

To complete the matching we also need to relate the parameter on the Coulomb branch to the energy of the Schrodinger system which is achieved with
\begin{equation}
    E= \frac{\nu^2 (1+m')(M^2 - u)}{M^2 (1+ m' + m'^2) + 2 m' M \tilde{M} - m' \tilde{M}^2 } \, . 
\end{equation}
 
 To close this section let us remark that along the critical line, parametrised by $\varkappa= \eta =\zeta $, we find very particular behaviour in the matching.  
 First we can note that the masses are related via $\tilde{M} =- (1+ 4\varkappa^2) M $, Using \eqref{eq:tau}, we find that the elliptic modulus of the torus is $\tau = i \mathbb{K}(\frac{1}{1+4\varkappa^2})/\mathbb{K}(\frac{4\varkappa^2}{1+4\varkappa^2})$.  When $\varkappa = \frac{1}{2\sqrt{3}}$ we encounter a point for which the complex uniton action has exactly twice the magnitude as the real uniton action, and at this point we have a relation between the masses  $\tilde{M}= -\frac{4}{3} M $.  At this point we have the $\tau = i \mathbb{K}(\frac{3}{4})/\mathbb{K}(\frac{1}{4})  $. Continuing to increase the deformation we arrive at $\varkappa = \frac{1}{2}$ when the complex uniton has the same magnitude as the real uniton for which we find $\tau=i$ and $\tilde{M} = - 2M $. At $\varkappa= \frac{\sqrt{3}}{2}$, for which $|S_I| = 2 |S_{CI}|$, we have $\tilde{M} = - 4M$ and $\tau = i \mathbb{K}(\frac{1}{4})/\mathbb{K}(\frac{3}{4})  $. 
 
 We remark that the previously discussed duality in \cite{basar2013resurgence} that sends $m\rightarrow m'$, results in an S-duality sending $\tau\rightarrow\frac{-1}{\tau}$. In addition we observe that $\varkappa=\frac{1}{2\sqrt{3}}$ and $\varkappa=\frac{\sqrt{3}}{2}$ are dual under this transformations, whereas $\varkappa=\frac{1}{2}$, corresponding to $\tau=i$, is self-dual.

\section{Conclusion and Outlook}
We thus conclude our study of the bi-Yang-Baxter deformed $SU(2)$ PCM. We have seen that the model harbours two types of solutions which we have dubbed the real and the finite uniton, both with a quantised finite action. By employing an adiabatic compactification \cite{demulder2016resurgence, cherman2015decoding} we obtained a reduced quantum mechanics whose non-perturbative behaviour is dominated by finite action configuration derived from the unitons. Moreover, we were able to find an $\mathcal{N}=2$ Seiberg-Witten theory that gives rise to the same WKB curve as that of our reduced quantum mechanics. By introducing a new example into the framework of resurgence, we hope to expand the non-perturbative discourse. In particular, we believe the complex saddle point in our system might elucidate more advanced structures of resurgence. Possible future directions of study could include:

\begin{itemize}
 
\item  In \cite{dunne2014uniform}, Equation (108), it was observed that the following relation holds for both the double well and the Sine-Gordon quantum mechanics
\begin{equation} \label{eq:108}
    \frac{\partial E}{\partial B}=-\frac{g^2}{S_I}\left( 2B + g^2\frac{\partial A}{\partial g^2} \right)\,.
\end{equation} 
$A(B, g^2)$ is a function that appears in the global boundary conditions of the uniform WKB and is determined by $u(\theta_\text{midpoint})$. It was first introduced by \cite{zinn2004multi1, zinn2004multi2} and for the Sine-Gordon model it reads
\begin{equation}
    A_{SG}(B, g^2) = \frac{4}{g^2} -\frac{g^2}{2}\left(B^2 + \frac{1}{4}\right) -\frac{g^4}{8}\left(B^3 + \frac{B}{4} \right) + \mathcal{O}(g^6)
\end{equation}
It was shown in \cite{basar2015resurgence} that this relation in the undeformed limit is a consequence of a generalised Matone's relation on the gauge theory side. In addition, \cite{dunne2014uniform} attribute a great deal of importance to this relation as it explain a lot of the resurgent behaviour. However, in the compactified Yang-Baxter deformed models studied in this paper, there does not appear to be a related identity. It would be interesting to understand how the relation \eqref{eq:108} can be modified in systems with a real and a complex saddle point. 

\item   In Figures \ref{fig:doublerealtwisteduniton} and \ref{fig:criticalcomplextwisteduniton} we observed that the uniton with twisted boundary conditions fractionates into 2 separate lumps. In the undeformed model \cite{cherman2015decoding}, it was shown how solutions for these individual constituent fractons can be constructed. These are not exact solutions to the equations of motions, but are rather \textit{quasi}-solutions, meaning that the equations of motion can be satisfied with parametrically good accuracy in some limit of the moduli $\lambda_i$. Critically, it was shown that the amplitude of a fracton-anti-fracton event carries an ambiguity that precisely cancels the Borel-resummation ambiguity of the perturbative sector given by Equation \eqref{eq:leadingambiguity}. It would be an impressive check for the resurgence programme to extent this analysis to the YB-deformed PCM which also harbours a complex uniton fractionation.

\item The  Thermodynamic Bethe Ansatz (TBA) is a powerful technique to study integrable field theories that exploits the exact scattering matrix of the model, for recent introductions see \cite{bombardelli2016s, tongeren2016introduction, levkovich2016bethe}. As was mentioned before, the scattering matrix of the $SU(2)$ YB deformed PCM, given by \eqref{eq:smatrix1} and \eqref{eq:smatrix2}, first appeared in \cite{fateev1996sigma}. In the TBA framework of $O(N)$ integrable sigma models, Volin \cite{volin2011quantum, volin2010mass} used integral resolvents to recover the mass gap of the theory and find an asymptotic expansion of the energy and particle densities. This approach was further expanded to study renormalon ambiguities in Gross-Neveu and PCM models by \cite{marino2019new, marino2019renormalons}. Moreover, it was shown that the resurgent structure of the Sine-Gordon quantum mechanics can be reinterpreted in terms of TBA equations \cite{grassi2019non}. It would be very interesting to extend these ideas to the bi-YB deformed PCM.

\item  The apparent connection between the 2d integrable theory and the $\mathcal{N}=2$ gauge theories provokes a number of questions.  First, is this simply coincidental? If not, is there a more fundamental way to make this connection we find (that doesn't required picking particular coordinate, adiabatic reduction etc.)?  Second, what significance do dualities exhibited on the gauge theory side hold for the integrable models? Third, how do the integrable Hitchin systems associate to the gauge theory \cite{dorey2002exact} compare to the integrable structures of the PCM (e.g. Maillet-brackets and twist funcitons)  and its deformations? If such a connection can be made it seems likely it is via  the use of affine Gaudin models \cite{vicedo2017integrable, zotov20111+1}.  A final  intriguing question here is to understand if the wall-crossing phenomena seen in the gauge theory have an interpretation and implication for the two-dimensional deformed sigma-model considered in this paper.

\item  In a recent series of papers \cite{costello2017gauge, Costello2018gauge, costello2019gauge}, it was shown how one can construct 2d integrable field theories from a certain four-dimensional holomorphic Chern-Simons type theory. It was later shown by \cite{delduc2020unifying} that Yang-Baxter deformations can also be incorporated in this framework. A direct question is to understand the more general resurgent structure of integrable 2d field theories from the perspective of this 4d gauge theory.  A rather concrete first question would be to understand the significance of the uniton, and its cousin in complex field space, within the gauge theory.  A potential root here would be to exploit the connection   with affine Gaudin models   established by \cite{vicedo2019holomorphic}.

\end{itemize}

\acknowledgments

We would like to thank S. Prem Kumar, Jan Troost, Daniele Dorigoni, Ben Hoare, Philip Glass and Saskia Demulder for useful conversations on this and related topics.

 DCT is supported by The Royal Society through a University Research Fellowship {\em Generalised Dualities in String Theory and Holography} URF 150185 and in part by STFC grant ST/P00055X/1 and in part by the ``FWO-Vlaanderen'' through the project G006119N and by the Vrije Universiteit Brussel through the Strategic Research Program ``High-Energy Physics'’.  LS is supported by The Royal Society through grant RGF\textbackslash R1\textbackslash 180087 Generalised Dualities, Resurgence and Integrability. 

\appendix
\section{Evaluating Uniton Actions} \label{ap:intaction}

Here, we dwell upon the following observation. When integrating the action, in both the real and the complex untion we switch to $w=f(z)$ coordinates. We transition to polar coordinates $w=r e^{i \theta}$. Because there is no $\theta$-dependence we integrate it out. When considering the real uniton we make the substition $\rho=r^2-1$, for the complex uniton we substitute $\rho=-r^2-1$. Remarkably, in both cases we obtain the following integrand:
\begin{equation}
    g(\rho)=\frac{-2(2+\rho)^2}{(4+4\rho+(1+(\zeta+\eta)^2)\rho^2)(4+4\rho+(1+(\zeta-\eta)^2)\rho^2)},
\end{equation}
with the only difference that for the real uniton we integrate $\rho$ from positive infinity to $-1$, for the complex uniton, we integrate $\rho$ from $-1$ to negative infinity. 

On the interval $(-1, \infty)$, we can construct a continuous (i.e. without branch cuts) anti-derivative:
\begin{equation}
    \frac{(\zeta+\eta)\text{arctan}(\frac{(\zeta+\eta)\rho}{2+\rho})- (\zeta-  \eta)\text{arctan}(\frac{(\zeta-\eta)\rho}{2+\rho})}{4\zeta\eta}.
\end{equation}
This can be used to evaluate the real uniton action \eqref{eq:realunitonaction}. It cannot be used to compute the complex uniton action because it is in particular discontinuous at $\rho=-2$. On the interval ($-\infty, -1)$, we can use
\begin{equation}
        \frac{(\zeta+\eta)\text{arccot}(\frac{(\zeta+\eta)\rho}{2+\rho})- (\zeta-  \eta)\text{arcot}(\frac{(\zeta-\eta)\rho}{2+\rho})}{4\zeta\eta}
\end{equation}
as an antiderivative to compute the complex uniton action.

Jointly, they can be used to reconstruct $\int_{-\infty}^\infty g(\rho)d\rho$. This integral can be easily computed using the Cauchy residue theorem. The integrand vanishes in all direction at infinity, and it has 4 poles, two in the upper half plane and two in the lower half plane. This yields (for $\eta,\zeta\in\mathbb{R}$)
\begin{equation}
    \int_{-\infty}^\infty g(\rho)d\rho = \frac{\pi}{4\zeta\eta}(|\zeta-\eta|-|\zeta+\eta|).
\end{equation}
Via the above explanation or by using the identity $2\,x\, \text{arctan}(x)+2\,x\,\text{arccot}(x) = \pi |x|$, we thus obtain the following relation between the complex and the real uniton actions for $\eta,\zeta\in\mathbb{R}$
\begin{equation}
    S_{CI}(\zeta,\eta)-S_I(\zeta,\eta)= \frac{\pi(1+(\zeta+\eta)^2)}{4\zeta\eta}(|\zeta-\eta|-|\zeta+\eta|).
\end{equation}

\bibliography{resurgencebib}

\end{document}